\documentclass[letterpaper,10pt]{article}
\usepackage[utf8]{inputenc}
\usepackage{graphicx}
\usepackage{graphics}
\usepackage{amsmath}
\usepackage{amssymb}
\usepackage{amsthm}
\usepackage{xcolor}
\usepackage{slashed}
\usepackage{bbm}
\usepackage{bm}
\usepackage{array}
\usepackage{authblk}
\usepackage{dcolumn}
\usepackage{enumitem}
\usepackage[mathlines]{lineno}
\usepackage{color}
\usepackage{afterpage}
\usepackage{hyperref}
\usepackage{appendix}
\usepackage{mathrsfs}

\newcommand{\nc}{\newcommand}
\nc{\ev}{\mathrm{eV}}
\nc{\mev}{\mathrm{MeV}}
\nc{\gev}{\mathrm{GeV}}
\nc{\kev}{\mathrm{keV}}
\nc{\tev}{\mathrm{TeV}}
\nc{\pev}{\mathrm{PeV}}
\nc{\eev}{\mathrm{EeV}}
\nc{\zev}{\mathrm{ZeV}}
\nc{\tv}{\mathrm{TV}}

\usepackage{vmargin}
\setmargins{2.5cm}      
{1.5cm}                 
{15.5cm}                
{23.42cm}               
{10pt}                  
{1cm}                   
{0pt}                   
{2cm}                   

\title{Cosmic ray spectrum of protons plus helium nuclei between $6 \, \tev$ and $158 \, \tev$ from HAWC data} 
\date{}

\author[1]{A.~Albert} 
\author[2]{R.~Alfaro}
\author[3]{C.~Alvarez}
\author[2]{J.R.~Angeles Camacho}
\author[4]{J.C.~Arteaga-Vel\'{a}zquez\thanks{Corresponding author: juan.arteaga@umich.mx}}
\author[5,6]{K.P.~Arunbabu}
\author[2]{D.~Avila Rojas}
\author[7]{H.A.~Ayala Solares}
\author[2]{E.~Belmont-Moreno}
\author[8]{C.~Brisbois}
\author[3]{K.S.~Caballero-Mora}
\author[9]{T.~Capistr\'{a}n}
\author[10]{A.~Carrami\~{n}ana}
\author[11]{S.~Casanova}
\author[4]{U.~Cotti}
\author[12]{J.~Cotzomi}
\author[13,14]{E.~De la Fuente}
\author[10]{R.~Diaz Hernandez}
\author[15]{M.A.~DuVernois}
\author[16]{M.~Durocher}
\author[13]{J.C.~Díaz-V\'{e}lez}
\author[2]{C.~Espinoza}
\author[9]{N.~Fraija}
\author[17]{J.A.~García-Gonz\'{a}lez}
\author[9]{F.~Garfias}
\author[9]{M.M.~Gonz\'{a}lez}
\author[8]{J.A.~Goodman}
\author[16]{J.P. Harding}
\author[18]{B.~Hona}
\author[19]{D.~Huang}
\author[3]{F.~Hueyotl-Zahuantitla}
\author[19]{P.~H\"{u}ntemeyer}
\author[9]{A.~Iriarte}
\author[20]{V.~Joshi}
\author[18]{D.~Kieda}
\author[1]{G.J.~Kunde}
\author[5]{A.~Lara}
\author[9]{W.H.~Lee}
\author[2]{H.~Le\'{o}n Vargas}
\author[21]{J.T.~Linnemann}
\author[9]{A.L.~Longinotti}
\author[22]{G.~Luis-Raya}
\author[1]{K.~Malone}
\author[12]{O.~Martinez}
\author[23]{J.~Martínez-Castro}
\author[24]{J.A.~Matthews}
\author[25]{P.~Miranda-Romagnoli}
\author[4]{J.A.~Morales-Soto}
\author[12]{E.~Moreno}
\author[7]{M.~Mostaf\'{a}}
\author[11]{A.~Nayerhoda}
\author[26]{L.~Nellen}
\author[18]{M.~Newbold}
\author[25]{R.~Noriega-Papaqui}
\author[27]{N.~Omodei}
\author[22]{E.G.~P\'{e}rez-P\'{e}rez}
\author[28]{C.D.~Rho}
\author[10]{D.~Rosa-Gonz\'{a}lez}
\author[12]{H.~Salazar}
\author[11, 29]{F.~Salesa Greus}
\author[2]{A.~Sandoval}
\author[2]{J.~Serna-Franco}
\author[8]{A.J.~Smith}
\author[18]{R.W.~Springer}
\author[30]{K.~Tollefson}
\author[10]{I.~Torres}
\author[31]{R.~Torres-Escobedo}
\author[10]{F.~Ure\~{n}a-Mena}
\author[12]{L.~Villase\~{n}or}
\author[19]{X.~Wang}
\author[8]{E.~Willox}
\author[31]{H.~Zhou}
\author[4]{C.~de Le\'{o}n}
\author[4]{J.D.~\'{A}lvarez\thanks{Corresponding author: science@juandedios.info}}
\author[32]{G.B.~Yodh}
\author[33]{A.~Zepeda}

\affil[1]{Physics Division, Los Alamos National Laboratory, Los Alamos, 87545 New Mexico, USA }
\affil[2]{Instituto de F\'{i}sica, Universidad Nacional Aut\'{o}noma de M\'{e}xico, 04510 Ciudad de Mexico, Mexico }
\affil[3]{Universidad Aut\'{o}noma de Chiapas, Tuxtla Guti\'{e}rrez,  29050 Chiapas, Mexico}
\affil[4]{Universidad Michoacana de San Nicol\'{a}s de Hidalgo, 58040 Morelia, Mexico }
\affil[5]{Instituto de Geof\'{i}sica, Universidad Nacional Aut\'{o}noma de M\'{e}xico, 04510 Ciudad de Mexico, Mexico }
\affil[6]{St. Albert’s College (Autonomous), Cochin, 682018 Kerala, India}
\affil[7]{Department of Physics, Pennsylvania State University, University Park, 16802 Pennsylvania, USA}
\affil[8]{Department of Physics, University of Maryland, College Park, 20742-4111 Maryland, USA}
\affil[9]{Instituto de Astronom\'{i}a, Universidad Nacional Aut\'{o}noma de M\'{e}xico, 04510 Ciudad de Mexico, Mexico }
\affil[10]{Instituto Nacional de Astrof\'{i}sica, \'{O}ptica y Electr\'{o}nica, Tonantzintla, 72840 Puebla, Mexico}
\affil[11]{Institute of Nuclear Physics Polish Academy of Sciences, PL-31342 IFJ-PAN, 31342 Krakow, Poland }
\affil[12]{Facultad de Ciencias F\'{i}sico Matem\'{a}ticas, Benem\'{e}rita Universidad Aut\'{o}noma de Puebla, 72570 Puebla, Mexico }
\affil[13]{Departamento de F\'{i}sica, Centro Universitario de Ciencias Exactas e Ingenierias, Universidad de Guadalajara, 44430 Guadalajara, Mexico}
\affil[14]{Institute for Cosmic Ray Research, University of Tokyo,  277-8582 Kashiwa, Kashiwanoha, Japan}
\affil[15]{Department of Physics, University of Wisconsin-Madison, Madison, 53706 Wisconsin, USA }
\affil[16]{Physics Division, Los Alamos National Laboratory, Los Alamos,87545 New Mexico, USA }
\affil[17]{Tecnologico de Monterrey, Escuela de Ingenier\'{i}a y Ciencias, Ave. Eugenio Garza Sada 2501, 64849 Monterrey, N.L., Mexico}
\affil[18]{Department of Physics and Astronomy, University of Utah, Salt Lake City, 84112 Utah, USA}
\affil[19]{Department of Physics, Michigan Technological University, Houghton, 49931-1295 Michigan, USA }
\affil[20]{Erlangen Centre for Astroparticle Physics, Friedrich-Alexander-Universit\"at Erlangen-N\"urnberg, 91058 Erlangen, Germany}
\affil[21]{Department of Physics and Astronomy, Michigan State University, East Lansing, 48824 Michigan, USA }
\affil[22]{Universidad Politecnica de Pachuca, Pachuca, 42083 Hgo, Mexico }
\affil[23]{Centro de Investigaci\'on en Computaci\'on, Instituto Polit\'ecnico Nacional, 07738 M\'exico City, Mexico.}
\affil[24]{Department of Physics and Astronomy, University of New Mexico, Albuquerque, 87131-0001 New Mexico, USA }
\affil[25]{Universidad Aut\'{o}noma del Estado de Hidalgo, 42184 Pachuca, Mexico }
\affil[26]{Instituto de Ciencias Nucleares, Universidad Nacional Aut\'{o}noma de Mexico, 04510 Ciudad de Mexico, Mexico}
\affil[27]{Department of Physics, Stanford University, Stanford, CA 94305–4060, USA}
\affil[28]{University of Seoul, 02504 Seoul, Republic of Korea}
\affil[29]{Instituto de Física Corpuscular, CSIC, Universitat de València, E-46980, Paterna, 46980 Valencia, Spain}
\affil[30]{Department of Physics and Astronomy, Michigan State University, East Lansing, 49931-1295 Michigan, USA }
\affil[31]{Tsung-Dao Lee Institute \& School of Physics and Astronomy, Shanghai Jiao Tong University, 200240 Shanghai, China}
\affil[32]{Department of Physics and Astronomy, University of California, Irvine, 92697 Irvine, CA, USA}
\affil[33]{Physics Department, Centro de Investigaci\'{o}n y de Estudios Avanzados del IPN, 07360 Ciudad de Mexico, Mexico}

\begin{document}

\maketitle

  \begin{abstract}
    A measurement with high statistics of the differential energy spectrum of light elements in cosmic rays, in particular, of primary H plus He nuclei, is reported. The spectrum is presented in the energy range from $6$ to $158 \, \tev$ per nucleus. Data was collected with the High Altitude Water Cherenkov (HAWC) Observatory between June 2015 and June 2019.  The analysis was based on a Bayesian unfolding procedure, which was applied on a subsample of vertical HAWC data that was enriched    to $82\%$ of events induced by light  nuclei.
    To achieve the mass separation, a cut on the lateral age of air shower data was set guided by predictions of CORSIKA/QGSJET-II-04 simulations. The measured spectrum is consistent with a broken power-law spectrum and shows a kneelike feature at around $E = 24.0^{+3.6}_{-3.1} \, \tev$, with a spectral index $\gamma = -2.51 \pm 0.02$ before the break and with $\gamma = -2.83 \pm 0.02$ above it. The feature has a statistical significance of $4.1 \, \sigma$. Within systematic uncertainties, the significance of the spectral break is  $0.8 \, \sigma$.
   \end{abstract}

\providecommand{\keywords}[1]
{
  \small	
  \textbf{\textit{Keywords---}} #1
}

\section{Introduction}

 Cosmic rays are mainly relativistic atomic nuclei that impinge nearly isotropically on Earth from outer space with energies that extend from a few $\mev$ to some $\zev$ \cite{Roulet17, kachelriess19, PDG19}. Above $10^{13} \, \ev$, cosmic rays can be studied indirectly by means of air shower techniques \cite{Gaisser17, Blumer09, Haungs03}, and below $10^{15} \, \ev$, with direct methods by using particle detectors on board of balloons, spaceships and satellites \cite{Maestro15, Picozza18}. 
 
 The energy region between $10^{13}$ and $10^{15} \, \ev$ is the frontier between the direct and indirect detection techniques of cosmic rays. Historically, data have been difficult to obtain in this energy interval due to limitations owing to both detection methods. In spite of that, early experiments have found out that the differential energy spectrum of cosmic rays in this energy regime can be roughly described by a power-law $E^{\gamma}$  with a spectral index $\gamma \sim -2.7$ and that it seems to be dominated by hydrogen and helium nuclei \cite{Roulet17, PDG19} at  least up to $700 \, \tev$ \cite{Bartoli15}. These observations appear to be consistent with theoretical models that assume the existence of a common type of galactic source for $\tev$ and $\pev$ cosmic rays \cite{Hoerandel04, Giacinti15, Thoudam16}, for example, supernova remnants \cite{Roulet17, kachelriess19, PDG19}. Yet, they cannot rule out more complex astrophysical scenarios that involve, for instance,  the presence of local cosmic ray sources \cite{Kachelriess15, Kachelriess18} or of a new population of cosmic ray accelerators with cutoff energies of $\tev$ \cite{Zatsepin,Zatsepin2,Ptuskin10, Stanev14}.  Such models usually predict fine structures in the energy spectrum of the all-particle and individual mass groups of cosmic rays at $\tev$ energies, whose existence can only be tested with 
 precise data and with high statistical power measurements on the energy and composition of cosmic rays.
 
 In this regard, recent data provided by the satellites DAMPE \cite{dampe19, dampe21} and NUCLEON \cite{nucleon19}, as well as the HAWC extensive air shower (EAS) Observatory \cite{Hawc17} seem to reveal that, in fact, the energy spectra of cosmic rays in the $\tev$ region show the presence of individual features that cannot be fitted by a single power law.

 First hints about the existence of fine structure in the energy spectra of cosmic rays came from the balloon-borne ATIC-2 \cite{atic09,  atic092} and CREAM \cite{cream17} experiments, and from early measurements carried out with the NUCLEON satellite observatory \cite{nucleon, nucleon2}. The data from these instruments seemed to point out the presence of spectral breaks between $10$ and $40 \, \tev$ in the spectra of  H and He nuclei. However, those results were not conclusive due to the lack of statistics. A clear indication of a feature in the $10 - 100 \, \tev$ range was provided later by the HAWC observatory,  which showed the existence of a break in the all-particle energy spectrum of cosmic rays at around $46 \, \tev$ \cite{Hawc17}. Just recently, the NUCLEON experiment, with more statistics, gave further support to the existence of breaking features in the proton and helium spectra at energies of around $Z \times 10 \, \tev$ \cite{nucleon19}, respectively, while the  DAMPE satellite experiment \cite{dampe19, dampe21} provided significant evidence for individual kneelike structures in the spectra of protons and helium nuclei at $\sim 14$ and $\sim 34 \, \tev$, respectively. In addition, the HAWC collaboration found a steepening in the energy spectrum of the light mass group (H$+$He) of cosmic rays close to $30 \, \tev$ \cite{Hawc19}. The relation between these structures and the spectral break in the all-particle energy spectrum at  $46 \, \tev$ is still not clear, but future studies on the different mass groups of cosmic rays, as in \cite{nucleon19},  may throw some light on the issue. 
 
  In the present paper, we have updated the analysis performed in \cite{Hawc19} on the energy spectrum of light primaries at tens of $\, \tev$. Since the appearance of \cite{Hawc19} further improvements have been included in the study, such as the employment of an updated set of Monte Carlo (MC) simulations of the HAWC  detectors \cite{Hawccrab19} and the usage of a bigger experimental dataset, which spans the period of time from June 11, 2015, to June 3, 2019. This reduced both statistical and photomultiplier tube (PMT) systematic uncertainties. As in \cite{Hawccrab19}, the analysis procedure in this work is based on an unfolding technique, which is applied on a large collection of data that has a high proportion of  H and He nuclei induced events ($>82 \%$ abundance).  Mass separation is done event-by-event using an energy dependent cut on the lateral shower age parameter,  derived from predictions of the QGSJET-II-04 hadronic interaction model \cite{qgsjetii4} for different primary nuclei. 
  
  We have targeted the mass group of light elements, as it is the most abundant component in the flux of cosmic rays in the energy region of interest  \cite{Roulet17, kachelriess19, PDG19} and because it is easier to separate with the present analysis technique. The paper is organized in the following way:
  in Sec. \ref{experiment}, we present the HAWC detector, the EAS reconstruction, and the methods for the calibration of the primary energy and estimation of the lateral shower age. In Sec. \ref{MCsection},  we discuss the MC simulations used in our analysis. Section \ref{selectioncuts} gives the event selection criteria.  Section \ref{Analysis} describes the data, the mass separation and the reconstruction of the spectrum. The unfolded result and a comparison with measurements of other experiments come in Sec. \ref{results}.
  Section \ref{discussion} discusses the result and Sec.  \ref{conclusions}  gives our conclusions from the work. 
  Appendix \ref{appmodels} contains a description of the composition models used in this analysis. Appendix \ref{apperrors} provides a  detailed list of the statistical and systematic error sources. Appendix \ref{appchecks} describes systematic checks carried out to verify our result.

\section{The HAWC observatory}
 \label{experiment}
  \subsection{Experimental setup}
  
  HAWC is a high altitude air shower observatory optimized for studying the gamma-ray sky in the  $500 \, \gev - 100 \, \tev$ energy range. However, it can also work as a cosmic ray detector at primary energies from a few $\tev$ up to $1 \, \pev$ \cite{Hawc17, Hawc18}. The observatory is located at $4100 \, \mbox{m}$ \textit{a.s.l.} on a plateau ($19^{\circ}$ N,  $97^{\circ}$ W) between the volcanoes Sierra Negra and Pico de Orizaba in the east-central part of Mexico \cite{Hawc17b}. Its location  (at an  atmospheric depth of $\sim 640 \, \mbox{g}/\mbox{cm}^2$) allows HAWC to have high sensitivity to hadronic EAS with energies in the $\tev$ range. Since the detector is close to the maximum of the air shower, $\langle X_{max} \rangle \sim 560 \, \mbox{g}/\mbox{cm}^2$ for H  ($425  \, \mbox{g}/\mbox{cm}^2$ for Fe) at  $1 \, \pev$ according to QGSJET-II-04 (see also \cite{sibyll23c}), the effects of fluctuations are reduced and it is possible to determine the primary energy with good precision. 

  For the detection of EAS, HAWC employs a dense array of $300$ water Cherenkov detectors (WCD), which covers a flat surface of $22000 \, \mbox{m}^2$ ($\approx 150  \, \mbox{m} \times 150  \, \mbox{m}$). Each WCD contains $4$ PMTs and almost $200000 \, \mbox{L}$ of water. The PMTs are anchored at the bottom of the WCDs and monitor the water above them.

  \subsection{Air shower reconstruction}

  During the passage of an EAS through the detector, the relativistic particles of the shower produce Cherenkov light in the WCDs, which induces pulses in the PMTs. The signals are digitized and an effective charge $Q_\mathrm{eff}$ is assigned to each pulse. The detector is calibrated to obtain uniform charge assignments and to correct for  time delays between detectors \cite{Hawc17b, Hawcdaq18}. The reconstruction software  uses data from   PMTs with $Q_\mathrm{eff}$ below a maximum calibrated value of  $ \approx 10^{4} \, PE$  to estimate various EAS observables of the event, such as  arrival direction, shower core position, the lateral distribution of deposited charge, lateral shower age  and primary energy \cite{Hawc17, Hawccrab19, Hawc17b}.  
  In the following subsections, we   detail  the estimation of the lateral shower age and  primary energy in HAWC.

  \subsubsection{Lateral shower age}
  
  The lateral shower age $s$ is related to the shape of the lateral distribution of an EAS and  gives a measure of its steepness. It is an important parameter for the study of air showers as it depends on the distance from the shower maximum to the observation point and  is sensitive to the primary mass. The lateral age was introduced through the Nishimura-Kamata-Greisen  lateral density distribution  in the context of pure electromagnetic cascades \cite{NKG1, NKG2, NKG3}. 
  
  According to their relative age value, EAS can be classified into old and young showers \cite{Kascade06}. Old showers have large $s$ values and flatter lateral distributions. They are characterized by shower maxima at small atmospheric depths. Young showers possess small values of $s$ and steeper lateral distributions, and are associated to EAS that penetrate deeper in the atmosphere. On average, heavy primaries tend to produce older showers than light nuclei, while high energy primaries create younger EAS than low energy ones.

  \begin{figure}[!t]
  \centering
  \includegraphics[width=3.3in]{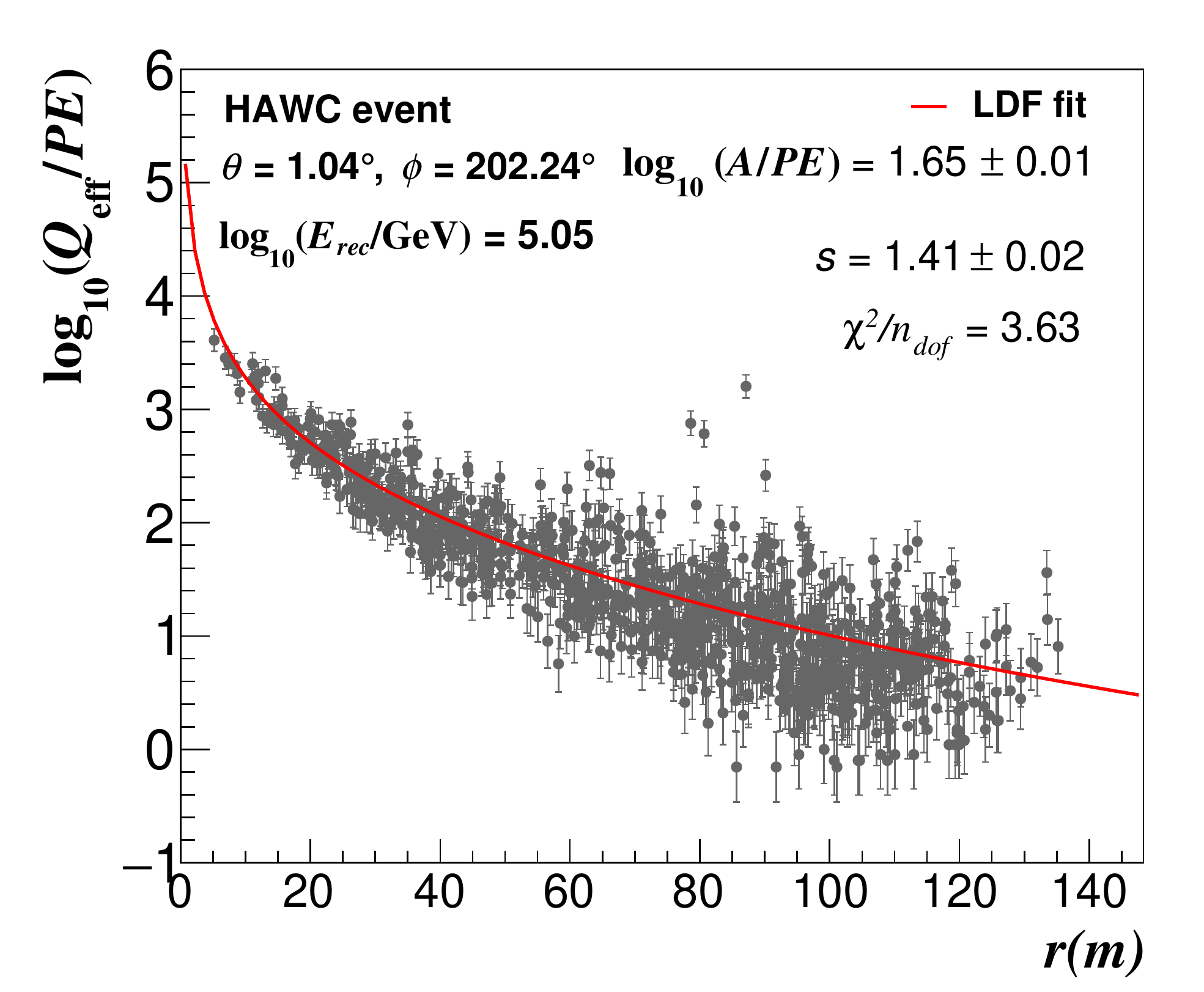}
  \caption{The lateral effective charge distribution of an EAS event measured with HAWC on June 2, 2019. The estimated energy, zenith angle and azimuth are $\log_{10}(E_{rec}/\gev) = 5.05$, $\theta = 1.04^{\circ}$ and $\phi = 202.24^{\circ}$, respectively. The gray dots represent the  measured $Q_\mathrm{eff}$ per PMT in $PE$ (photoelectron) units. The vertical errors are the systematic uncertainties. The result of the fit with Eq. (\ref{eq1}) is shown with a red line. The corresponding fit parameters are shown; the number of degrees of freedom is 1018.}
  \label{FigLDF}
  \end{figure}

  In HAWC, the lateral age of EAS is obtained event by event from a $\chi^2$ fit with a modified Nishimura-Kamata-Greisen  function,
  \begin{equation}
    f(r) = A \left(\frac{r}{r_0}\right)^{s-3}  \left(1 + \frac{r}{r_0}\right)^{s-4.5},
    \label{eq1}
  \end{equation} 
  to the lateral charge distribution measured by the PMTs $Q_\mathrm{eff}(r)$ \cite{Hawccrab19}. Here, $r$ is the radial distance to the EAS axis in the shower plane, $r_0 \sim 124 \, \mbox{m}$ is the Molière radius at the HAWC site and $A$ is the amplitude of the function, which is also a fit parameter. This lateral distribution function,   originally proposed for describing EAS initiated by gamma rays, also gives  a reasonable description of the measured lateral distribution of hadron-induced showers \cite{Hawcldf19}. This is illustrated in Fig.~\ref{FigLDF}, where we show a fit of Eq. (\ref{eq1}) to the measured lateral distribution of a typical hadronic event that arrived with a zenith angle $\theta = 1.04^{\circ}$ and an azimuth $\phi = 202.24^{\circ}$,  which had a reconstructed primary energy of $\log_{10}(E_{rec}/\gev) = 5.05$. The rather young shower age  was $s = 1.41 \pm 0.02$. 
  The result of the fit gave a reduced $\chi^2$ of $3.63$ for $n_{dof} = 1018$ degrees of freedom. This is a large value for $\chi^2/n_{dof}$, which is due to the natural width of the lateral distribution of hadronic air showers, which is bigger than the experimental error on  $Q_\mathrm{eff}(r)$. 
  
  Note, in Fig.~\ref{FigLDF}, the presence of outliers in the measured lateral distribution. These features are usually present in hadronic induced EAS and are mainly associated to large and localized charged depositions in the detectors from shower muons \cite{HAWC_2013}. In general, gamma rays create EAS with smoother lateral distributions than  those from cosmic rays. This difference is employed in HAWC for gamma/hadron separation \cite{Hawc17b, Hawccrab19}.
  The outliers produce a small bias on the fitting parameters of the order of a few percent. In particular, for the example presented in Fig.~\ref{FigLDF}, they induce an increment on $s$ and $\log_{10}(A)$ of $3\%$ and $2\%$, respectively.
  
  It is worth to point out that the reduced  $\chi^2$ of the lateral distributions of the measured data is similar to the predictions of MC simulations up to $\log_{10}(E_{rec}/\gev) = 4.2$ for a mixed composition scenario using our reference composition model, which will be described in the next section, and QGSJET-II-04. Meanwhile, at higher energies the experimental mean of $\chi^2/n_{dof}$ tends to be larger than the MC expectations for the mixed composition assumption. In particular, for $\log_{10}(E_{rec}/\gev) > 5.3$ the values of the reduced  $\chi^2$ of the data are above  the MC predictions for pure proton and iron nuclei, which implies that in this energy regime the width of the measured lateral distributions of hadronic EAS is larger than expected from QGSJET-II-04 simulations. Further studies are needed to understand the origin of such differences.

  \subsubsection{Primary energy}

  The primary energy of the shower event is estimated from a maximum log-likelihood procedure \cite{Hawc17}, which computes and compares the probabilities that the measured lateral distribution of PMT signals from a given shower with reconstructed zenith angle $\theta$ is produced by proton primaries of different energies, $E$. The calculation also includes the probability of observing active PMTs with no signals during the event. In the algorithm, the probability values of the operational PMTs are extracted from probability tables, which are generated using proton-induced EAS simulations with a number of hit PMTs ($nHit$) greater than $75$ and with EAS cores and arrival directions successfully reconstructed. The tables are obtained from CORSIKA/QGSJET-II-04 simulations for $\log_{10}(E/\gev) = [1.85, \, 6.15]$ and $\theta \leq 60^\circ$. 

   \section{Monte Carlo simulations}
    \label{MCsection}
    
   Air shower simulations initiated by cosmic rays in HAWC were carried out using CORSIKA v7.40 \cite{Heck:1998vt} without the thinning option and with the hadronic interaction models FLUKA \cite{Fluka} and QGSJet-II-04 \cite{qgsjetii4}. FLUKA is employed for hadron energies of  $E_{lab} < 80 \, \gev$, while QGSJet is used at higher energies.
   
   Simulations were conducted for eight primary species, in particular, H, He, C, O, Ne, Mg, Si and Fe, using an energy spectrum  $E^{-2}$ for the energy interval  $5 \, \mbox{GeV} - 2 \, \mbox{PeV}$. The MC data cover the  zenith angle range $\theta = [0^{\circ},  65^{\circ}]$ with a $\cos\theta \sin\theta$ distribution. Shower cores are thrown flat in radius up to $1 \, \mbox{km}$ from the center of the array, but 
   reweighted to simulate a distribution uniform in area.
   
\begin{figure}[!t]
  \centering
  \includegraphics[width=3.3in]{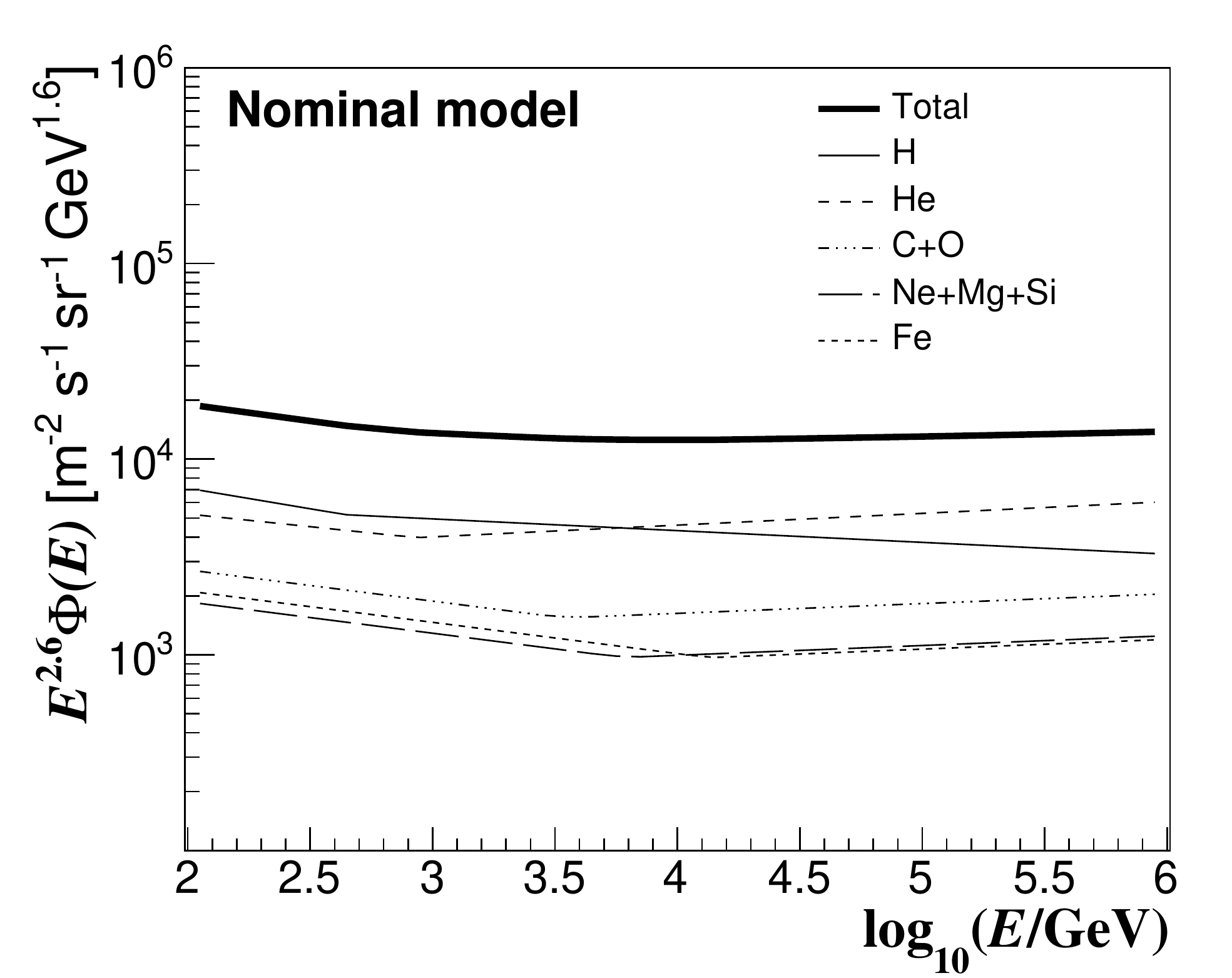}
  \caption{Nominal composition model \cite{Hawc17} used for the analysis in this work. The model was obtained from fits to the AMS-2 \cite{ams14, ams15}, CREAM I-II \cite{cream09, cream11}, and PAMELA \cite{pamela} cosmic ray data. The black bold line represents the  all-particle energy spectrum, the thin continuous line and the short-dashed line, the fitted spectra of H and He nuclei. The sum of the C and O energy spectra is indicated by the dashed-dotted line, and the combination of the spectra of Ne, Mg and Si primaries, by the long-dashed line. The dotted line correspond to the fit spectrum of Fe nuclei. 
  }
  \label{FigCREAMmodel}
  \end{figure}

   The HAWC detector response was simulated using software based on GEANT4 \cite{Geant4}.  Both MC and measured  events were reconstructed with the same algorithm in order to study the influence of experimental systematic uncertainties on the estimated EAS parameters.
   
  MC events were weighted to reproduce the nominal composition model introduced in \cite{Hawc17}. This model gives a fair description of the cosmic ray elemental spectra measured  by the direct experiments AMS-2 \cite{ams14, ams15}, CREAM I-II \cite{cream09, cream11}, and PAMELA \cite{pamela}  in the energy interval from $100 \, \gev$ to $\sim 200 \, \tev$. The data are fit with a broken power law, which is extrapolated up to a few $\pev$.  Figure ~\ref{FigCREAMmodel} illustrates the cosmic ray intensities in our nominal composition model, with  the  predictions for  light ($Z \leq 2$), intermediate ($ 3 \leq Z \leq 14$) and heavy cosmic ray nuclei. The  expressions and fit parameters\footnote{There is a typo in the value of the normalization energy $E_{0}$ of the broken power-law functions in the nominal model of \cite{Hawc17} that is corrected here. The  parameter $E_0$  should have the values $1200$, $1600$, $2000$, $2400$, $2800$, and $5600$ in $\gev$ units, for C, O, Ne, Mg, Si and Fe nuclei, respectively.} are taken from \cite{Hawc17}. The total number of simulated EAS in the full zenith angle range and the whole  energy interval for protons and helium primaries were $3 \times 10^{10}$ and  $1.3 \times 10^{10}$, respectively, while for the rest of elemental nuclei, we simulated $10^{9}$ MC events per mass group. For vertical events with $\theta \lesssim 16^{\circ}$ and primary energies greater than $10 \, \tev$, the number of simulated events is reduced by a factor of $2.2 \times 10^{4}$. Appendix \ref{appmodels} gives other composition models used to estimate systematic errors.
   
   \begin{table}[t]
    \begin{center}
    \caption{Effects of the selection criteria on the datasets. The cuts are shown on the left column. The central columns represent the fraction of events from the previous cut (in percent) which pass  the cut. The second column was obtained for measured data, and the third column, for MC simulations in the framework of the nominal composition model used in this work. Calculations start with datasets which satisfy $N_{hit} > 10$, the minimum for which the reconstruction saves data. As in \cite{Hawc17}, the cosmic ray detection rates in HAWC are also computed. 
    }
   \begin{tabular}{lcccccc}
   \hline 
   \hline
    Selection cut &&& \multicolumn{2}{c}{\% of remaining}&&  Measured rate\\ 
     &&& \multicolumn{2}{c}{events respect}&& \\ 
     &&& \multicolumn{2}{c}{to previous cut}&& ($kHz$)\\  
    &&& Data & MC  && \\ 
   \hline
   Trigger &&& $100.00$ & $100.00$ && $24.61$ \\
   Passed angle and core &&&&&&\\
   reconstruction &&& $95.18$  & $100.00$ && $23.42$\\ 
   $N_{hit} \geq 75$ &&& $23.65$  & $26.95$ && $5.54$\\
   $N_{r40} \geq 40$ &&& $26.70$  & $28.39$  && $1.48$\\ 
   Zenith angle &&& $27.82$  & $29.35$ && $0.41$\\
   Fraction hit  &&& $36.02$  & $31.11$ && $0.15$\\
   Primary energy &&& $93.48$  & $92.94$  && $0.14$\\
   \hline
   \hline
   \end{tabular}
   \label{tab1}
   \end{center}
  \end{table}

 \section{Selection cuts}
 \label{selectioncuts}
 
  A set of selection criteria were applied to both data and MC simulations for the reconstruction of the energy spectrum with the main purpose of  reducing the influence of systematic uncertainties in the final result.  
  The selection criteria were chosen after a detailed MC study of their effect on the core position, arrival angle, primary energy of air showers and on the HAWC effective area. 
  
  The first cut discards  EAS events that have not successfully passed the core and arrival direction reconstruction or have less than $75$  hit PMTs. To reduce the uncertainty on the core position, we selected events with at least 40 hit PMTs within a radius of $40 \, \mbox{m}$ from the reconstructed EAS core ($N_{r40} \geq 40$). According to MC simulations, this cut only leaves data with reconstructed shower cores on the array or within a distance of $20 \, \mbox{m}$ from the boundary of HAWC. A tighter selection would reduce the efficiency of HAWC for cosmic rays with energies close to $10 \, \tev$ and increase the uncertainties on both the effective area and the reconstructed spectrum: requiring event  cores inside the array increases the  uncertainty on the energy spectrum up to $50 \%$ around $10 \, \tev$ respect to the value with our standard cut. 

  To decrease systematic errors associated with inclined showers, we keep only near-vertical EAS with $\theta < 16.7^\circ$. This value is close to the upper limit of the zenith angle range corresponding to the table used for energy calibration of vertical EAS.  We also removed showers with low efficiency by requiring an estimated shower energy $\log_{10}(E_{rec}/\gev) > 3.5$   and $f_{hit} \geq 0.2$, where
  $f_{hit}$ is the fraction of active PMTs with hits in the event \cite{Hawc17b}.  Finally, we applied an upper cut of $\log_{10}(E_{rec}/\gev) < 5.5$ 
  to focus our analysis in the region where the uncertainties in the composition studies of cosmic rays due to the PMT systematic effects are smaller.

  The effects of the successive application of the selection cuts on the total number of events of both MC and experimental datasets are seen in Table \ref{tab1}. 
  Large reductions in the total number of selected events are associated with constraints  on $N_{hit}$, $\theta$, and $N_{r40}$, as in \cite{Hawc17}. There is also an important decrease in the data and in MC simulations due to the cut on $f_{hit}$, which removes shower events below a few $\tev$. These are low energy events that trigger the detector much  more frequently than EAS at higher energies, which is why such a cut strongly affects both the data and MC samples.

     \begin{figure}[!tp]
     \centering
     \includegraphics[width=3.3in]{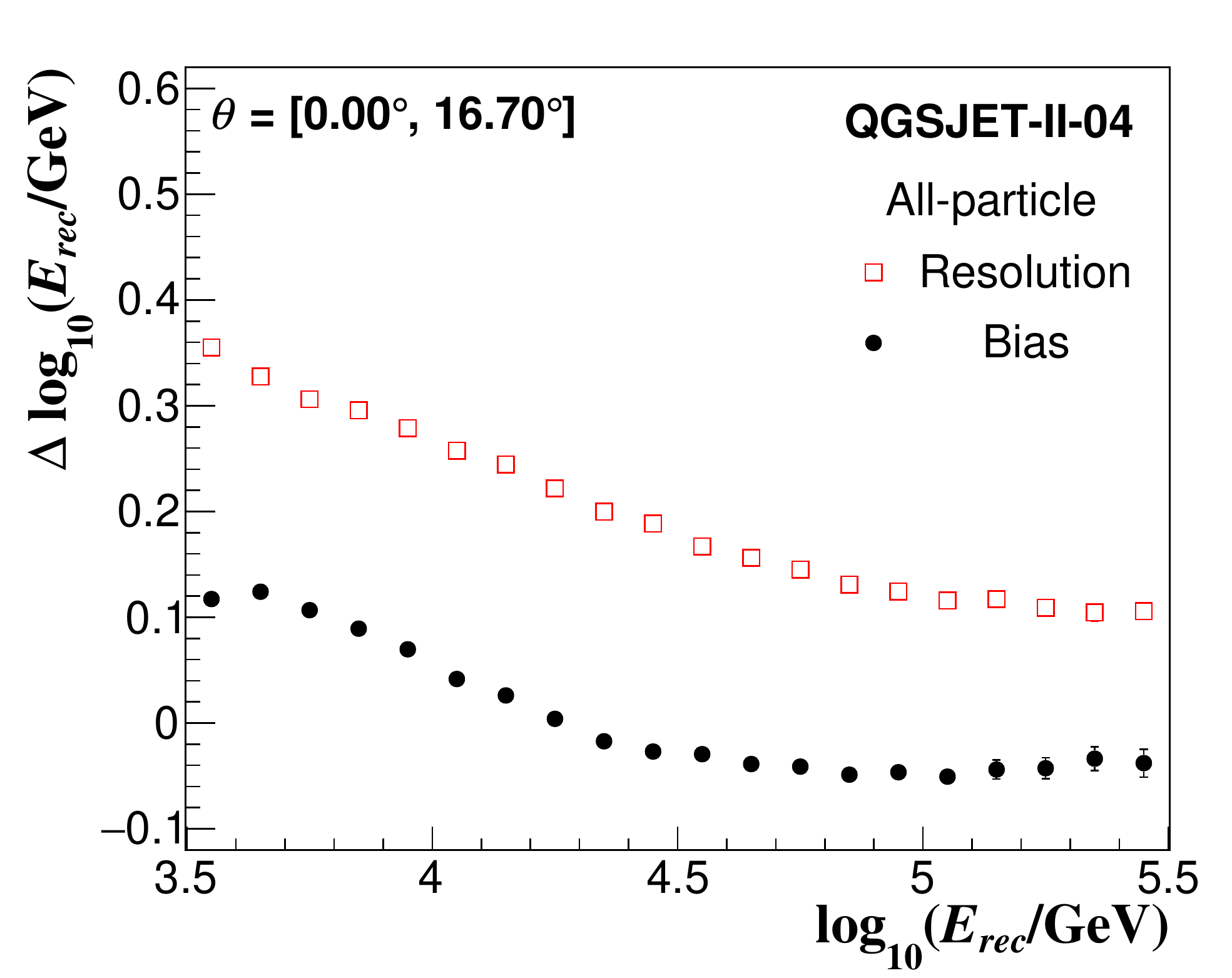} 
    \caption{The mean bias (black circles) and resolution (red open squares) of the primary energy   of cosmic ray induced EAS as a function of the estimated energy in HAWC according to MC predictions with QGSJET-II-04. The plots were obtained for events with $\theta < 16.7^\circ$ and using the all-particle spectrum described in our nominal composition model. The selection cuts discussed in the paper were also applied. The energy bias is defined as  $\Delta \log_{10}(E_{rec}) = \log_{10}(E_{rec}) - \log_{10}(E)$, where $E_{rec}$ and $E$ are the reconstructed and true primary energies of the EAS. The energy resolution is defined as the standard deviation of the $\Delta \log_{10}(E_{rec})$ distribution. }
  	\label{SystE}
    \end{figure}

  The selection criteria have almost the same effect on both MC and experimental data samples according to Table \ref{tab1}. 
  There are, however, some differences between the selection efficiencies of simulations and measurements. One of the largest ones is found when applying the cut on the fraction hit. In this case, the observed difference between MC and data can be mainly attributed to the fact that the energy spectrum in our nominal composition model of cosmic rays is softer than the actual one below $\sim 20 \, \tev$. An increment of $\Delta \gamma = 0.1$ in the magnitude of the spectral index in the  simulations with the nominal composition model for $E <  20 \, \tev$ increases the selection efficiency due to the cut on $f_{hit}$ up to approximately $35 \%$ in the MC sample, which is closer to the corresponding selection efficiency for HAWC data. The value of $\Delta \gamma$ used to perform the previous calculation was derived by comparing the histograms of $E_{rec}$ for the measured data and for the nominal composition model. On the other hand,  the selection efficiency in MC data due to  the $f_{hit}$ cut can be further incremented by  $\sim 1 \%$  taking also into account in the simulations the observed difference between the nominal composition model and the measured data regarding the relative abundance of light primaries. It is worth to mention that the estimated systematic errors for the energy spectrum of H$+$He performed in this work consider the contributions from uncertainties in the composition model and the energy spectrum of cosmic rays (see Appendix \ref{apperrors}).

   According to MC simulations, for $\log_{10}(E_{rec}/\gev) \geq 3.8$  the mean systematic uncertainties of the shower core position and the arrival direction of EAS in the selected data  are below $17 \, \mbox{m}$ and $0.5^{\circ}$, respectively. The bias and resolution of the primary energy are $|\Delta \log_{10}(E_{rec}/\gev)| \leq 0.09$ and $\sigma \log_{10}(E_{rec}/\gev) \leq 0.3$, correspondingly, above $\log_{10}(E_{rec}/\gev) = 3.8$.  As an example, the expected mean bias and the resolution of the primary energy are shown in Fig.~\ref{SystE} as a function of $E_{rec}$. The energy estimation and the pointing accuracy of the detector have been verified independently in \cite{Hawc17} using measurements of the position of the Moon shadow as a function of the reconstructed energy.

   \section{Description of the analysis}
   \label{Analysis}
   
   \subsection{Experimental dataset}
   
   In the present analysis, we have used data collected with the central detector of HAWC
   from June 11, 2015, to June 3, 2019. The total effective time amounts to  $T_{eff} = 3.74 \, \mbox{yr}$, which corresponds to an experimental livetime of $94 \%$. The
   data sample contains
   $2.9 \times 10^{12}$ EAS. After applying the selection criteria, we kept 
   $1.6 \times 10^{10}$ showers.

  \subsection{Analysis technique}

  The reconstruction of the energy spectrum of proton and helium primaries
  applies an unfolding analysis  
  to a subsample of  events enriched in  light elemental nuclei
  by a cut on the shower age. 
  We correct for  contamination by heavy nuclei, and triggering and reconstruction efficiency.
  We give details of the reconstruction chain in the following subsections.

  \subsubsection{Extraction of an enriched subsample of light elements}

\begin{figure}[!t]
  \centering
  \includegraphics[width=3.3in]{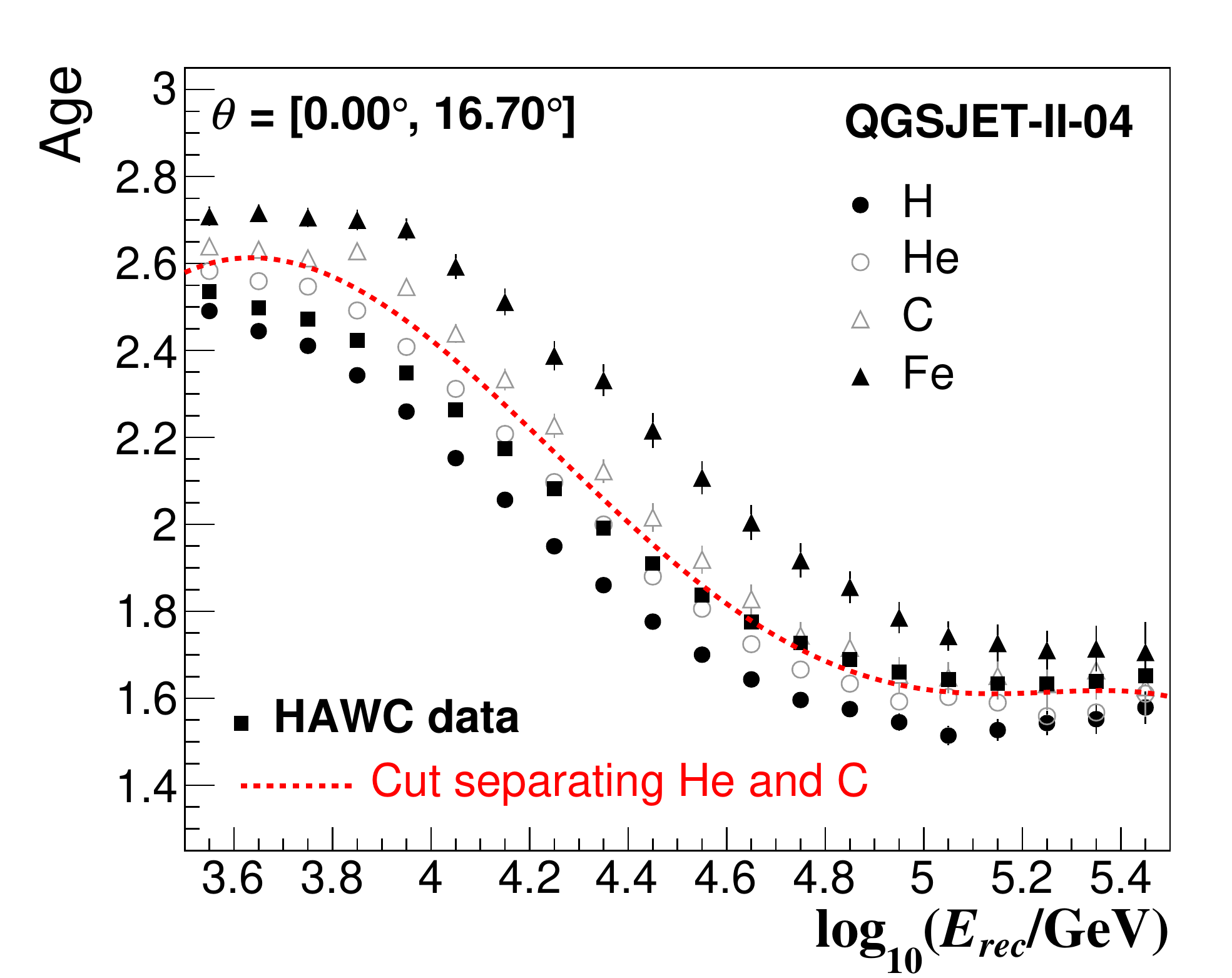}
  \caption{Predictions of the QGSJET-II-04 model for the energy dependence of the mean lateral age  in vertical air showers initiated by four cosmic ray species at HAWC. From top to bottom, the MC points correspond to Fe (solid triangles), C (hollowed triangles), He (hollowed circles) and H (solid circles) primaries, respectively. For clarity, not all the elemental nuclei simulated in this work were included in the plot. HAWC data has also been added to the figure. They are shown with  black squares.
  The $s_{He-C}$ cut employed to extract the enriched subsample of light nuclei is  plotted using a dashed line in red.}
  \label{FigAge}
  \end{figure}

   The lateral shower age is sensitive to the mass composition of cosmic rays in HAWC, as  can be seen in Fig.~\ref{FigAge}. The plot shows QGSJET-II-04 predictions for the mean $s$ of EAS caused by different mass cosmic rays as a function of the estimated energy $E_{rec}$.
   The age parameter defined in Eq.~(\ref{eq1}) decreases for light  nuclei and for high energy cosmic rays, since these primaries produce more penetrating EAS with shower maxima closer to HAWC. 
   The age increases slightly above $E_{rec} = 10^{5} \, \gev$,  due to 
   the maximum calibrated charge of the PMTs  
   and the finite sampling area of the detector.

   \begin{figure}[!b]
  \centering
  \includegraphics[width=3.3in]{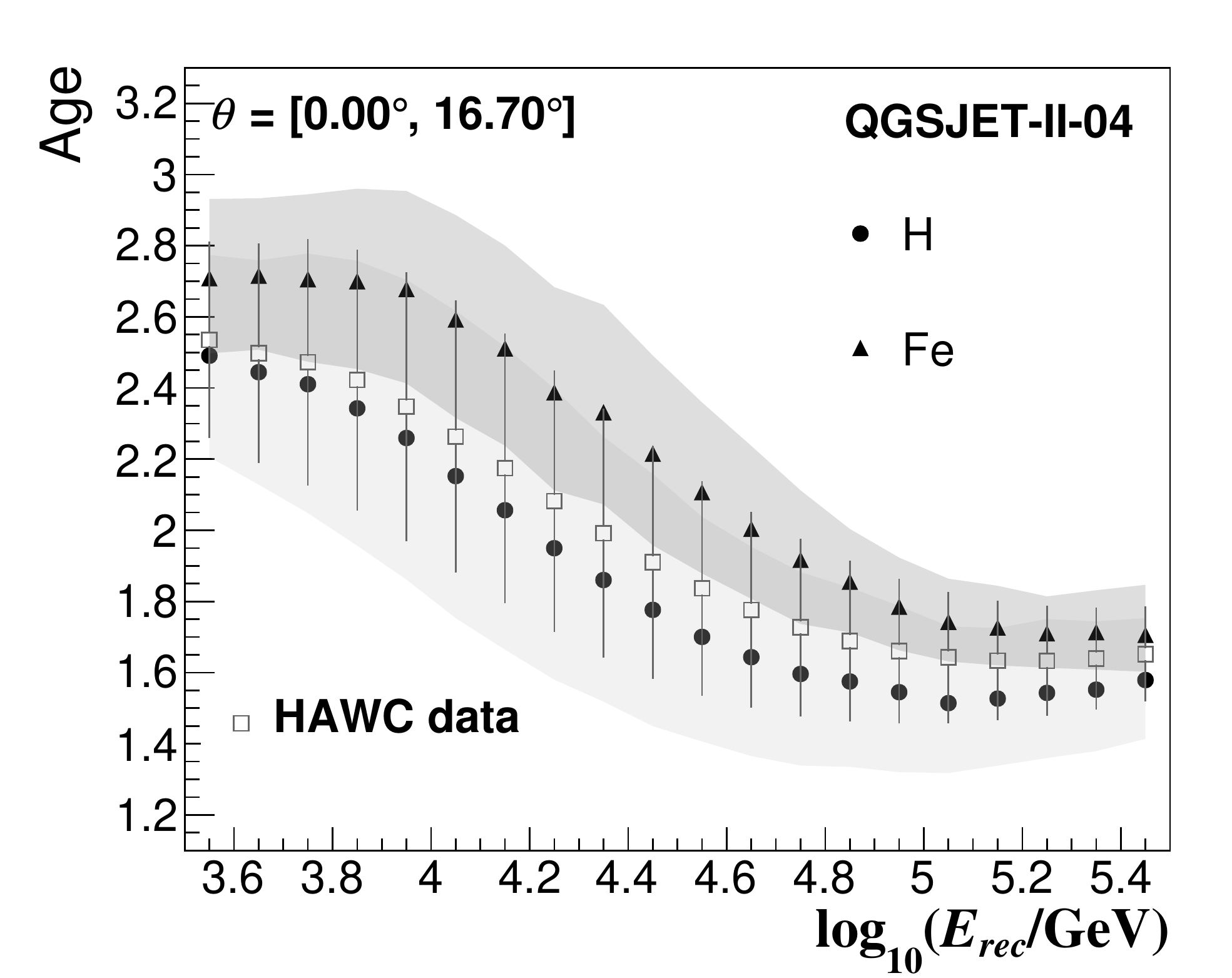}
  \caption{The average lateral age as a function of the estimated energy for MC simulations and HAWC data. The curves are shown with their respective $1\sigma$ statistical errors: for MC, error bands are used, while for measured data, vertical error bars. Mean results for HAWC data are shown with open black squares. Meanwhile, expected values for proton and iron primaries are represented by circles (lower curve) and triangles (upper curve), respectively. They were obtained for EAS with $\theta < 16.7^\circ$ using QGSJET-II-04. }
  \label{FigAgeStat}
  \end{figure} 
  
   On the other hand, in Fig.~\ref{FigAge}, we have also compared model predictions against the average HAWC shower age. The comparison  shows overall agreement between data and expectations up to $E = 3.2 \times 10^{5} \, \gev$ for vertical EAS.  The measured shower age lies between the predictions
   for H and Fe primaries.
   For $E < 2 \times 10^{4} \, \gev$, the mean shower age of the data lies between the predictions  for pure H and He nuclei,  suggesting that light cosmic rays dominate in this energy range.
   At higher energies,  the data is between the expectations for He and C primaries, which may indicate that  heavier cosmic ray nuclei become more important from $2 \times 10^{4} \, \gev$ to $3.2 \times 10^{5} \, \gev$. We will return to this point later, in Sec. \ref{discussion}. 
 
   In order to extract our data subsample for the analysis, we apply an age cut at $s_{He-C}$, which lies between He and C nuclei. The selection criterion keeps showers with low age which are most likely produced by protons and He primaries. We choose this simple cut as we looked for a separation criterion with minimal complications that allows to get a subsample dominated by light primaries and with large statistics. The current age cut does not maximize the purity of the sample. However, it provides an energy spectrum for H plus He nuclei with a similar shape (within $1 \%$ and $8 \%$) to the one obtained with the criterion based on the maximization of the purity of the subsample (see Appendix \ref{appchecks}). Besides its simplicity, the  age cut  $s_{He-C}$  has the advantage that it provides an effective area that is flatter than the one derived from a maximum purity criterion. This is another reason of our preference for the cut $s_{He-C}$. In any case, we have included the contribution to the systematic error of the spectrum due to variations in the purity of the subsample by moving upwards and downwards our age cut (see Appendix \ref{apperrors}). In particular, we have put the selection cut at the curves for the  mean shower age predictions of C and He, 
   respectively.
   
   We must point out that the spectrum of the  light mass group of cosmic rays can also be estimated without applying a cut on the measured data, for example, by fitting the bidimensional histogram for the measured shower age and $E_{rec}$ with MC distributions for the light and heavy cosmic ray nuclei using  unfolding methods as those applied in \cite{Antoni05}. These procedures have the advantage that they allow to estimate independently the background of heavy cosmic ray nuclei in the data sample but they introduce a larger correlation with the light cosmic ray spectrum than in the case of the simple approach with the age cut, where the influence of the heavy nuclei is expected to be reduced. A small dependence  of the result on the composition model is, however, introduced in the simple approach with the age cut trough the estimation of the contamination of the heavy primaries  in the selected data subsample. Each procedure has its own systematic errors. Therefore it is important to confirm the results with different techniques. In this paper, we have adopted the analysis using the age cut, however, alternative analyses with unfolding methods like in \cite{Antoni05} are under way. In this regard, preliminary results were presented in \cite{HAWC_CR_2021}. They are very encouraging, as they confirm the main findings in this paper about the existence of a break at $\tev$ energies in the H+He energy spectrum of cosmic rays. 
   
   According to MC simulations with our nominal composition model, the fraction of light nuclei in the  subsample selected with the shower age cut varies from roughly $97 \%$  at  $E_{rec} = 3.2 \times 10^{3} \, \tev$ down  to $82 \%$ at $3.2 \times 10^{5} \, \tev$. 
   About $\sim 64 \%$ of hydrogen and helium primaries pass the cut,  almost independent of the estimated energy.  After using  the age cut on the measured data, we retained $9.9 \times 10^{9}$ events.
   The separation of  light and heavy nuclei is imperfect, as fluctuations event by event of the shower age are comparable to the average separation of  light and heavy nuclei, as shown in Fig.~\ref{FigAgeStat}.  
   
   On the other hand, MC simulations also predict that  the systematic uncertainties in the energy interval $\log_{10}(E_{rec}/\gev)$ $> 3.8$ for the arrival direction and the core position of the selected data subset are smaller than $0.44^\circ$ and $13 \, \mbox{m}$, respectively, and that the energy resolution $\sigma \log_{10}(E_{rec}/\gev)$ is  not larger than $0.26$ and decreases with the reconstructed primary energy. In particular for $\log_{10}(E_{rec}/\gev) = 4.0$ and $5.0$, $\sigma \log_{10}(E_{rec}/\gev)$ is equal to $0.23$ and $0.10$, respectively.

   \subsubsection{Measured energy histogram}
 
\begin{figure}[!b]
   \centering
  \includegraphics[width=3.3in]{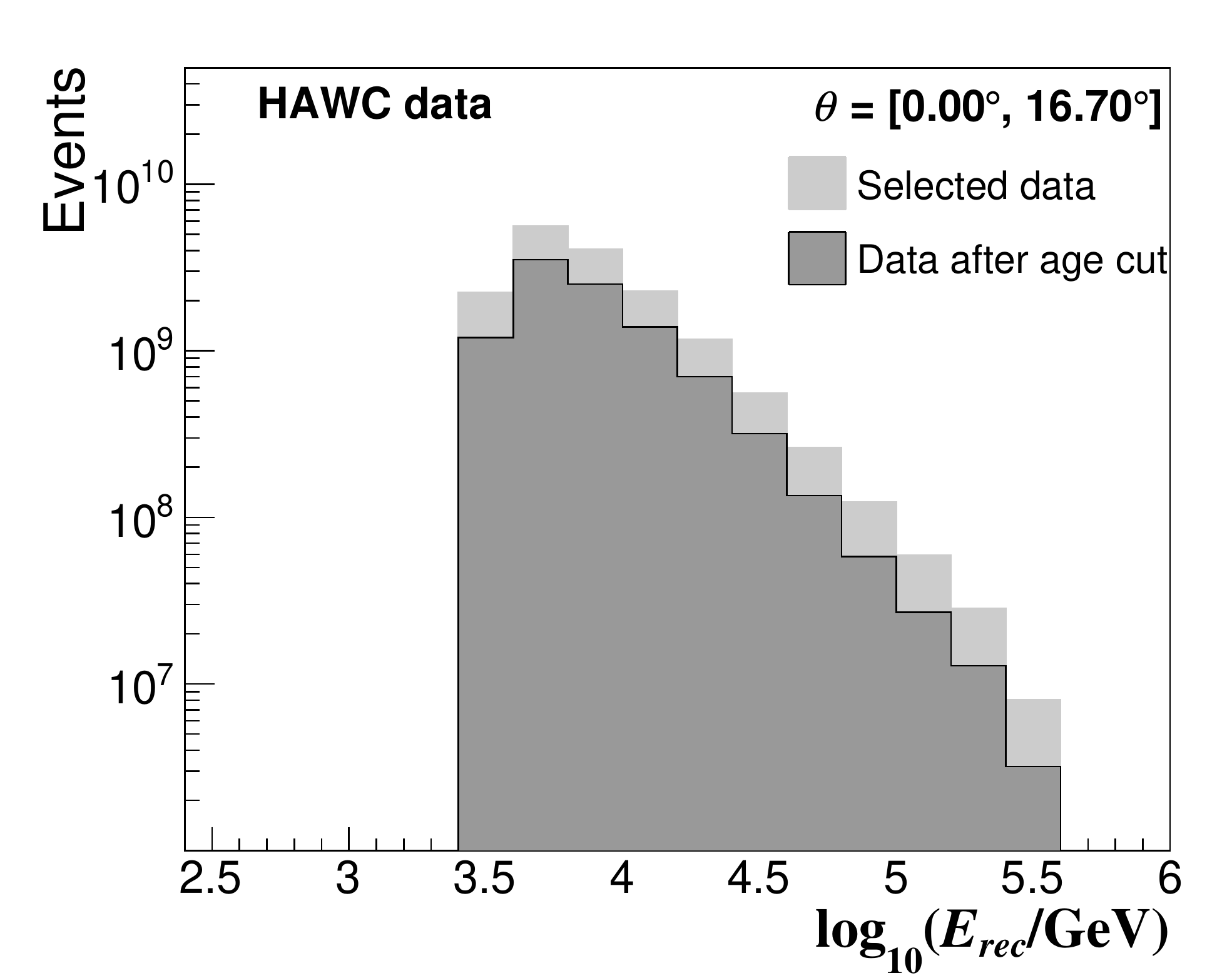}
  \caption{The raw energy distribution of the subsample of HAWC data enriched to events initiated by light primaries (dark gray) compared with the distribution previous to the employment of the age cut (light gray). The plots are not corrected for energy  bin resolution effects.}
  \label{FigEnergyHisto}
  \end{figure}

   The next step in the reconstruction procedure is to build the energy histogram $N(E_{rec})$ for the selected subsample of EAS obtained after the shower age cut. This is shown in figure \ref{FigEnergyHisto}, where a bin size of $\Delta \log_{10}(E_{rec}/\gev) = 0.2$ has been used, which is of the order of magnitude of the energy resolution in the selected subsample. 
 
    \begin{figure}[!t]
    \centering
    \includegraphics[width=3.3in]{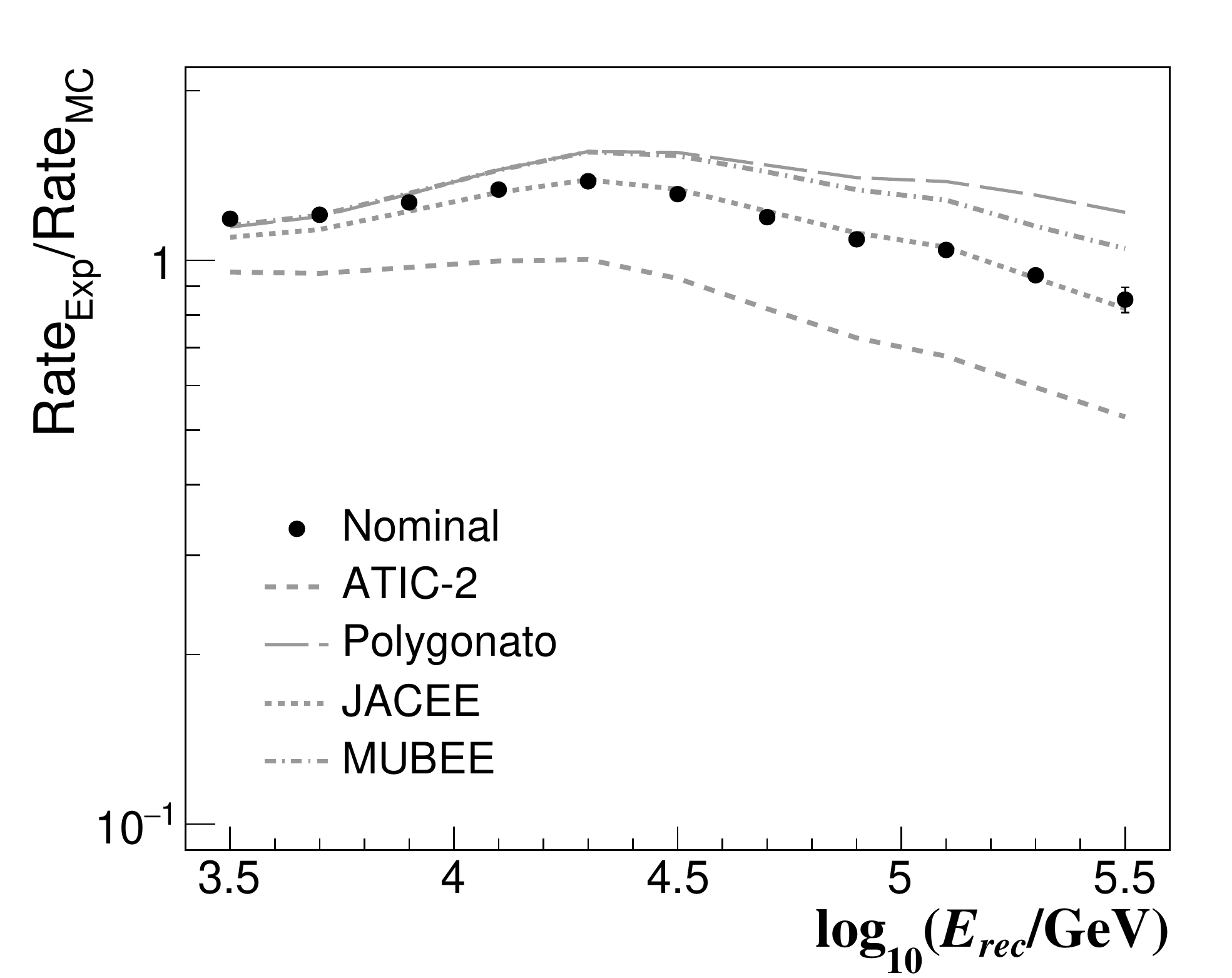}
    \caption{The ratio between the event rates for measured data and MC simulations using QGSJET-II-04 and different cosmic ray composition models  (cf.  Appendix \ref{appmodels}) after applying the shower age cut. The ratios are plotted against the reconstructed primary energy.  The composition models used for each curve are the nominal one (data points), ATIC-2 (dashed line), Polygonato (long dashed line), JACEE (dotted line) and MUBEE (dashed-dotted line). Statistical errors are displayed as vertical error bars for the case of the nominal model.}
    \label{Fig_Ratio_Exp_MC}
  \end{figure}  
  
\begin{figure}[!t]
   \centering
  \includegraphics[width=3.3in]{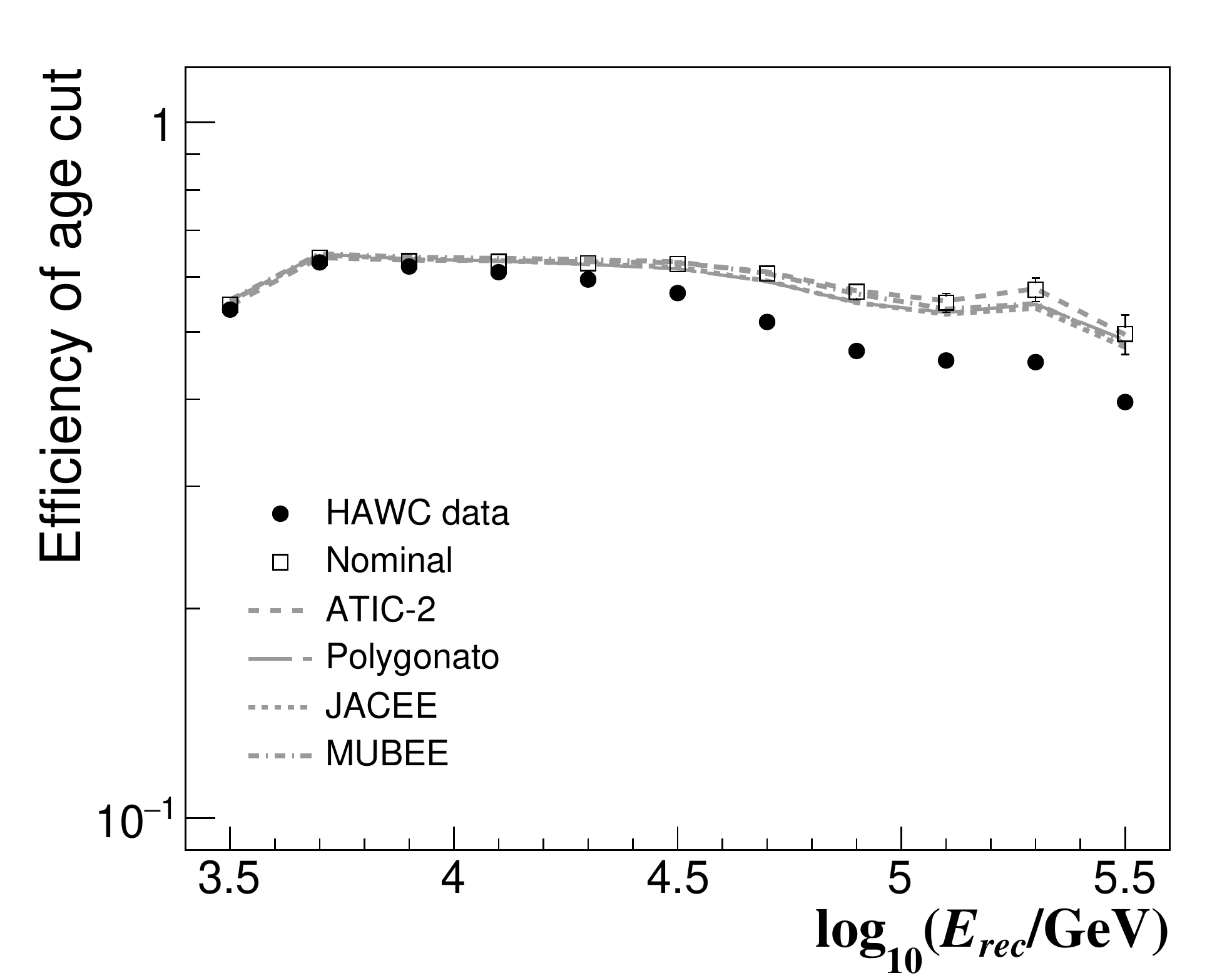}
  \caption{The efficiency of the shower age cut for HAWC data and MC simulations versus the reconstructed primary energy. The efficiency is estimated from its effect on the event samples selected with the criteria of Sec.  \ref{selectioncuts}. The shown curves represent the  age cut efficiencies for HAWC data (black circles) and QGSJET-II-04 simulations with the nominal (open squares), ATIC-2 (dashed line), Polygonato (long dashed line), JACEE (dotted line) and MUBEE (dashed-dotted line) composition models of cosmic rays. The vertical error bars represent statistical errors.}
  \label{Fig_Eff_agecut}
  \end{figure}

   In Fig.~\ref{Fig_Ratio_Exp_MC}, we have estimated the ratio between the  measured rate of events and the predicted ones using MC simulations with our nominal composition model and alternative ones, described in Appendix \ref{appmodels},  after applying the shower age cut of Fig.~\ref{FigAge}. From Fig.~\ref{Fig_Ratio_Exp_MC}, we observe that the ratios have values between $1.6$ and $0.5$ and  they vary with the reconstructed energy. All of them exhibit a maximum at around $\log_{10}(E_{rec}/\gev) = 4.3$. The measured rates are larger than the expectations with  the nominal, Polygonato, JACEE and MUBEE models, but smaller than the predictions with the ATIC-2 model. Therefore, albeit of the individual differences between the data and the models, the experimental rates are within the expectations from the cosmic ray composition models. On the other hand, the energy evolution of the ratio curves implies that the  energy distribution behind the measured data does not follow a single power law like in MC simulations. The results of   Fig.~\ref{Fig_Ratio_Exp_MC} seem to hint the existence of a break in the measured energy distribution at around the position of the maximum in the ratio curves. We will come to this point later in Sec. \ref{results}.

  Finally, in Fig.~\ref{Fig_Eff_agecut}, we have calculated the efficiency of the shower age cut or the fraction of remaining events after applying the age cut over the selected HAWC data. The computation was carried out by dividing  the contents of the energy histograms of  Fig.~\ref{FigEnergyHisto} for the subsample of young EAS and for the selected data sample that does not contain the shower age cut. The efficiency of the age cut in measured data is compared with the corresponding efficiency for QGSJET-II-04  simulations using different cosmic ray composition models, including the nominal one. From the plots of Fig.~\ref{Fig_Eff_agecut}, we see that the fraction of remaining events in the experimental subsample of young EAS is smaller than expected from MC simulations mainly at high energies. This discrepancy seems to point out that in the framework of QGSJE-II-04 the relative abundance of heavy nuclei in HAWC data is larger than predicted by the cosmic ray composition models used in this work. Such difference between data and MC simulations reduces the magnitude of the intensity of protons and helium nuclei estimated with the present procedure, but it does not change the main conclusions about its shape. The effect of the discrepancy in the final result was estimated and included as a systematic error (see Appendix \ref{apperrors}).

 \subsubsection{Unfolding procedure}
   
   Now, in order to correct the measured distribution for migration effects.  we must apply an unfolding procedure. For this aim, we employed the Bayesian algorithm \cite{richardson, lucy, agostini}. However, the final result has been verified using the Gold's unfolding procedure (see Appendix \ref{apperrors}) \cite{Gold64, Antoni05, Ulrich01}. In the Bayesian method, the unfolded distribution, $N(E)$ is found iteratively from the measured histogram by means of the calculation of a matrix $P(E|E_{rec})$ which provides the conditional probability that a given event with energy in the bin $E_{rec}$ is due to an EAS with true energy in the interval around $E$.  The smearing matrix is computed using the Bayes’s theorem
   \begin{equation}
 	 P(E|E_{rec}) = \frac{P(E_{rec}|E) \cdot P(E)}{\sum_{E^\prime} P(E_{rec}|E^\prime) \cdot P(E^\prime)},
 	\label{eq2}
   \end{equation}    
   where $P(E) = N(E)/\sum_{E} N(E)$ is the previous approximation to the probability of the unfolded distribution and $P(E_{rec}|E)$ is the response matrix of the detector. The response matrix is generally estimated from MC simulations. It represents the probability that a shower with a primary energy  $E$ is reconstructed with an energy $E_{rec}$ in the experiment.  The smearing matrix is then substituted into the equation
    \begin{equation}
 	 N(E) = \sum_{E_{rec}} P(E|E_{rec}) N(E_{rec}),
 	\label{eq3}
   \end{equation}  
   from which the unfolded distribution is obtained.
   
   The unfolding procedure starts with a first guess at the probability $P(E)$. This is used to estimate a more accurate $N(E)$  by means of Eqs.~(\ref{eq2}) and (\ref{eq3}), which is employed to calculate $P(E)$ for the next iteration.
   The stopping criterion is described below.
 
\begin{figure}[!t]
    \centering
    \includegraphics[width=3.3in]{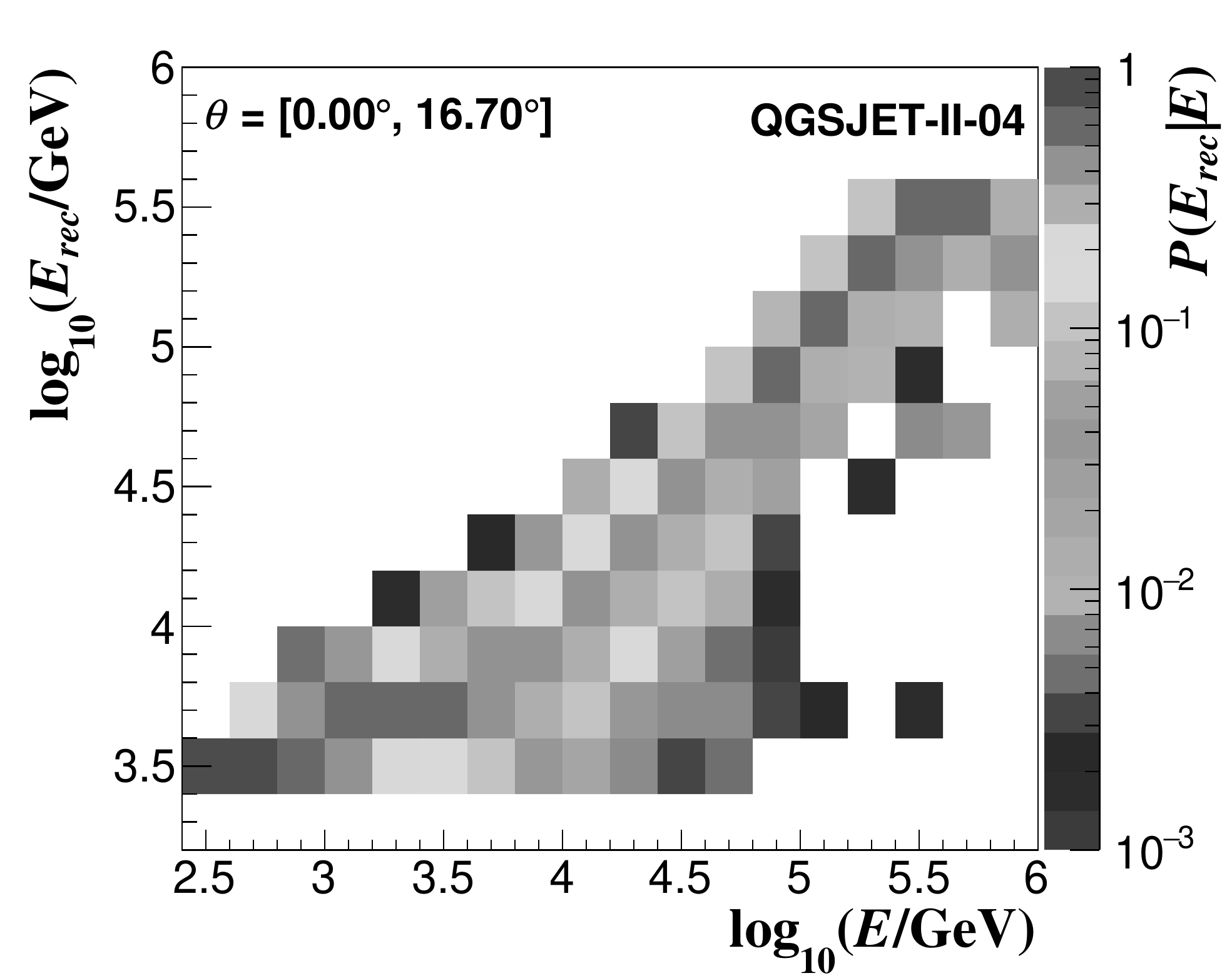}
    \caption{The response matrix $P(E_{rec}|E)$  for the subset of EAS enriched to events induced by protons and helium nuclei as estimated from MC simulations using our nominal composition model.}
    \label{Responsematrix}
  \end{figure}

   To begin with, we computed the response matrix for the subset of selected events by using our nominal MC dataset. The matrix was built in the $\log_{10} (E_{rec}/\gev)$ vs $\log_{10} (E/\gev)$ phase space for the ranges from $2.4$ to $6.0$, which were both divided in bins of width $0.2$ as for the measured energy histogram. The resulting response matrix is shown in Fig.~\ref{Responsematrix}.  
   For the initial guess of $N(E)$, we used a power-law distribution with spectral index as predicted by the nominal composition model. In addition, to eliminate the propagation of statistical fluctuations from the response matrix, a smoothing procedure was applied to $N(E)$ at the end of each iteration, but not in the final result. The procedure was carried out by smoothing the unfolded distribution with a broken power-law function \cite{BPL} inside the range from $E = 10^{3}$ to $3.2 \times 10^{5} \, \gev$. However, we have cross-checked the unfolded result using as a smoothing function a polynomial of degree $5$ and the smoothing 353HQ-twice algorithm \cite{smoothing} as installed in the ROOT package \cite{root}. Employing as a smoothing function a single power-law formula  produces an unfolded energy distribution whose forward-folded histogram is flatter than the original distribution  $N(E_{rec})$. For this reason, we avoided to use this approach in our unfolding analysis.  
  
   As the stopping criterion, we  look for a minimum in the weighted mean squared error ($WMSE$) \cite{Cowen98, Antoni05} at which the squared sum of the average statistical uncertainties and the systematic biases of the result are smallest. The $WMSE$ is defined as:
   \begin{equation}
       WMSE = \frac{1}{n} \sum^{n}_{i = 1} \frac{\bar{\sigma}^{2}_{stat, i} + \bar{\delta}^{2}_{bias, i}}{N(E_i)},
    \label{eqn4}
   \end{equation}
   where $n$ is the number of energy bins in the unfolded distribution, $\bar{\sigma}_{stat, i}$ is the average statistical uncertainty of $N(E_i)$, while $\bar{\delta}_{bias, i}$ is the mean bias of $N(E_i)$ introduced by the unfolding algorithm. For the estimation of $\bar{\sigma}_{stat}$  and $\bar{\delta}_{bias}$, the bootstrap method \cite{Rice10} was implemented as described in \cite{Antoni05, Fuhrmann12}. In particular, at a given iteration level, a set of $m = 60$ toy distributions are produced from $N(E_{rec})$ for the estimation of $\bar{\sigma}_{stat}$  and $\bar{\delta}_{bias}$.
   
   \subsubsection{Reconstruction of the energy spectrum}

   Once the unfolded spectrum $N(E)$ for the data subsample is obtained, the energy spectrum for protons and helium nuclei is calculated from the formula
   \begin{equation}
 	 \Phi(E) = \frac{N(E)}{A_{eff}(E) \, \, \Delta E \, \,  T_{eff} \, \, \Delta \Omega},
 	\label{eq6}
  \end{equation}
  where $\Delta E$ is the size of the bin at $E$,  $T_{eff} = 1.18 \times 10^{8} \, \mbox{s}$ is the effective livetime for the collected data,  $\Delta \Omega = 0.27 \, \mbox{sr}$ is the solid angle interval covered by the measurements and $A_{eff}(E)$ is a corrected effective area defined as 
   \begin{equation}
 	 A_{eff}(E) = f_{corr} (E) \, A^{H+He}_{eff}(E).
 	 \label{eq7}
  \end{equation} 
  In the above expression, $f_{corr}(E)$  is the factor introduced to correct the unfolded result for the contamination of heavy elements ($Z \geq 3$);  $A^{H+He}_{eff}(E)$ is the effective area (defined below) of the instrument for  detection of protons and helium nuclei in the enriched subsample of young EAS, which correct the unfolded result for the loss of light primaries after using the different selection cuts.
 
\begin{figure*}[!t]
     \centering
     \footnotesize
     \includegraphics[width=3.0in]{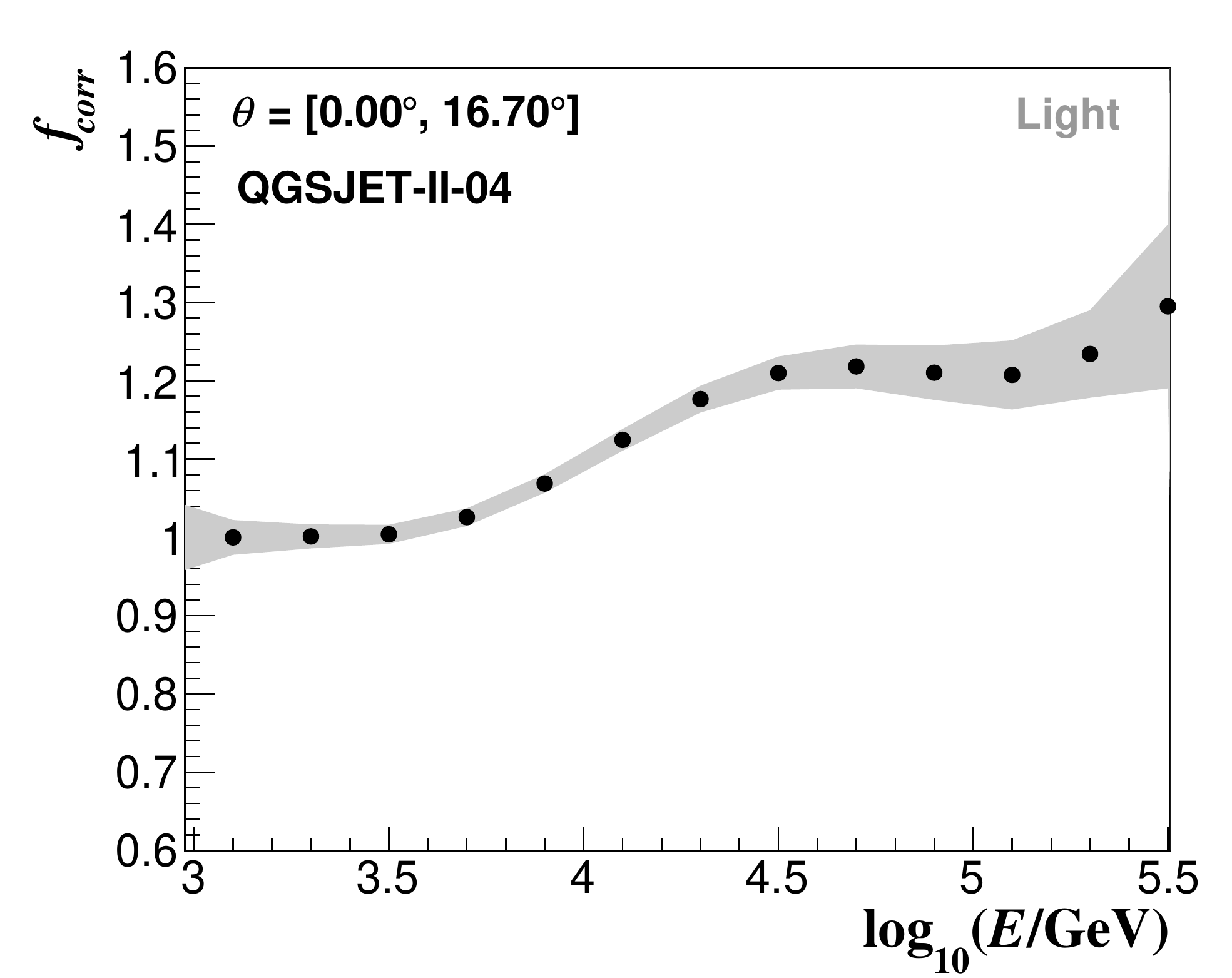} 
     \includegraphics[width=3.0in]{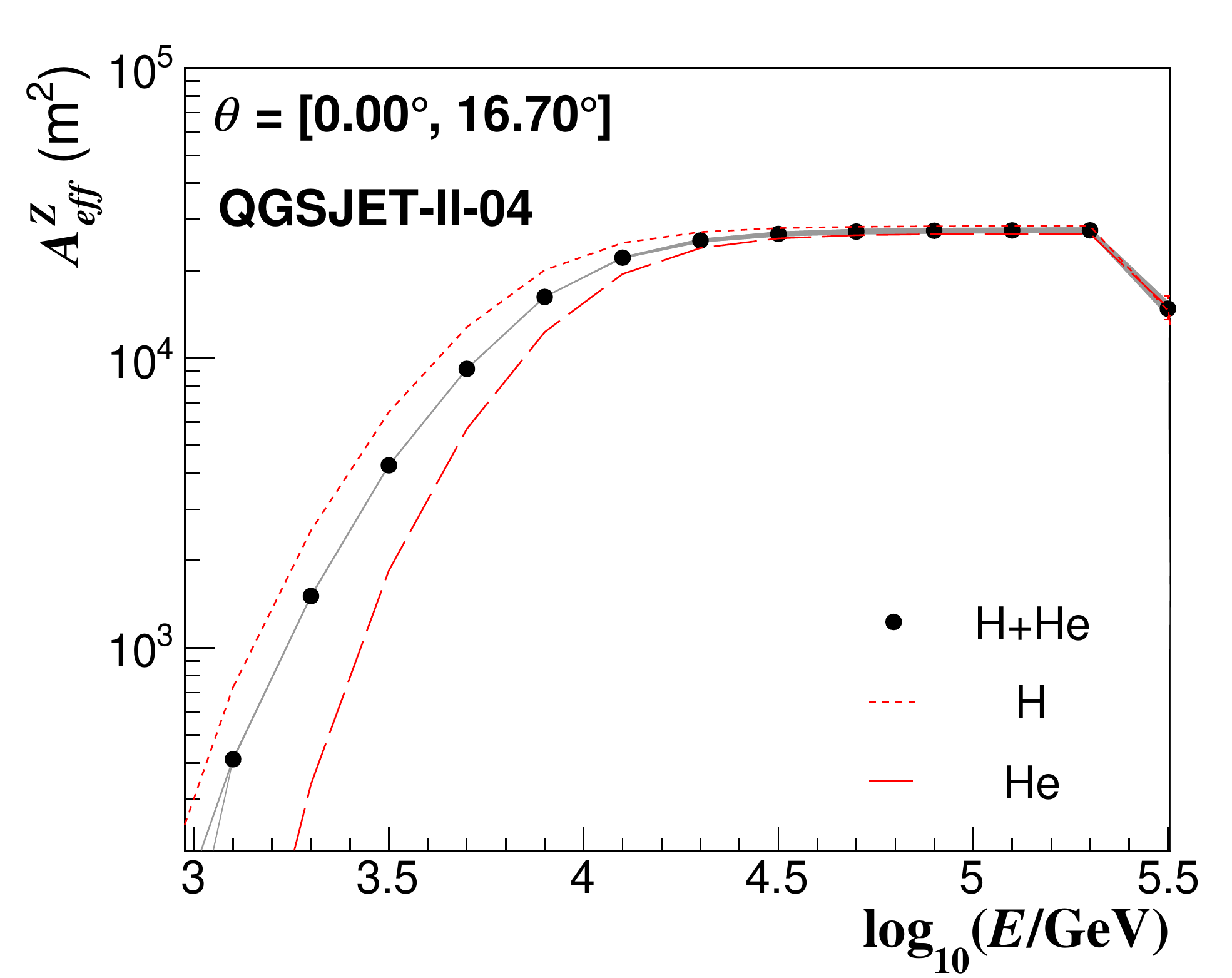} 
    \caption{Left: the correction factor applied to the energy spectrum of the light mass group of cosmic rays for the contamination of heavy nuclei vs the true primary energy $E$. Right: the effective area for 
    light primaries (black data points) compared to the effective area  for protons (upper dotted line) and helium (lower dashed line) primaries and plotted against the true primary energy $E$. The gray bands represent the statistical uncertainties in both panels. In these plots, we used our MC simulations with the nominal cosmic ray composition model.}
  	\label{Effarea}
    \end{figure*}

  The factor $f_{corr} (E)$ is estimated as the inverse of the proportion of light primaries in the aforementioned subsample at the energy $E$, calculated from MC simulations  with our nominal cosmic ray composition model. Specifically, $f_{corr} (E)$  is just the ratio $N^{MC}(E)/N^{MC}_{H+He}(E)$ between the number of selected events after using the age cut $N^{MC}(E)$  and the number of H and He events $N^{MC}_{H+He}(E)$ in this subsample at the true energy E.  The result is shown in Fig.~\ref{Effarea}  (left panel). From this plot, we observe that $f_{corr} (E)$ grows at high energies, due to the increasing relative abundance of the heavy mass nuclei in the subsample. It has a feature at around $\log_{10} (E/\gev) = 4.5$, because of  the effect of the trigger and selection efficiency on the energy distribution of the heavy mass group in the MC data subset. Below this energy, the corresponding efficiency decreases rapidly producing a fast reduction in $f_{corr} (E)$ at low energies. At higher energies, the efficiency for heavy primaries starts to reach its maximum value, which reduces the rate of increment of $f_{corr} (E)$. 
 
  The shape of the correction factor is almost similar for the different cosmic ray composition models employed in this work, with some differences in the slopes and the magnitudes of the curves due to the distinct abundances of the heavy elements in each model. Since the light mass group of cosmic rays is the dominant component in the subsample of young EAS, the effect of the uncertainties of the relative abundance of the heavy component on $f_{corr} (E)$ is reduced and, hence, the corresponding error on the shape of the reconstructed energy spectrum. Therefore, we can use $f_{corr} (E)$ as estimated with the MC simulations for the present analysis independently of the shape of the experimental spectrum for H$+$He in this energy regime. This point is demonstrated in Appendix \ref{appchecks}, where we have performed systematic checks with MC simulations and different composition models that show that our analysis method allows to reconstruct the shape of the spectrum of light primaries without previous knowledge about the existence of features in the spectrum under analysis. Even more, the study of systematic uncertainties performed in Appendix \ref{apperrors} points out that the shape of the reconstructed spectrum for light primaries is the same whether we use our nominal composition model or the alternative models described in Appendix \ref{appmodels}.

  The effective area for the light primaries is defined by \cite{Hawc17} 
  \begin{equation}
 	 A^{H+He}_{eff}(E) = A_{thrown}  \frac{\cos \theta_{max} + \cos \theta_{min}}{2} 
 	  \epsilon^{H+He}(E).
 	\label{eq8}
  \end{equation}
  Here $A_{thrown}$ is the total area at ground level where the core of the MC events were thrown, the cos term gives the projection of the area averaged on the solid angle within  the zenith angle range from $0^\circ$ to $ 16.7^\circ$, 
  and  $\epsilon^{H+He}$ is the probability that an EAS event induced by a light primary (hydrogen or helium nuclei) triggers the detector and passes all the selection cuts for the  young shower subsample. $A^{H+He}_{eff}(E)$ from our nominal MC simulations is plotted in Fig.~\ref{Effarea} (right panel) against the true primary energy compared to the effective area for pure hydrogen and helium nuclei. The maximum efficiency is achieved between $\log_{10}(E/\mbox{GeV}) \sim 4$ and $5.4$. 
  At lower energies, the decrease is due to the trigger and the selection cuts, while above $\log_{10}(E/\mbox{GeV}) = 5.4$, it is caused by the cut on the reconstructed energy.
  The effective areas for pure H and He nuclei are not equal.  For $\log_{10}(E/\mbox{GeV}) < 3.8 \, \gev$ they differ by more than $30 \%$ with respect to the central value for $H+He$, for this reason and due to the reduction of the effective area as well as the increment of the correlations at lower energies, we report the spectrum only above $6 \, \tev$.

  \section{Results}
  \label{results}
  
     \begin{figure}[!t]
    \centering
    \includegraphics[width=3.3in]{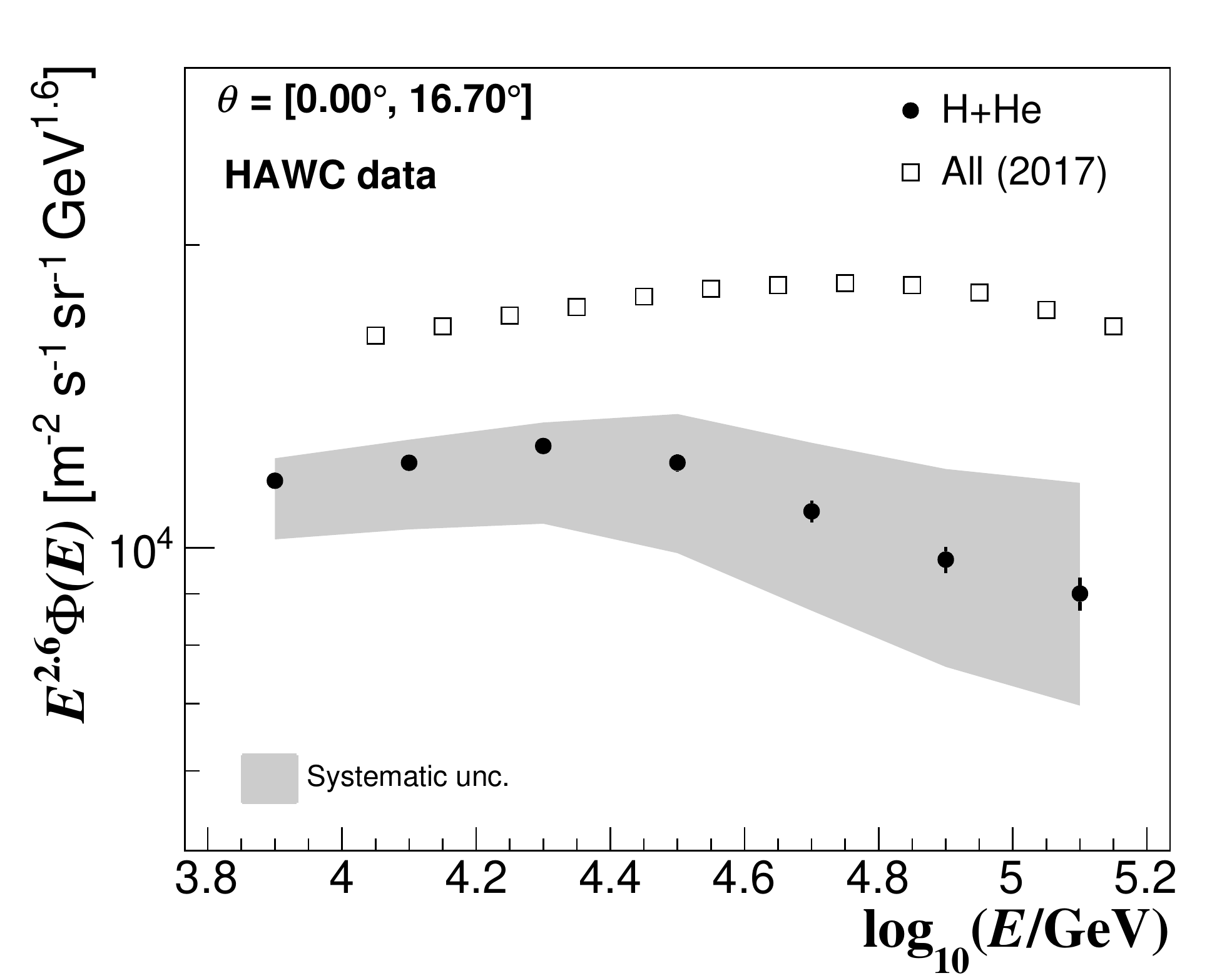}
    \caption{The reconstructed cosmic ray energy spectrum for protons plus helium primaries
    in the present analysis with HAWC (black points).
    The gray error band and the error bars represent the systematic and statistical uncertainties, respectively. The all-particle energy spectrum for cosmic rays measured by HAWC and presented  in \cite{Hawc17} is also shown (open squares). 
    Statistical errors smaller than the data points are not shown.}
    \label{spectrum}
  \end{figure}

  \subsection{Energy spectrum of H plus He cosmic ray nuclei}

  The energy spectrum of light cosmic ray nuclei estimated from this analysis is presented in Fig.~\ref{spectrum} and Table \ref{tabspectrum} for  $\log_{10}(E/\mbox{GeV}) = [3.8, 5.2]$ along with its corresponding systematic and statistical errors. The result has been constrained to $\log_{10}(E/\mbox{GeV}) < 5.2$ due to a rapid increase of the systematic uncertainties at higher energies (as we will see in the next subsection). Figure ~\ref{spectrum} seems to reveal a 
  slope change around a few tens of $\tev$ in the spectrum of $\mbox{H}+\mbox{He}$ primaries. 
  We  compared 
  two fits to the data with a single power law
  \begin{equation}
        \Phi (E)= \Phi_0 E^{\gamma_1},
    	\label{eq9}
 \end{equation}
 and a broken power law  \cite{BPL, DanielKG2013}
 \begin{equation}
 	 \Phi (E)= \Phi_0 E^{\gamma_1} \left[1 + \left(\frac{E}{E_0}\right)^\varepsilon  \right]^{(\gamma_2 - \gamma_1)/\varepsilon},
 	\label{eq10}
  \end{equation}
  where $E_0$ is the energy position of the break, $\gamma_1$ and $\gamma_2$ are the spectral indexes before and after the break in the spectrum,  
  while $\varepsilon$ measures the sharpness of the feature. The fits were done by chi-squared minimization 
  for correlated data points \cite{PDG19stat}, 
  taking into account the correlation  
  from the unfolding. The covariance matrix  
  has the contributions from the statistics of MC and experimental data
  (see the next subsection and Appendix \ref{apperrors} for details). The contributions
  were calculated according to \cite{Adye11a, Zig17} and added
  to obtain the total covariance matrix, $V_{stat}$,  used for the fit.

\begin{table}[!b]
    \begin{center}
    \caption{Values of the energy spectrum $\Phi(E)$ for the light mass group of cosmic rays as derived in this analysis using HAWC data calibrated with the QGSJET-II-04 model. The width of the energy bins employed in this study is $\Delta \log_{10}(E/\gev) = 0.2$. The statistical ($\delta \Phi_{stat}$) and systematic ($\delta \Phi_{syst}$) errors of the spectrum are also given.}

   \begin{tabular}{cccccccccc}
   \hline 
   \hline 
    $E$  && $\Phi(E)$ && $\pm$ & $\delta \Phi_{stat}$ & & $+ \delta \Phi_{syst}$& & $- \delta \Phi_{syst}$\\
    $[\gev]$ && \multicolumn{8}{c}{$[\mbox{m}^{-2} \, \mbox{s}^{-1}  \, \mbox{sr}^{-1}  \, \mbox{GeV}^{-1}]$} \\
    \hline\\
$7.94\times 10^{3}$ && $(8.44$ && $\pm$ & $0.07$ & & $+0.45$& &$-1.06)\times 10^{-7}$ \\
$1.26\times 10^{4}$ && $(2.66$ && $\pm$ & $0.03$ & & $+0.14$& &$-0.38)\times 10^{-7}$ \\
$2.00\times 10^{4}$ && $(8.34$ && $\pm$ & $0.12$ & & $+0.46$& &$-1.36)\times 10^{-8}$ \\
$3.16\times 10^{4}$ && $(2.42$ && $\pm$ & $0.05$ & & $+0.29$& &$-0.45)\times 10^{-8}$ \\
$5.01\times 10^{4}$ && $(6.55$ && $\pm$ & $0.16$ & & $+1.11$& &$-1.33)\times 10^{-9}$ \\
$7.94\times 10^{4}$ && $(1.77$ && $\pm$ & $0.05$ & & $+0.41$& &$-0.39)\times 10^{-9}$ \\
$1.26\times 10^{5}$ && $(4.95$ && $\pm$ & $0.19$ & & $+1.43$& &$-1.12)\times 10^{-10}$ \\
   \hline
   \hline 
   \end{tabular}
   \label{tabspectrum}
   \end{center}
  \end{table}
   
\begin{figure}[!t]
    \centering
    \includegraphics[width=3.3in]{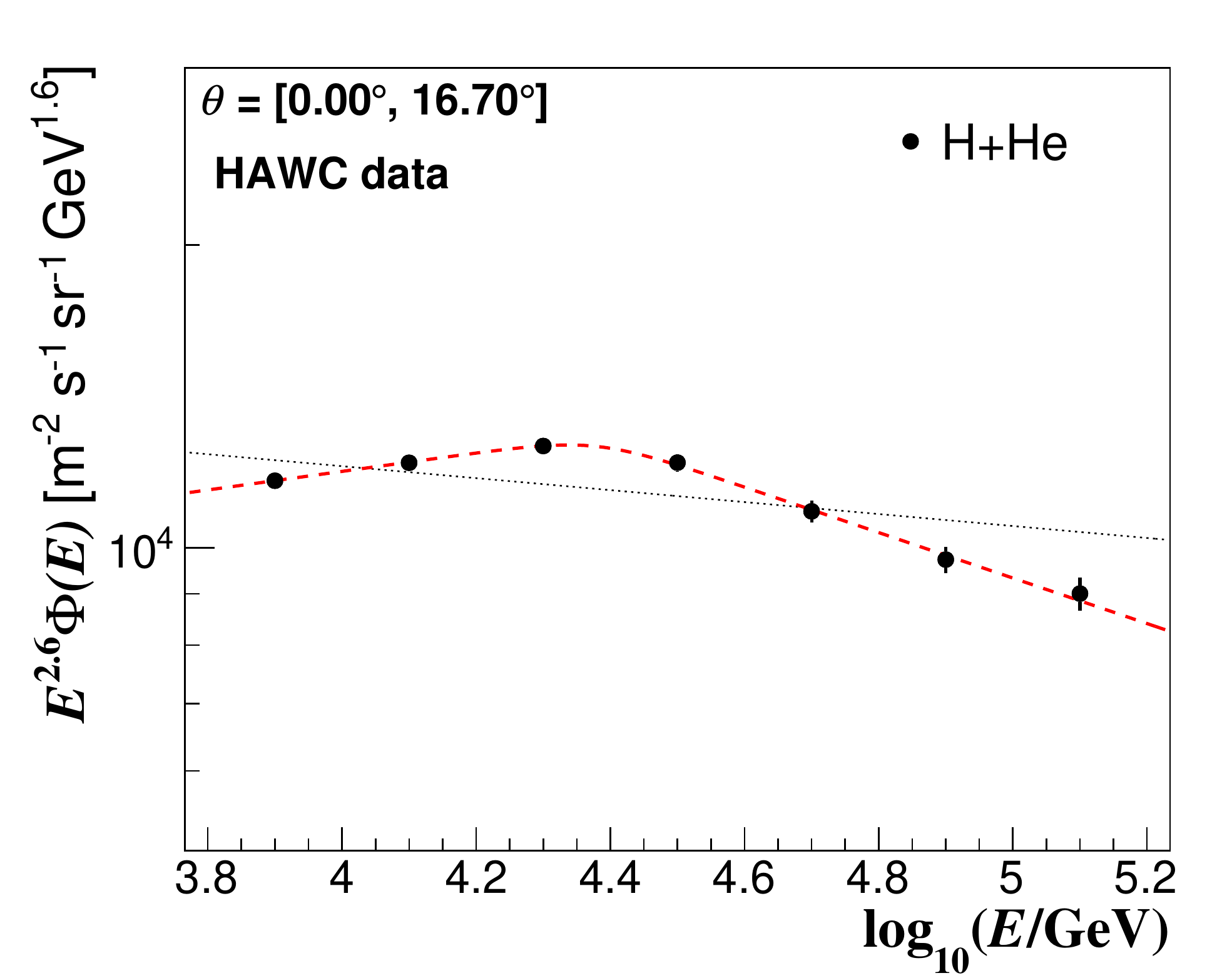}
    \caption{The fits to the HAWC energy spectrum for H plus He nuclei (black circles) in the range $\log_{10}(E/\mbox{GeV}) = [3.8, 5.2]$. The black dotted line shows the fit with the single power-law function of Eq.~(\ref{eq9}) and the red dashed line, the fit with the broken power-law expression (\ref{eq10}). Only statistical uncertainties are shown, which are represented by vertical error bars. At low energies, the diameter of the data points is larger than the error bars.}
    \label{fitspectrum}
  \end{figure}

  By fitting the spectrum with Eq.~(\ref{eq9}), we obtained 
  \begin{eqnarray}
   \Phi_0 &=& 10^{4.32 \pm 0.02} \, \mbox{m}^{-2} \, \mbox{s}^{-1}  \, \mbox{sr}^{-1}  \, \mbox{GeV}^{-1}, \nonumber \\
   \gamma_1 &=& -2.66 \pm 0.01, \nonumber
  \end{eqnarray}
  with $\chi_0^2 = 177.51$, for $\nu_0 = 5$ degrees of freedom. 
  The fit with the broken power-law formula of Eq.~(\ref{eq10})  yielded 
 \begin{eqnarray}
   \Phi_0 &=& 10^{3.71 \pm 0.09} \, \mbox{m}^{-2} \, \mbox{s}^{-1}  \, \mbox{sr}^{-1}  \, \mbox{GeV}^{-1}, \nonumber \\
   \gamma_1 &=& -2.51 \pm 0.02, \nonumber \\
   \gamma_2 &=& -2.83 \pm 0.02, \nonumber \\
   E_0 &=& 10^{4.38 \pm 0.06} \, \gev, \nonumber \\
   \varepsilon &=& 9.8 \pm 4.1. \nonumber
  \end{eqnarray}
  The resulting chi-squared was $\chi_1^2 =  0.26$ and the number of degrees of freedom were $\nu_1 = 2$. The fitted functions are shown in Fig.~\ref{fitspectrum}. 
   We will use now the  test statistic 
  \begin{equation}
      TS = -\Delta \chi^2 =  -(\chi_1^2 - \chi_0^2)
      \label{eqTS}
  \end{equation}
  to compare the scenarios.
  From the fits, we found 
  $TS_{obs} = 177.25$. We translated this into a $p-$value using $49 \times 10^{3}$ toy MC spectra with correlated data points,
  assuming that the data is best described by the single power-law formula.
  Following \cite{PDG19MC} we used  a multivariate Gaussian as a probability distribution for the data and the covariance matrix $V_{stat}$. In the resulting $TS$ values, we found just one case with $TS \geq TS_{obs}$, which implies a $p$-value equal to $2 \times 10^{-5}$. We also observed that $0.5 \%$ of the  MC toy spectra have a $\chi^2$  smaller than $\chi_1^2 =  0.26$ when using  formula (\ref{eq10}) in the fits.
  Thus, from the test statistic, the broken power-law hypothesis is favored by the data with a significance  of $4.1 \, \sigma$.  We also performed several ``sanity 
  checks'' (see Appendix \ref{appchecks}) to rule out the kink 
  being produced by systematic effects. 
  Our result confirms the kink that HAWC \cite{Hawc19} previously reported at tens of $\tev$ in the cosmic ray  energy spectrum for protons and helium nuclei and the hints found in Fig.~\ref{Fig_Ratio_Exp_MC} in favor of a spectral break in the spectrum of this mass group.
  
\begin{center}
 \begin{figure*}[!t]
    \centering
    \includegraphics[width=5.0in]{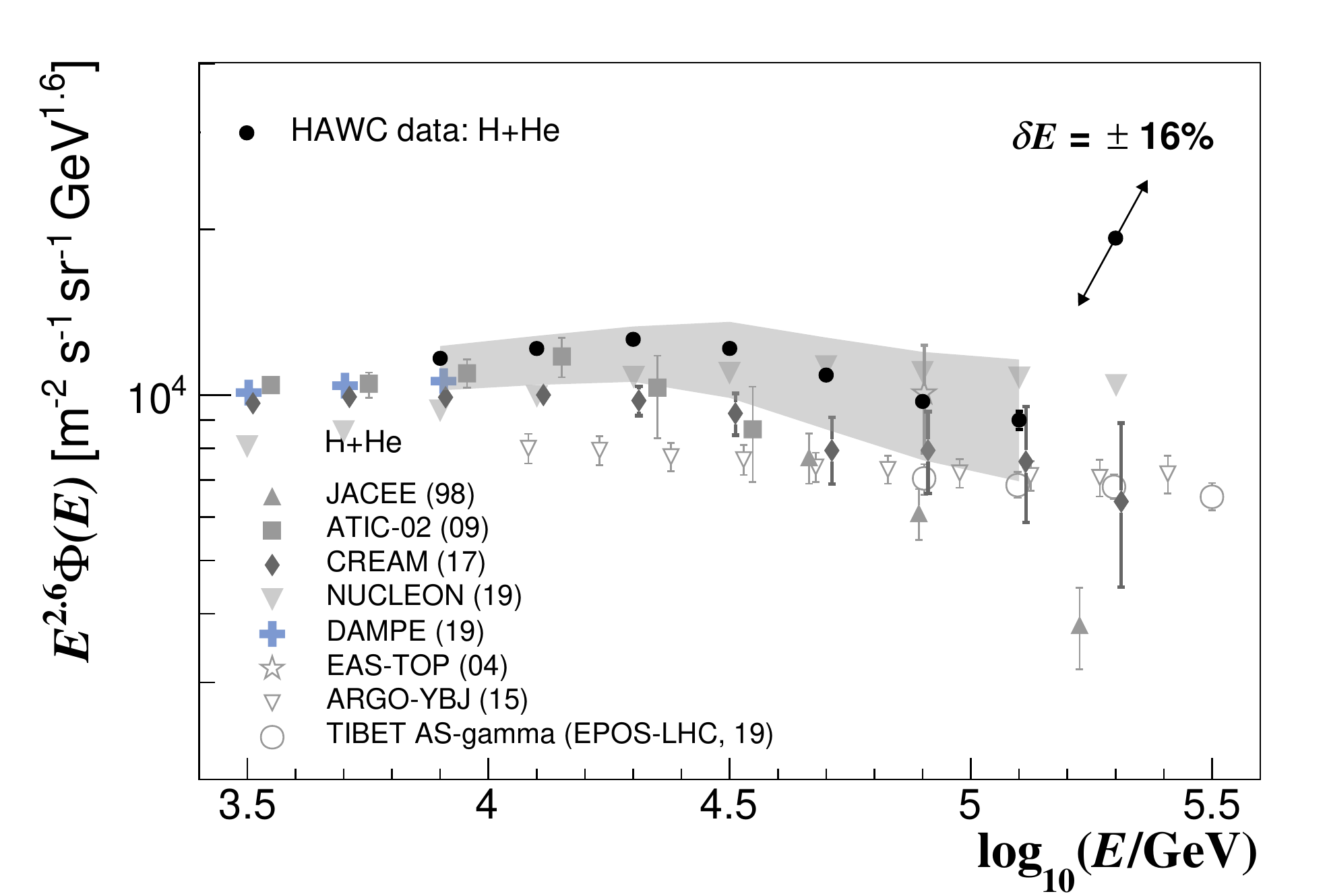}
    \caption{The spectrum for H$+$He cosmic ray nuclei as measured by HAWC (black circles) and calibrated with the post-LHC hadronic interaction model QGSJET-II-04 in comparison with similar measurements of the spectrum from direct and indirect experiments. In particular, the spectra from the direct cosmic ray detectors  ATIC-2 (squares) \cite{atic09,  atic092}, CREAM I-III (diamonds) \cite{cream17}, NUCLEON (downward solid triangles) \cite{nucleon19}, JACEE (upward triangles) \cite{jacee} and DAMPE (crosses) \cite{Dampe19a} are presented.
    Indirect measurements are also shown from the EAS observatories ARGO-YBJ (downward hollowed triangles) \cite{Argo15}, Tibet AS-gamma (open circles) \cite{Tibetasgamma19} and EAS-TOP (hollowed star) \cite{Eastop04}. The gray band around the HAWC data points represents 
    systematic uncertainties. The statistical uncertainties of the HAWC measurements are shown with vertical error bars. The magnitude of the systematic uncertainty in the spectrum after varying the energy scale within systematic errors $\delta E = \pm 16 \%$ is shown in the upper right corner of the plot with arrows.}
    \label{spectrumcompilation}
  \end{figure*}
\end{center}

  In Fig.~\ref{spectrumcompilation},  this work  is compared with other experiments. We have included measurements from the direct cosmic ray detectors ATIC-2 \cite{atic09,  atic092}, CREAM I-III  \cite{cream17}, NUCLEON  \cite{nucleon19}, JACEE  \cite{jacee} and DAMPE \cite{Dampe19a} along with data from the air shower observatories ARGO-YBJ  \cite{Argo15}, Tibet AS-gamma \cite{Tibetasgamma19} and EAS-TOP  \cite{Eastop04}. 
  Close to $E = 10 \, \tev$, we see good agreement of HAWC data with ATIC-2  within systematic uncertainties.
  Between $20$ and $126 \, \tev$,  the HAWC measurement is in a fair agreement with the NUCLEON spectrum. In general, the HAWC result  is higher than the CREAM I-III and  ARGO-YBJ data below $80 \, \tev$. However, close to $100 \, \tev$, the CREAM I-III and  ARGO-YBJ  spectra are in agreement within systematic uncertainties with the HAWC spectrum. On the other hand, HAWC data is above JACEE and Tibet AS-gamma measurements. The HAWC spectrum is not in agreement with the single power law behavior  reported by ARGO-YBJ in this energy interval \cite{Argo15}, while above $24 \, \tev$ the slope of the HAWC spectrum is harder than that of JACEE, but softer than the one of NUCLEON. Finally, at higher energies, our result is in agreement with the single data point from the EAS-TOP experiment at $\sim 80 \, \tev$. 
  
  \subsection{Uncertainties in the magnitude of the spectrum}

\begin{figure}[!t]
    \centering
    \includegraphics[width=3.3in]{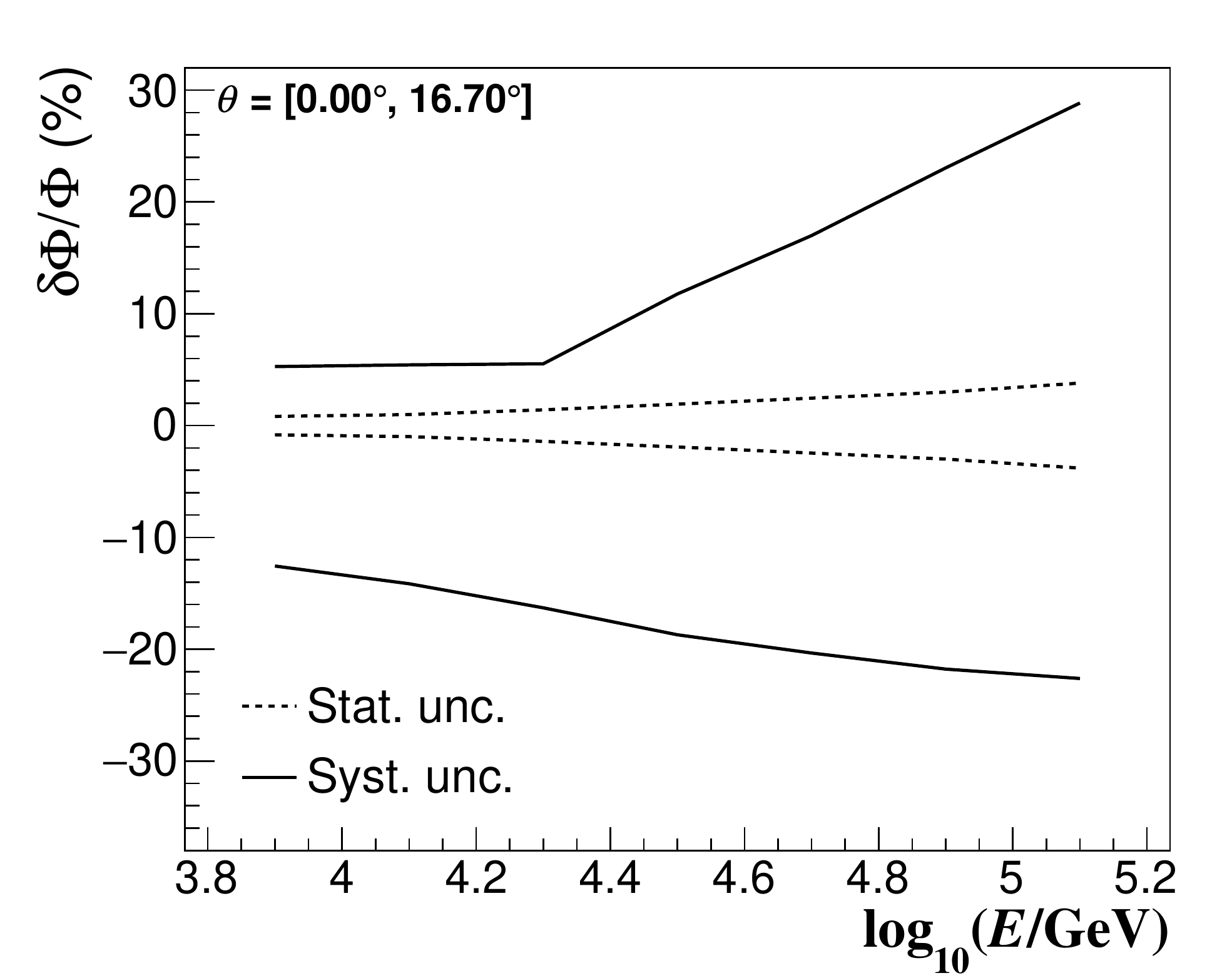}
    \caption{The systematic (continuous line) and statistical (dotted line) relative uncertainties of the cosmic ray spectrum for protons plus helium nuclei measured with HAWC, which is shown in Fig.~\ref{spectrum}.}
    \label{spectrum_error1}
  \end{figure}

  The total uncertainty (the sum in quadrature of the systematic and statistical uncertainties) of the unfolded spectrum is between $+29.1 \%$ and $-22.9  \%$ for energies $ \log_{10}(E/\mbox{GeV})< 5.2$. Figure~\ref{spectrum_error1} shows that the uncertainties decrease for low energies. The statistical uncertainty (dominated by MC statistics) rises from  $\pm 0.8 \%$ to $\pm 3.8  \%$ between 
  $\log_{10}(E/\gev) = 3.8$ and 
  $\log_{10}(E/\gev) = 5.2$.
  The systematic uncertainties over the same range vary from 
  $+5.3   \%/$$-12.6   \%$ to  $+28.9   \%/$$-22.6 \%$: 
  the systematic uncertainties dominate the total error.  
  For energies $ \log_{10}(E/\mbox{GeV}) > 5.2$, the systematic error grows
  rapidly to $+41.8 \%/-25.1  \%$ at $ \log_{10}(E/\mbox{GeV}) \sim 5.4$, so 
  we report the spectrum only up to $ \log_{10}(E/\mbox{GeV}) = 5.2$.

  We evaluated a number of sources of systematic uncertainty.
  The most important ones involve uncertainties on the PMT performance ($+28.5   \%/-10.6   \%$ effect on the deconvolved spectrum); the cosmic ray composition model ($+2.1  \%/-17.2  \%$); and the hadronic interaction model ($-10.9  \%$  to $ -3.7 \%$).  The rest of the systematic uncertainties, added in quadrature, contribute
  $+5.2  \%/-7.0  \%$.
  Appendix \ref{apperrors} presents the systematic error evaluation in detail.
  
  The feature observed in the energy spectrum of light primaries does not disappear under the effect of these systematic sources, although we observe some variations in the intensity of the spectrum and the value of the change of the spectral index $\Delta \gamma$ around the break. One of the dominant systematic sources in the spectrum is the uncertainty on the PMT performance, which is dominated by the PMT-late-light systematics (cf.~ Appendix \ref{apperrors}). The late light effect dominates the upper limit of the total systematic error and introduces an energy dependent variation in the spectrum, which grows from $+4.5 \%$ up to $+28.3\%$ and reduces $\Delta \gamma$.  If we apply this systematic shift on the energy spectrum and repeat the fit with formula (\ref{eq10}) as well as the corresponding statistical analysis of the feature, we find that the significance of the break is reduced up to $3.8 \,  \sigma$. Hence, in spite of the flattening of the spectrum, the feature is still significant. 

 On the other hand, we have also investigated, whether the measured spectrum is consistent with a break in the spectrum for light primaries at $24 \, \tev$ even after considering the systematic uncertainties due to the PMT-late-light effect in the analysis of the spectral feature. For this aim, we have fitted the HAWC spectrum below the break with a power-law function, like equation (\ref{eq9}), including the systematic uncertainties due to the PMT late light in the procedure. Then, we extrapolated the data up to higher energies and use it to predict the number of expected EAS in the subsample of enriched EAS for the energy interval $\log_{10}(E/\gev) = [4.4, 5.2]$ under this simple hypothesis. For the estimation we used the expression 
   \begin{equation}
   N(E_i, E_f) = A_{eff}(\bar{E}) \, \,  T_{eff} \, \, \Delta \Omega \, \, \Phi_0 
   \frac{(E_f^{\gamma_1+1} - E_i^{\gamma_1+1})}{\gamma_1+1}, 
   \label{eqIntegralEvents}
   \end{equation}
   where $[E_i, E_f]$ is the energy interval of integration, $\log_{10}(\bar{E}) = [\log_{10}(E_i) + \log_{10}(E_f)]/2$ and $\Phi_0$ and $\gamma_1$ are the parameters of the fitted power-law function used in the extrapolation.   This calculation gave $(7.57^{+0.62}_{-0.30}) \times 10^{8}$ events. Next, we calculated the amount of events in the HAWC unfolded energy distribution inside the same energy interval within the systematic errors from the PMT-late-light simulations. This procedure resulted in $(6.15^{+0.65}_{-0.13}) \times 10^{8}$ events. To end, we compared the expected and the observed amount of events in the true primary energy range from $\log_{10}(E/\gev) = 4.4$ up to $5.2$, which gave a deficit of $\sim 2.0 \, \sigma$ in the data. This result points out that the break can not be explained just by systematic effects of the PMT late light.
   
   We now include the remaining sources of systematic uncertainties in the analysis. A fit with a power-law function, see Eq.~(\ref{eq9}), to the energy spectrum between $\log_{10}(E/\gev) = 3.8$ and $4.4$ gave the following results:  $\log_{10} (\Phi_0/\mbox{m}^{-2} \, \mbox{s}^{-1}  \, \mbox{sr}^{-1}  \, \mbox{GeV}^{-1}) = 3.72 \pm 0.08 \, (\mbox{stat})  {}^{+0.57}_{-0.43}\,  (\mbox{syst})$ and $\gamma_1 = -2.51 \pm 0.02  \, (\mbox{stat})  \, {}^{+0.11}_{-0.14} \,  (\mbox{syst})$. In order to get the uncertainties in these parameters, first, we have fitted the spectrum with the power-law function considering all systematic sources but the uncertainties in the relative cosmic ray composition. Then we fitted the energy spectrum obtained for each cosmic ray composition model and quoted the maximum and minimum variations of the fitting parameters as the uncertainties due to the cosmic ray composition models. Finally, we added in quadrature these uncertainties with the corresponding ones obtained with the fit including the other systematic sources. We proceeded in this way because we noticed that not all values of $\gamma_1$ allowed within the error band associated to the uncertainty in the relative abundances of cosmic rays  provide an event energy distribution in agreement with the plots of figs. ~\ref{Fig_Ratio_Exp_MC} and   ~\ref{Fig_Eff_agecut}. Now, by extrapolating the fitted power law within systematic uncertainties to higher energies and employing Eq. (\ref{eqIntegralEvents}), we should expect to observe $(7.57^{+1.08}_{-1.65}) \times 10^{8}$  events in the interval $\log_{10}(E/\gev) = [4.4, 5.2]$ and $(5.1^{+0.95}_{-1.6}) \times 10^{7}$ events for $\log_{10}(E/\gev) = [5.0, 5.2]$. However, we measured $(6.15^{+0.93}_{-1.21}) \times 10^{8}$ and $(3.03^{+0.87}_{-0.69}) \times 10^{7}$ events, respectively. Therefore, the differences between the expectations and the measurements are  $0.8 \, \sigma$ and $1.1 \, \sigma$, correspondingly. They are small, however, they seem to indicate a tension between the power-law scenario and HAWC measurements even after considering all systematic uncertainties. We quoted the difference of $0.8 \, \sigma$ above obtained  for the energy range  $\log_{10}(E/\gev) = [4.4, 5.2]$  as the significance of the observed kink in the spectrum when all the systematic uncertainties are included.

  \subsection{Uncertainties in the energy scale}
  
\begin{figure}[!t]
    \centering
    \includegraphics[width=3.3in]{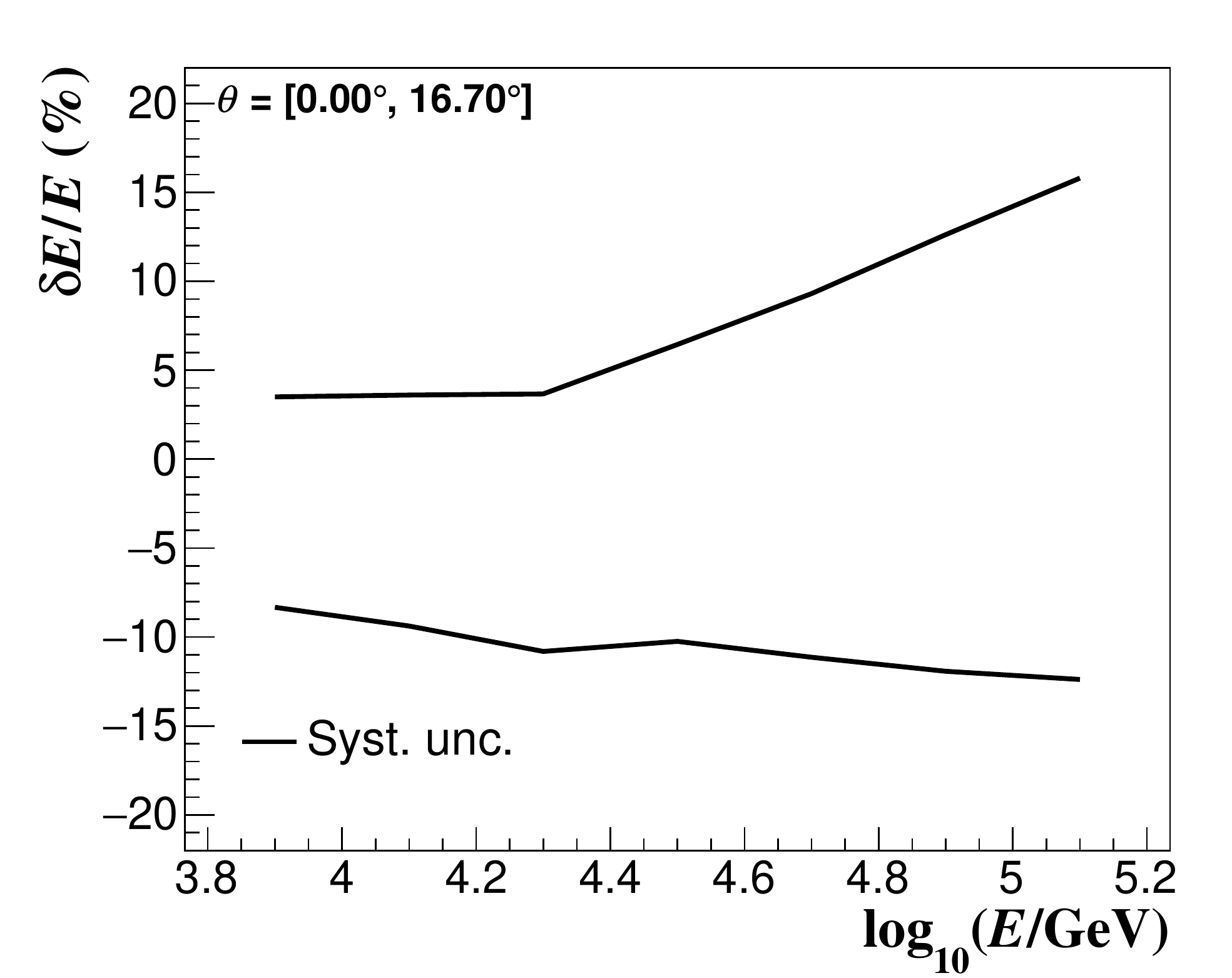}
    \caption{The total systematic uncertainties of the energy scale as a function of the primary energy $E$.}
    \label{Fig_Syst_E}
  \end{figure}

   Associated with the systematic uncertainties of the spectrum there are uncertainties on the energy scale $\delta E$, which can be roughly estimated from the following relation: $\delta \Phi/\Phi = -(\gamma + 1) \delta E/E$ \cite{PAO2021}, where $\gamma$ is the local value of the spectral index of the energy spectrum.  This procedure gives a total systematic uncertainty in the energy scale between  $-8.3 \%$ and $+3.5 \%$ at the low energy bin  [$ \log_{10}(E/\mbox{GeV}) \sim 3.9$], which evolves up to  $-12.4 \%$ and $+15.8 \%$, respectively, at high energies [bin $\log_{10}(E/\mbox{GeV}) \sim 5.1$], as it can be seen in Fig.~\ref{Fig_Syst_E}. A detailed estimation of the contribution of each systematic source to the total uncertainty in the energy scale is presented in Appendix \ref{apperrors}.

   \section{Discussion}
   \label{discussion}
   
   HAWC's observation of a spectral break in the cosmic ray spectrum of protons plus helium nuclei at $24 \, \tev$ provides further support to previous results from ATIC-2 \cite{atic09,  atic092}, CREAM I-III \cite{cream17} and NUCLEON \cite{nucleon, nucleon2, nucleon19} in favor of  fine structure in the spectra of light primaries in the $10-100 \, \tev$ regime.  The case is strengthened by recent  DAMPE \cite{dampe19, dampe21} measurements on the spectra of H and He nuclei exhibiting significant kinks at energies close to $14 \, \tev$ and $34 \, \tev$, respectively. These results imply that the break in the energy spectrum of H$+$He presented in Fig.~\ref{spectrum}  has its origin in breaks in the individual spectra of hydrogen and helium nuclei between $10$ and $40 \, \tev$. 
  
\begin{figure}[!t]
    \centering
    \includegraphics[width=3.3in]{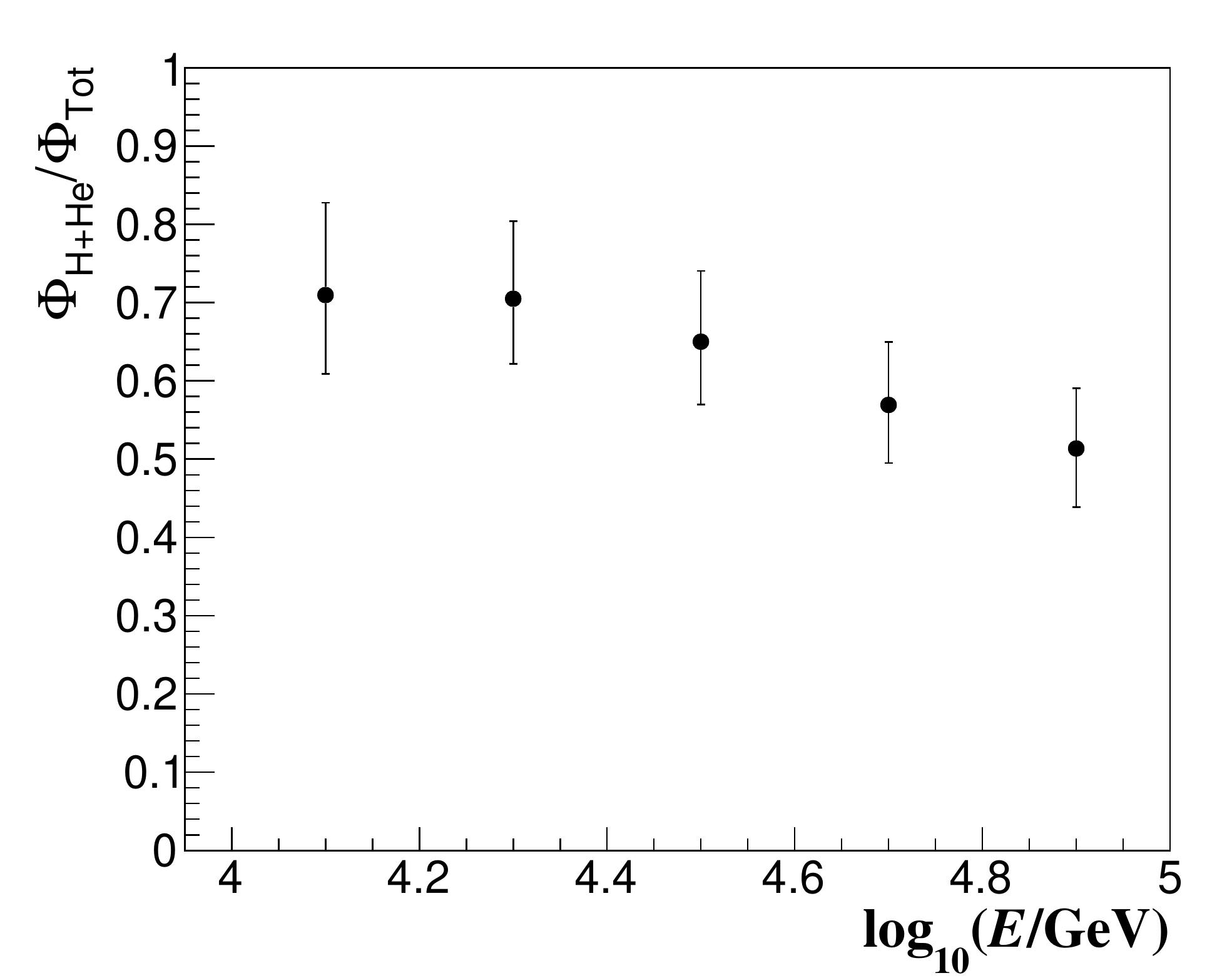}
    \caption{The intensity of the light cosmic ray nuclei obtained in this work divided by the all-particle spectrum of cosmic rays obtained with HAWC in \cite{Hawc17} plotted as a function of the primary energy. The vertical error bars represent the total uncertainty.}
    \label{ratio-light-all}
   \end{figure}

   The $\tev$ structure in the light component of cosmic rays may be connected with the break in the all-particle energy spectrum observed  at approximately $46 \, \tev$ with HAWC \cite{Hawc17} and confirmed by NUCLEON in \cite{nucleon19}.  The presence of both features in the same energy interval
   suggests that the referred feature in the intensity of H+He 
   could contribute to the structure observed at $\tev$ energies in the all-particle spectrum.
   In Fig.~\ref{spectrum}, we have compared the total spectrum of HAWC \cite{Hawc17} with our result for H+He. 
   There are two major differences between the features: the all-particle spectrum feature
   is wider and it is shifted to higher energies.  Further research is needed to find out the reasons. Nevertheless, these facts may suggest an increasing influence of the heavy component ($Z > 2$) of cosmic rays close to $100 \, \tev$, which seems consistent with the heavy element data from  NUCLEON
   \cite{nucleon, nucleon2, nucleon19}, 
   the measurements of HAWC on the mean shower age (cf. Fig.~\ref{FigAge}) and the analysis of the efficiency of the age cut (see Fig.~\ref{Fig_Eff_agecut}). The ratio $\Phi_{H+He}/\Phi_{Tot}$ between the spectrum of the light nuclei and the total intensity of cosmic rays  measured with HAWC  also seems to support such possibility. As observed in Fig.~\ref{ratio-light-all},  $\Phi_{H+He}/\Phi_{Tot}$ decreases from $10$ to $158 \, \tev$, which suggests an increase in the  relative abundance of  heavy nuclei in the total spectrum of cosmic rays close to $100 \, \tev$.

   The physical interpretation of our result is not yet clear, but it seems to require nonconventional models of production, acceleration and propagation of galactic cosmic rays. In general, it is thought that cosmic rays with energies from TeV to PeV are of galactic origin and that their acceleration and transport in the Galaxy occur through diffusive processes driven by magnetic fields. Acceleration up to PeV energies is assumed to take  place through the first order Fermi acceleration mechanism in shocked astrophysical plasmas \cite{ Axford1977, Krymsky1977, Bell1978, Blandford1978} of supernova remnants \cite{Baade34, Bell13} and propagation is believed to occur through scattering on random fluctuations in the interstellar magnetic field  \cite{Cesar1980, Ptuskin1993, Strong97}. The presence of magnetic fields in these processes implies a maximum confinement energy either at the source or at the Galaxy and hence, the presence of rigidity dependent cuts in the primary spectra at PeV energies \cite{Peters61}, while diffusive shock acceleration predicts a power-law behavior for the energy spectrum of cosmic ray nuclei from $\tev$  to $\pev$ \cite{Roulet17, Bell13}. As a consequence, it is difficult to understand the HAWC result within this scenario. 

   Some nonconventional models 
   predict  
   features in the spectra of different nuclei
   in the $\tev$ energy range, like the one observed by HAWC, and invoke the existence of new kinds of cosmic ray accelerators, nearby sources, or modifications to the standard mechanism of particle acceleration in astrophysical shocks. For instance, in \cite{Kachelriess15, Kachelriess18} an old supernova remnant (age $\sim 2 - 3 \, \mbox{Myr}$) located close to the Earth at a distance of  $\sim \mathcal{O}(100 \, \mbox{pc})$ is postulated as the dominant source of measured $10 - 100 \, \tev$ cosmic rays. Its maximum achievable energy is assumed to be around $Z \times 10 \, \tev$. The model leads naturally to breaks in the spectra of cosmic ray nuclei at tens of $\tev$ and  could explain the observed feature in the spectrum of proton plus helium primaries. On the other hand, in \cite{Zatsepin,Zatsepin2} a phenomenological model is proposed based on the hypothesis that the measured all-particle cosmic ray  spectrum from $100 \, \gev$ to $100 \, \pev$ could be described by assuming the contribution of three different types of sources, each
   characterized by a power-law cosmic ray spectrum but with distinct magnetic rigidity cutoffs: one at $200 \, \mbox{GV}$, another at $50 \, \mbox{TV}$ and the other at $4 \, \mbox{PV}$. The authors associate the first class of cosmic ray accelerators to nova explosions, the second  to supernova remnants (SNRs) expanding in the interstellar medium, and the last one to superbubbles. In the model, the second population produce H and He spectra with a kneelike feature in the energy range explored in this work. 
   In \cite{Yue20}, the possibilities that the above structure is due to the existence of a new population of $\tev$ accelerators or just to a single local source of $\tev$ cosmic rays are explored.  
   They favor a local source, following arguments from \cite{Savchenko15, Liu19}, 
   supported on data of the phase and dipole anisotropy of galactic cosmic rays. Finally, in \cite{Ptuskin13}, $\tev$ features in the spectrum of light cosmic rays appear as a consequence of two new  H and He  contributions with hard spectra  accelerated at reverse shocks of SNRs of types II and I, respectively. In this scenario,  
   the new sources contribute to the hardening of the energy spectra of protons and helium at rigidities of $240 \, \mbox {GV}$ and to the increase in the $\Phi_{He}/\Phi_{H}$ ratio between $100 \, \gev$ and $1 \, \tev$ as observed by the ATIC-2 \cite{atic09, atic092}, CREAM \cite{cream11, cream10} and PAMELA \cite{pamela} detectors.

     In order to distinguish among the predictions of different models that may explain the physical origin of the feature in question, we also must look at the details of the energy spectra of heavier cosmic ray elements in the interval $10 \, \tev - 1 \, \pev$.
   NUCLEON, in a recent study \cite{nucleon19}, has also provided evidence in favor of the existence of rigidity dependent breaks at $\sim 10 \, \mbox{TV}$ in the individual spectra of primary cosmic rays with $Z \geq 3$. Further research with CALET \cite{calet19}, DAMPE, HAWC, TAIGA-HiSCORE \cite{taiga19} and LHAASO \cite{lhaaso19} will soon test NUCLEON's observations 
   and, in turn,  provide an opportunity to understand the systematic uncertainties inherent in the direct and indirect cosmic ray detection techniques.

 \section{Conclusions}
 \label{conclusions}

  The HAWC observatory has measured with high statistics and precision the cosmic ray energy spectrum of H plus He in the energy interval from $6$ to $158$ TeV and  confirmed previous hints from the   ATIC-2, CREAM I-III and NUCLEON direct detectors that the energy spectrum of light primaries deviates from a plain power-law behavior in the $10 - 100 \, \tev$ energy interval. HAWC result also agrees with DAMPE recent measurements that point out the existence of individual softenings in the spectra of protons and helium nuclei at tens of TeV. Hence, HAWC results does not support previous observations 
  from the ARGO-YBJ air shower detector between $3$ and $300 \, \tev$, whose 
  spectrum agrees with a simple power-law form  \cite{Argo15}. HAWC results find a break in the H + He 
  spectrum of cosmic rays close to $24.0^{+3.6}_{-3.1} \, \tev$, which is produced by what it seems to be a smooth decrease in the spectral index from $\gamma = -2.51 \pm 0.02$ to $\gamma = -2.83 \pm 0.02$. Such a structure was previously reported by HAWC in \cite{Hawc19}. Now, it has been confirmed 
  using a larger EAS dataset with improved MC simulations of the detector. The break is observed with a statistical significance of $4.1 \, \sigma$. Under systematic uncertainties,  the feature has a significance of $0.8 \, \sigma$. This study demonstrates  that research on the composition  of cosmic rays is possible with the HAWC detector and  opens the door to deeper investigations in the $\tev$ range not only in HAWC but also in other present/future high altitude water Cherenkov observatories.

  \section*{Acknowledgments}

  We acknowledge the support from the US National Science Foundation (NSF); the US Department of Energy Office of High-Energy Physics; the Laboratory Directed Research and Development (LDRD) program of Los Alamos National Laboratory; Red HAWC (Mexico); Consejo Nacional de Ciencia y Tecnolog\'ia (CONACyT), M\'exico, Grants 271051, 232656, 260378, 179588, 254964, 258865, 243290, 132197, A1-S-46288, A1-S-22784, c\'atedras 873, 1563, 341, 323; DGAPA-UNAM Grants IG101320, IN111315, IN111716-3, IN111419, IA102019, IN112218; VIEP-BUAP; Programa
  Integral de Fortalecimiento Institucional (PIFI) 2012 and 2013, Programa de Fortalecimiento de la Calidad en Instituciones Educativas (PROFOCIE) 2014 and 2015; the University of Wisconsin Alumni Research Foundation; the Institute of Geophysics, Planetary Physics, and Signatures at Los Alamos National Laboratory; Polish Science Centre Grant, DEC-2017/27/B/ST9/02272; Coordinaci\'on de la Investigaci\'on Cient\'ifica de la Universidad Michoacana; Royal Society - Newton Advanced Fellowship 180385; Generalitat Valenciana, Grant CIDEGENT/2018/034.  We also acknowledge the significant contributions over many years of Stefan Westerhoff, Gaurang Yodh and Arnulfo Zepeda Domínguez, all deceased members of the HAWC collaboration. Thanks to Scott Delay, Luciano D\'iaz and Eduardo Murrieta for technical support.

 \appendix  
  
 \section{Composition models}
  \label{appmodels}

  To study the impact of
  the uncertainties in the composition of cosmic rays, we  used four additional composition models in our study. The first  is the  Polygonato model, described in \cite{Polygonato}. The others were obtained from fits to the spectra of the different mass groups of cosmic rays measured by the ATIC-2 \cite{atic07}, JACEE \cite{jacee} and MUBEE \cite{mubee} collaborations. For the individual fits, we  used a broken-power law expression similar to the one employed for the nominal composition model in \cite{Hawc17}. This formula is
  \begin{equation}
   \Phi(E/\gev) = \left\{ 
   \begin{array}{l@{\quad, \quad}l}
      \Phi_0  (E/E_0)^{\gamma_1}   & E < E_b \vspace{2pc}\\
      \Phi_0 (E_b/E_0)^{\gamma_1-\gamma_2} (E/E_0)^{\gamma_2}  & E \geq E_b.
   \end{array}
   \right.
   \label{eqdpl}
  \end{equation}
  Here, $\Phi_0$ is a normalization factor at the reference energy $E_0$ and $E_b$ is the energy at which appears the break; 
  $\gamma_1$ and $\gamma_2$ are the spectral indexes of the function before and after the kink. For all models, we  chose  $E_0(\gev) =  100$ for light primaries and $1200 \, \gev$ for heavy ones.  The results 
  are shown in Table \ref{tabCompositionModels}. The first model, denoted as ATIC-2, was derived from ATIC-2 data 
  \cite{atic07} between $E = 49 \, \gev$ and $31 \, \tev$. In this case, to get the parameters of the  C$-$Si heavy mass groups, a joint fit was performed to 
  a single value for $\gamma_1$, $\gamma_2$ and $E_b$ in the spectra.
  The individual normalization factors for each of the heavy elements were treated as free parameters during the fit. To obtain the spectrum for Fe nuclei, the previously fitted $\gamma_2$ and $E_b$ values were substituted in the corresponding broken-power law formula and were fixed during the fit. 

  \begin{table}[!t]
    \begin{center}
    \caption{Values of the parameters of three composition models used in the present analysis. The models were derived from fits with expression (\ref{eqdpl}) to the ATIC-2 \cite{atic07}, JACEE \cite{jacee} and MUBEE \cite{mubee} measurements on the elemental spectra of cosmic rays.}
    \footnotesize
   \begin{tabular}{lcccc}
   \hline 
   \hline 
     Model & $\Phi_0$ &  $\gamma_1$ & $\gamma_2$ & $E_b$\\
     & $[10^{-6} \, \mbox{m}^{-2} \, \mbox{s}^{-1}  \, \mbox{sr}^{-1}  \, \mbox{GeV}^{-1}]$ &&&[$\gev$]\\
     \hline\\
     ATIC-2 &&&&\\ 
     H &$4.40 \times 10^{4}$ &$-2.86$& $-2.60$&$159.6$\\
     He &$2.59 \times 10^{4}$ &$-2.61$&$-2.45$&$1093.6$\\
     C & $6.61$ &$-2.64$&$-2.48$&$11125.5$\\
     O & $10.73$ &$-2.64$&$-2.48$&$11125.5$\\
     Ne & $2.78$ &$-2.64$&$-2.48$&$11125.5$\\
     Mg & $4.72$ &$-2.64$&$-2.48$&$11125.5$\\
     Si & $5.34$ &$-2.64$&$-2.48$&$11125.5$\\
     Fe & $13.10$ &$-2.61$&$-2.48$&$11125.5$\\
    &&&&\\
     MUBEE &&&&\\ 
     H &$4.44 \times 10^{4}$ &$-2.72$& $-2.72$&$-$\\
     He &$2.66 \times 10^{4}$ &$-2.60$&$-2.63$&$732.7$\\
     C & $6.41$ &$-2.64$&$-2.56$&$31693.5$\\
     O & $10.46$ &$-2.64$&$-2.56$&$31693.5$\\
     Ne & $2.46$ &$-2.64$&$-2.00$&$31693.5$\\
     Mg & $4.13$ &$-2.64$&$-2.00$&$31693.5$\\
     Si & $4.58$ &$-2.64$&$-2.00$&$31693.5$\\
     Fe & $13.10$ &$-2.61$&$-3.00$&$4283.8$\\
     &&&&\\
     JACEE  &&&&\\
     H &$4.39 \times 10^{4}$ &$-2.80$& $-2.69$&$109.44$\\
     He &$2.67 \times 10^{4}$ &$-2.60$&$-2.59$&$1586.4$\\
     C & $6.35$ &$-2.64$&$-2.24$&$13106.9$\\
     O & $10.35$ &$-2.64$&$-2.24$&$13106.9$\\
     Ne & $2.46$ &$-2.64$&$-2.48$&$31693.5$\\
     Mg & $4.13$ &$-2.64$&$-2.48$&$31693.5$\\
     Si & $4.58$ &$-2.64$&$-2.48$&$31693.5$\\
     Fe & $13.10$ &$-2.61$&$-2.51$&$80717.9$\\
     &&&&\\
   \hline
   \hline 
   \end{tabular}
   \label{tabCompositionModels}
   \end{center}
  \end{table}

  The second fitted model, named  MUBEE, was obtained from a fit to combined ATIC-2 \cite{atic07} and MUBEE \cite{mubee} measurements. Below $E = 10 \, \tev$ only data from ATIC-2 was used, and above, just the measurements from MUBEE. To obtain $\gamma_1$ and  $E_b$ for the spectra of C$-$Si nuclei, we proceeded as before with the difference that to find $\gamma_2$ we decided to divide the heavy data in two mass groups:  C$-$F and Ne$-$K, and to perform different fits for each of them. For the model called JACEE, we  proceeded similarly,
  except that ATIC-2 data was kept up to  $E = 25 \, \tev$ and at higher energies, JACEE measurements were employed \cite{jacee}. In addition, we also fitted $E_b$ separately for each distinct mass group.

\begin{figure*}[!t]
              \includegraphics[width=3.0in]{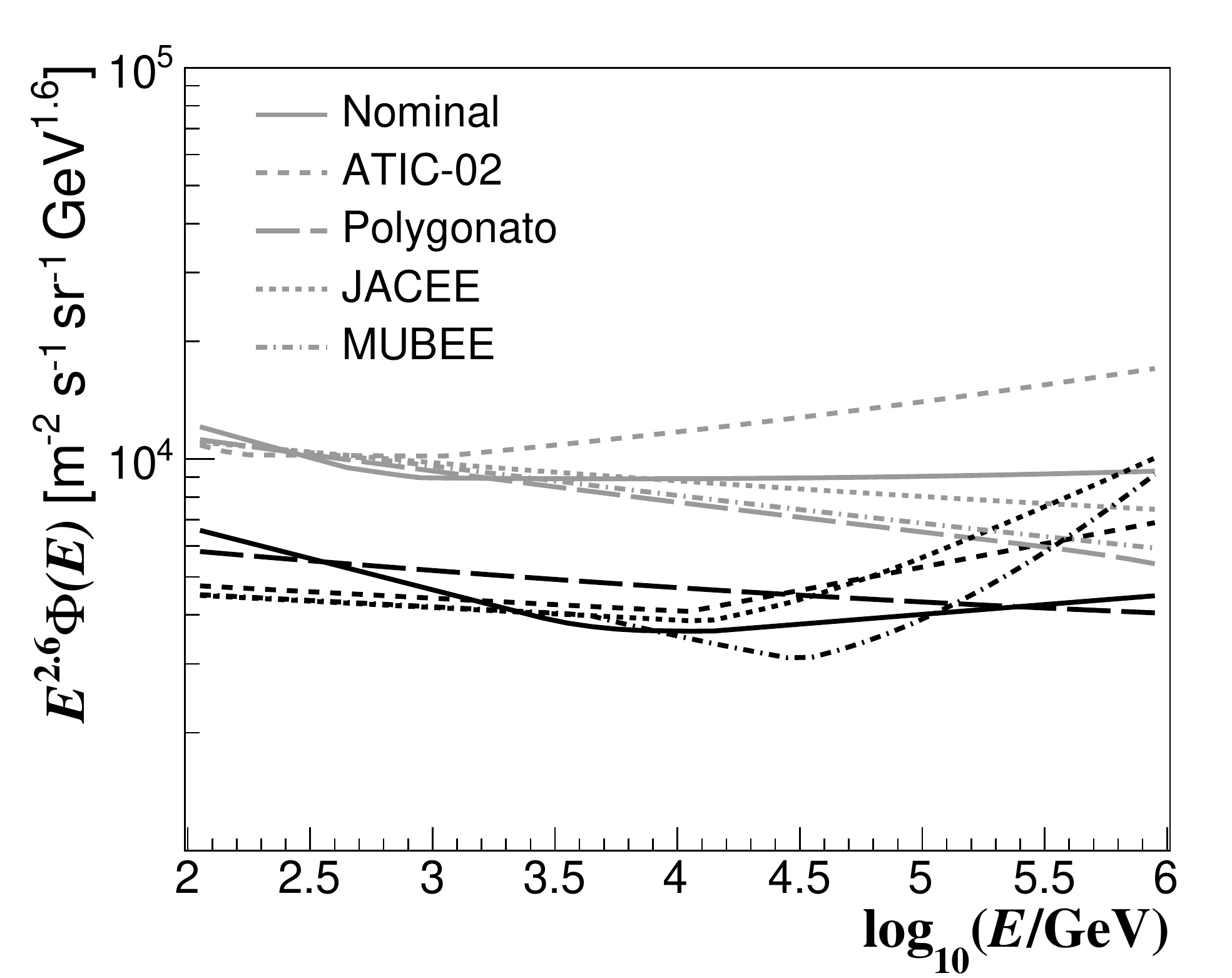} 
              \includegraphics[width=3.0in]{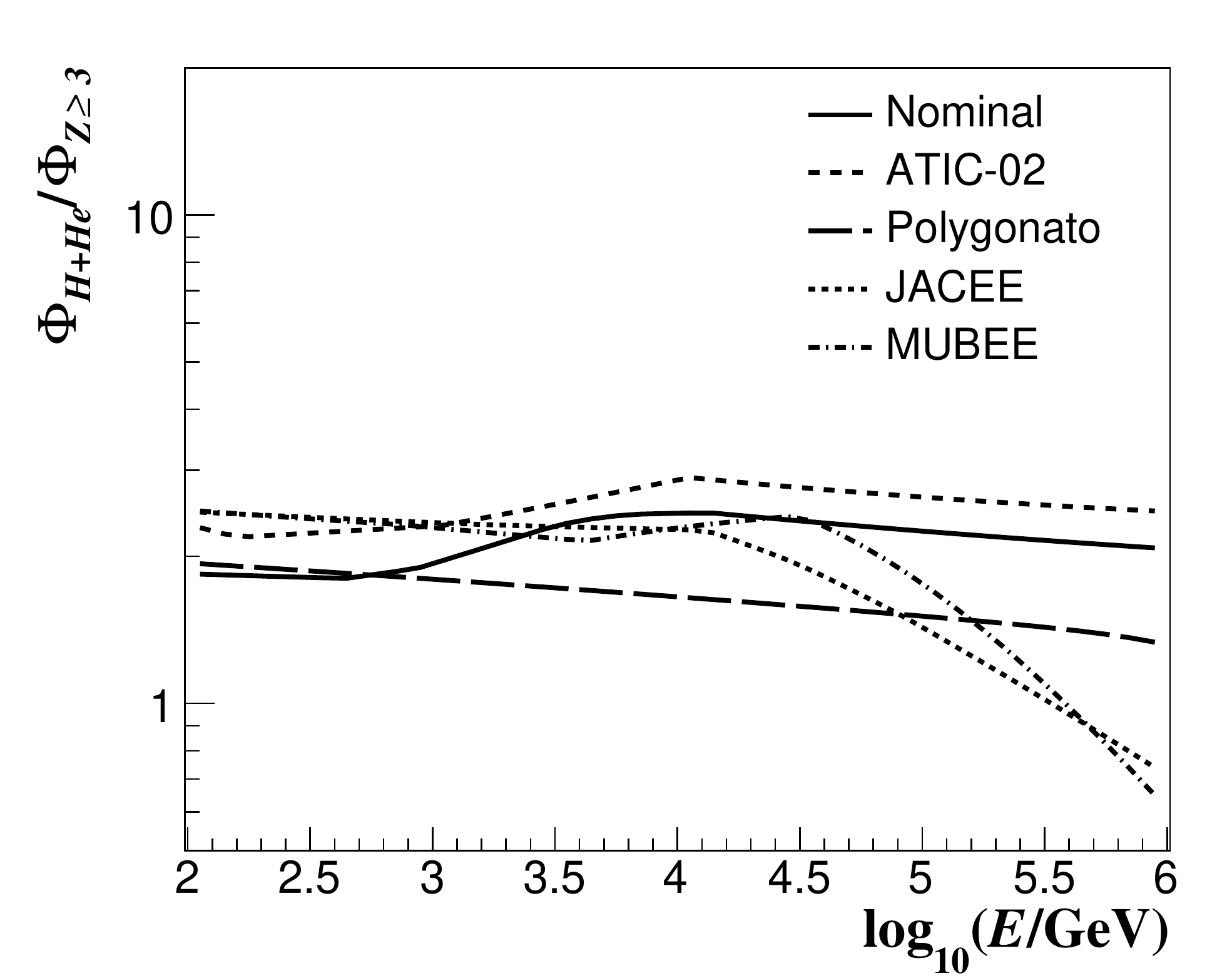}
    \caption{\textit{Left}: the energy spectra for light (gray, H$+$He) and heavy (black, $Z \geq 3$) cosmic ray primaries in the $100 \, \gev- 1 \, \pev$ regime as predicted with the different composition models used in this work. The nominal one (continuous line) was taken from \cite{Hawc17}. The Polygonato model (long-dashed line) was obtained from \cite{Polygonato}, and the other models, from fits with broken power-law formulas to data from  ATIC-2 \cite{atic07} (short-dashed line), JACEE \cite{jacee} (dotted line) and MUBEE \cite{mubee} (dashed-dotted line). \textit{Right}: the relative abundance of the light component of cosmic rays in the interval $100 \, \gev- 1 \, \pev$  as predicted with the above composition models. The same line style conventions used  on the left panel are used here.}
    \label{FigAbundancesModels}
  \end{figure*}

  To illustrate the results of the fits, in Fig. \ref{FigAbundancesModels} we display the predictions for the energy spectra of the light and heavy mass groups of cosmic rays and for the ratio of light to heavy cosmic ray nuclei $\Phi_{H+He}/\Phi_{Z\geq 3}$ according to each of the composition models, including the nominal one and the Polygonato scenario.

  \section{Statistical and systematic errors}
  \label{apperrors}
  
  For the calculation of the statistical error we have considered the following sources of uncertainty: 
  
  \textit{Statistics of the data.} The magnitude of this error is less than $0.03  \%$ due to the large number of events in the data. It was calculated by propagating the statistical error from  $N(E_{rec})$ according to \cite{Adye11a, Zig17}.  We assume that the reconstructed energy bin contents are independent and Poisson-distributed. Let us define 
  \begin{eqnarray}
     N_\mu        &=&   N(E_\mu), \\ 
     N_{rec, j}   &=&   N(E_{rec, j}),  \\ 
     M_{\mu j}   &=&  P(E_\mu|E_{rec, j}).
  \end{eqnarray}
  Then the covariance matrix $V_{stat}^{data}$ that provides the statistical errors and the correlation between the unfolded bins can be expressed as
    \begin{eqnarray}
    V_{stat}^{data}[N^{i}_\mu, N^{i}_\nu] &=&   \sum_{j, k} 
    \frac{\partial N^{i}_\mu}{\partial  N_{rec, j}}
     Cov[ N_{rec, j},  N_{rec, k}] 
     \frac{\partial N^{i}_\nu}{ \partial   N_{rec, k}},   \nonumber \\ 
  \end{eqnarray}
  with
  \begin{eqnarray}
     \frac{\partial N^{i}_\mu}{\partial  N_{rec, j}} &=&  M_{\mu j} + 
     \frac{ N^{i}_\mu}{ N^{i-1}_\mu}\frac{\partial N^{i-1}_\mu}{\partial  N_{rec, j}}   \nonumber \\ 
     &-&  \sum_{\sigma, k} 
     \frac{ N_{rec, k}}{N^{i-1}_\sigma}
     M_{\mu k}
     M_{\sigma k}
     \frac{\partial N^{i-1}_\sigma}{ \partial  N_{rec, j}} ,  \nonumber \\ 
  \end{eqnarray}
  where the superscript $i$ denotes the iteration level of the unfolded spectrum $N(E)$ and  
   \begin{equation}
      Cov[ N_{rec, j},  N_{rec, k}]  = N_{rec, j} \delta_{j,k}\\ 
  \end{equation} 
  is the covariance matrix for the bins of the measured spectrum.
  For $i = 0$, we have $\partial N^{0}_\mu/ \partial  N_{rec, j}= 0$.

 \textit{Limited statistics of the MC simulations.} The finite size of the MC data contributes with a statistical error within $\pm 3.8   \%$ through the response matrix. The uncertainty  was computed by error propagation  following \cite{Adye11a, Zig17}. The covariance matrix $V^{MC}_{stat}$ from which these errors are derived is calculated from the following expression:
  \begin{eqnarray}
      V_{stat}^{MC}[N^{i}_\mu, N^{i}_\nu] =  
     \sum_{\lambda, j} 
     \sum_{\rho, k}
      \frac{\partial N^{i}_\mu}{\partial  P_{j \lambda}} 
     Cov[P_{j \lambda}, P_{k \rho}] 
      \frac{\partial N^{i}_\nu}{\partial  P_{k \rho}},   \nonumber \\ 
     \label{covmc}
  \end{eqnarray}
  where we have defined
  \begin{eqnarray}
     P_{j \mu}   &=&  P(E_{rec, j}|E_\mu).
  \end{eqnarray}
  and
    \begin{eqnarray}
     \frac{\partial N^{i}_\mu}{\partial  P_{j \lambda}}  &=& 
     \left[ 
      N^{i-1}_\mu \delta_{\mu \lambda} 
     -  M_{\mu j} N^{i-1}_\lambda
      \right] 
      \frac{N_{rec,j}}{\sum_{\sigma} P_{j\sigma} N^{i-1}_\sigma}  \nonumber \\ 
       &+& \frac{N^{i}_\mu}{N^{i-1}_\mu}
       \frac{\partial N^{i-1}_\mu}{\partial   P_{j \lambda}}  \nonumber \\ 
        &-& \sum_{\sigma, k} 
        \left[
        \frac{N_{rec, k}}{N^{i-1}_\sigma}
        M_{\mu k}
        M_{\sigma k} \frac{\partial N^{i-1}_\sigma}{\partial P_{j \lambda}}
        \right].  \nonumber \\ 
  \end{eqnarray}
  In formula (\ref{covmc}), $ Cov[P_{j \lambda}, P_{k \rho}] $ represents
  the covariance matrix for the bins of the response matrix. It is different from zero in the following cases \cite{agostini}:
  \begin{eqnarray}
       Cov[P_{j \lambda}, P_{k \lambda}] 
       &=&\left\{ 
       \begin{array}{r@{\quad ; \quad}l}
         P_{j \lambda}\left[1 - P_{k \lambda}\right]/\Tilde{N}_{MC, \lambda} &   j = k   \nonumber \\ 
            - P_{j \lambda} P_{k \lambda}/\Tilde{N}_{MC, \lambda} &    j \not=  k  \nonumber \\ 
       \end{array}
       \right.,
  \end{eqnarray} 
  where $\Tilde{N}_{MC, \lambda} =  (\sum_{k=1} w_{k \lambda})^2/(\sum_{k=1} w^2_{k \lambda})$ is the equivalent number of unweighted events inside the bin $E_\lambda$ of the true MC energy distribution used in the construction of the response matrix. Here, the sum runs over the number of simulated events in the bin and $w_{k \lambda}$ denotes the weight of each of these events. In case that $i = 0$, we have $ \partial N^{0}_\mu/\partial  P_{j \lambda}  = 0$.
  
   On the other hand, the systematic uncertainty was calculated by summing in quadrature the error sources below; in each case we cite the effect on the intensity of the H + He spectrum, and took the systematic error for each source as the size of the observed effect: 
  
   \textit{Unfolding algorithm.} Its uncertainty is found between $-1.1  \%$ and $+1.2  \%$. It was evaluated by comparing the experimental result obtained with the Bayesian procedure with that using the Gold's algorithm \cite{Gold64} as implemented in \cite{Antoni05, Ulrich01}. 
      
      In Gold’s  procedure \cite{Gold64}, a real diagonal matrix $D$ is found iteratively, which allows to estimate the unfolded histogram $N(E)$ from the data by using the equation 
      \begin{equation}
       N(E) = D N(E_{rec}), 
       \label{eqgold1}
      \end{equation}
      with
      \begin{equation}
       diag(D)  = \frac{P(E)}{\sum_{E^\prime} P(E_{rec}|E^\prime) P(E^\prime)}. 
       \label{eqgold2}
      \end{equation}
      Here, $P(E)$ is the probability distribution of the unfolded histogram at the previous iteration level. 
      
      Now, in order to guarantee real positive  solutions and to take into account the statistical uncertainties of the data \cite{Antoni05, Ulrich01}, we replaced $N(E_{rec})$ by 
      \begin{equation}
       N^\prime(E_{rec}) = [P(E_{rec}|E)]^{T} C C N(E_{rec}),  
       \label{eqgold3}
      \end{equation}
      in Eq.(\ref{eqgold1}) and $P(E_{rec}|E)$ by the matrix
      \begin{equation}
       P^\prime = [P(E_{rec}|E)]^{T}CC P(E_{rec}|E), 
       \label{eqgold4}
      \end{equation}
       in Eq.(\ref{eqgold2}), where $C$ is a diagonal matrix, whose matrix elements on the main diagonal are equal to the inverse of the statistical uncertainties of the measurements $N(E_{rec})$.

       \textit{Seed for the unfolding method.} To estimate the error of the unfolded result due to  the initial energy distribution used in the Bayesian algorithm, we have repeated the unfolded procedure with two distinct priors: a uniform distribution, as suggested by \cite{agostini}, and an $E^{-1.5}$ distribution. The power-law choice that was employed matches the all-particle energy histogram measured with HAWC for $10  \, \tev \leq E \lesssim 46 \, \tev$ \cite{Hawc17}, which may be expected to be closer to the true distribution of H plus He primaries, since this region seems to be dominated by the light mass group of cosmic rays  \cite{atic09, atic092, cream17, nucleon19}. By comparing the resulting spectra with the one of reference obtained in this work (see Fig. \ref{spectrum}), we found a bias in the intensity   within $-1.4  \%$  and $+0.7  \%$. We used that as the systematic error due to the seed in the unfolding algorithm.
      
       \textit{Smoothing procedure in unfolding algorithm.} The employment of a broken power law in the smoothing procedure of the unfolding analysis was done to achieve a fast convergence of the result. However, this procedure may introduce a systematic error in the unfolded spectrum. To compute this uncertainty, we smoothed the unfolded distributions with two alternative functions: a fifth degree polynomial and the 353HQ-twice smoothing algorithm \cite{smoothing} as installed in ROOT \cite{root}.  The resulting spectra were then compared with the one of reference obtained with the Bayesian method. We found differences that range from $-2.5  \%$ to $+3.7  \%$, which we  used as the corresponding systematic error.

        \textit{Corrected effective area.} The systematic uncertainty in $A_{eff}(E)$ was evaluated using MC simulations. By varying $A_{eff}(E)$  inside its  allowed limits, we observed a variation in the energy spectrum from $-2.1  \%$ to $+2.2  \%$.

       \textit{Position of the age cut.} We moved the cut to  
     the expected line for 
     the mean age of He events and then, to the mean age of C events 
     (see Fig.~\ref{FigAge}). Using the  cut on $s_{C}$, the spectrum is increased by at most $+ 3.7  \%$, while by setting the cut on $s_{He}$, a decrease of up to $- 6.6  \%$ is observed. 
    
       \textit{Cosmic ray composition uncertainty.} The dependence on the primary cosmic ray abundances enters through the response matrix and the corrected effective area, which are computed  with MC simulations. 
     We replaced our nominal model with the four alternative composition scenarios of Appendix \ref{appmodels} and repeated the analysis in each case. In addition, we have also considered the uncertainty in the relative abundance of heavy nuclei observed in the analysis of the efficiency of the age cut (c.f Fig.~\ref{Fig_Eff_agecut}) presented in Sec. \ref{Analysis}. To estimate its effect on the unfolded spectrum, we have multiplied by a factor of two the intensity of heavy elements in our cosmic ray composition models, since, in this case, model predictions for the efficiency of the shower age cut are closer to the measured value. Then we have repeated the unfolding analysis with the new models. 
     The maximum and minimum differences of the results obtained with the different scenarios for the composition of cosmic rays with respect to the experimental  spectrum of reference were recorded as systematic uncertainties. At the bin $\log_{10}(E/\gev) = 3.9$, the systematic error has a value of $-11.6  \%$, while for the bin  $\log_{10}(E/\gev) = 5.1$,  it lies within $+2.1  \%/$$-17.3 \%$. In general, when using models with heavier (lighter) relative abundances than the nominal one, the magnitude of the spectrum decreases (increases) due to the larger (smaller) correction factors. In case of the ATIC-2 model, which have a lighter abundance, we observe a small decrement close to $10 \, \tev$ that is due to a compensation from the response matrix. This effect is related to the hard spectrum of the light component employed to construct the above matrix. 

       \textit{PMT charge resolution.}  In the nominal MC dataset, a PMT charge uncertainty of $10 \%$ is used \cite{Hawccrab19}. To evaluate the impact of this parameter on the final spectrum, the PMT charge resolution was varied within its allowed interval $[0 \%, 15 \%]$ (see \cite{Hawccrab19}). Changes of the H + He spectrum   were between  $-2.6  \%$ and $+1.8  \%$. 
     
      \textit{PMT-late-light simulation.} 
     Air showers have a broader time distribution than the laser pulses employed for calibration in HAWC \cite{Hawc17b, Hawccrab19}, consequently there appears a systematic error associated with the  calibration of the effective charge produced by the late light during the EAS event, which is important for high values of $Q_\mathrm{eff}$ ($> 50 \, PE\mbox{s}$). The effect of this systematic source leads to an overestimation of the charge, since broader pulses have a longer Time-over-Threshold. In simulations, to take into account the effect of the PMT late light, a linear correction is added in logarithmic space to the effective charge. The value of the correction is the same for all PMTs and it increases from zero at $\log_{10}(Q_\mathrm{eff}) = 1.25$ up to  $0.1$ at around $\log_{10}(Q_\mathrm{eff}) = 2.25$. To estimate the uncertainty of the PMT late light, the value of the correction at the upper limit of $\log_{10}(Q_\mathrm{eff})$ is varied between $0.5$ and $1.25$.  The limits are quite conservative and they are chosen in such a way that allow us to describe the measured charge distribution for triggered events. On the other hand, we have also included the case for zero correction in the study, assuming that the difference between the MC simulations and the data is due to deficiencies of the hadronic interaction model. A smaller value of the correction of the PMT-late-light effect in the simulation tends to flatten the energy spectrum, however, even with zero correction the spectral feature can be observed (cf. Sec. \ref{results}).
     We varied the late light effect within the above range and found that it produces a systematic effect from $-9.2  \%$ to $+28.3 \%$ in the energy spectrum.

      \textit{PMT threshold.}
    The impact of the uncertainty in the minimum detectable charge at the PMTs of HAWC has been also evaluated. For this aim, the nominal value (which is of the order of $0.2 \, \mbox{PE}$) was varied within the error interval of $\pm 0.05 \,  \mbox{PE}$, which was found from a HAWC calibration study based on vertical muon data \cite{Hawccrab19}. Correspondingly, we observed a bias of $-1.96  \%/$$+2.3  \%$ in the magnitude of the spectrum.

      \textit{PMT efficiency and its temporal evolution.}
     We estimated this uncertainty by 
     using distinct MC simulations, which incorporate the measurements of the individual PMT efficiencies in HAWC at different moments during the data taking period \cite{Hawccrab19} (in particular, on September 2015, April and July 2016,  February and June 2017 and February 2018).  This procedure allowed us to determine the uncertainties in the efficiency of the PMTs, which can change with the time due to possible aging effects. The detector layouts of the active PMTs in HAWC registered during these sampling epochs were also incorporated into these simulations in order to study the influence of PMTs that are removed for maintenance or are not active in HAWC during the data taking periods, because the MC simulations used in our nominal analysis were carried out for an ideal situation where all PMTs are working. Using these MC datasets, we found that the corresponding error in the energy spectrum ranges from $-4.1 \%$ to $+5.0  \%$.

\begin{figure*}[!t]
     \includegraphics[width=3.0in]{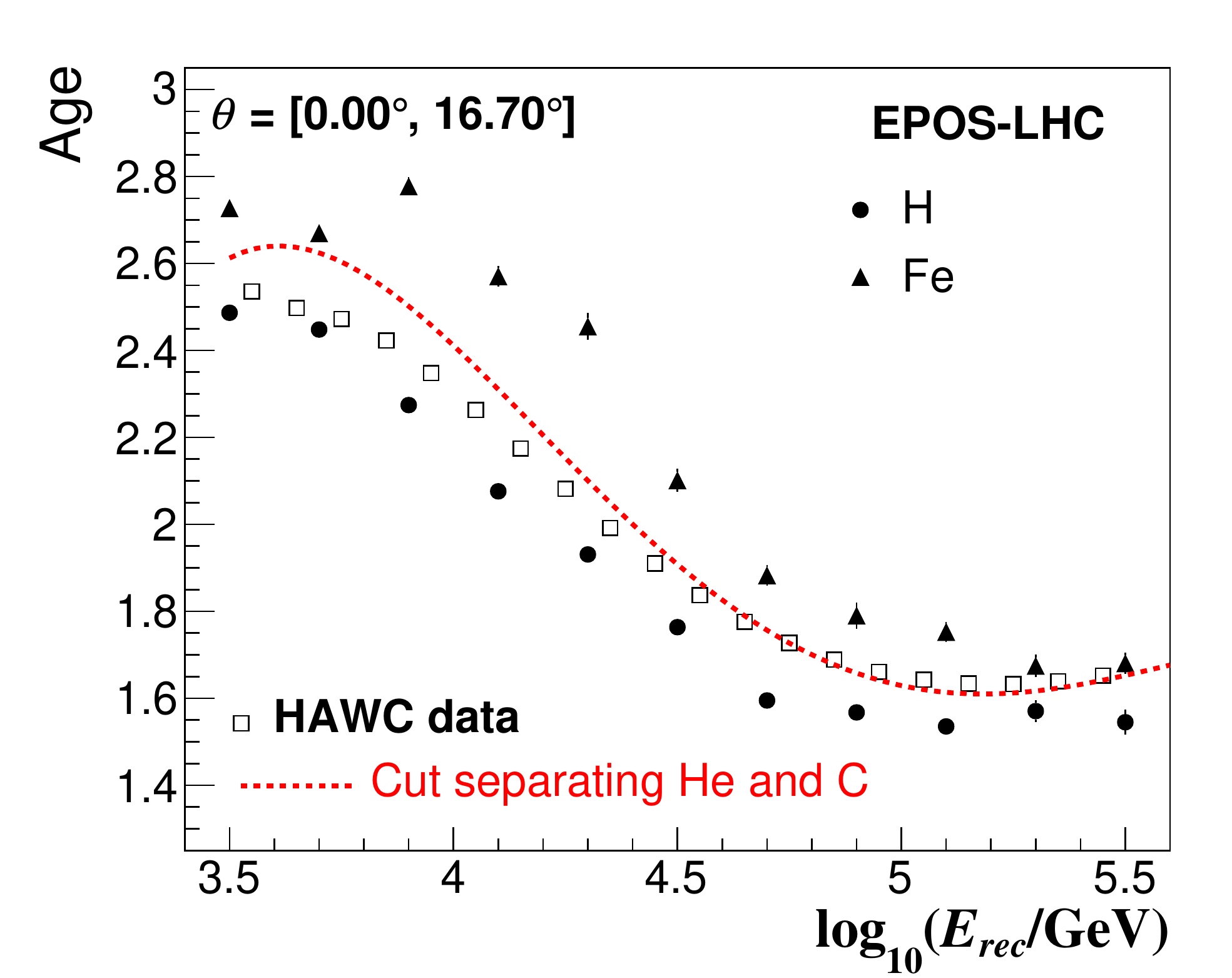} 
     \includegraphics[width=3.0in]{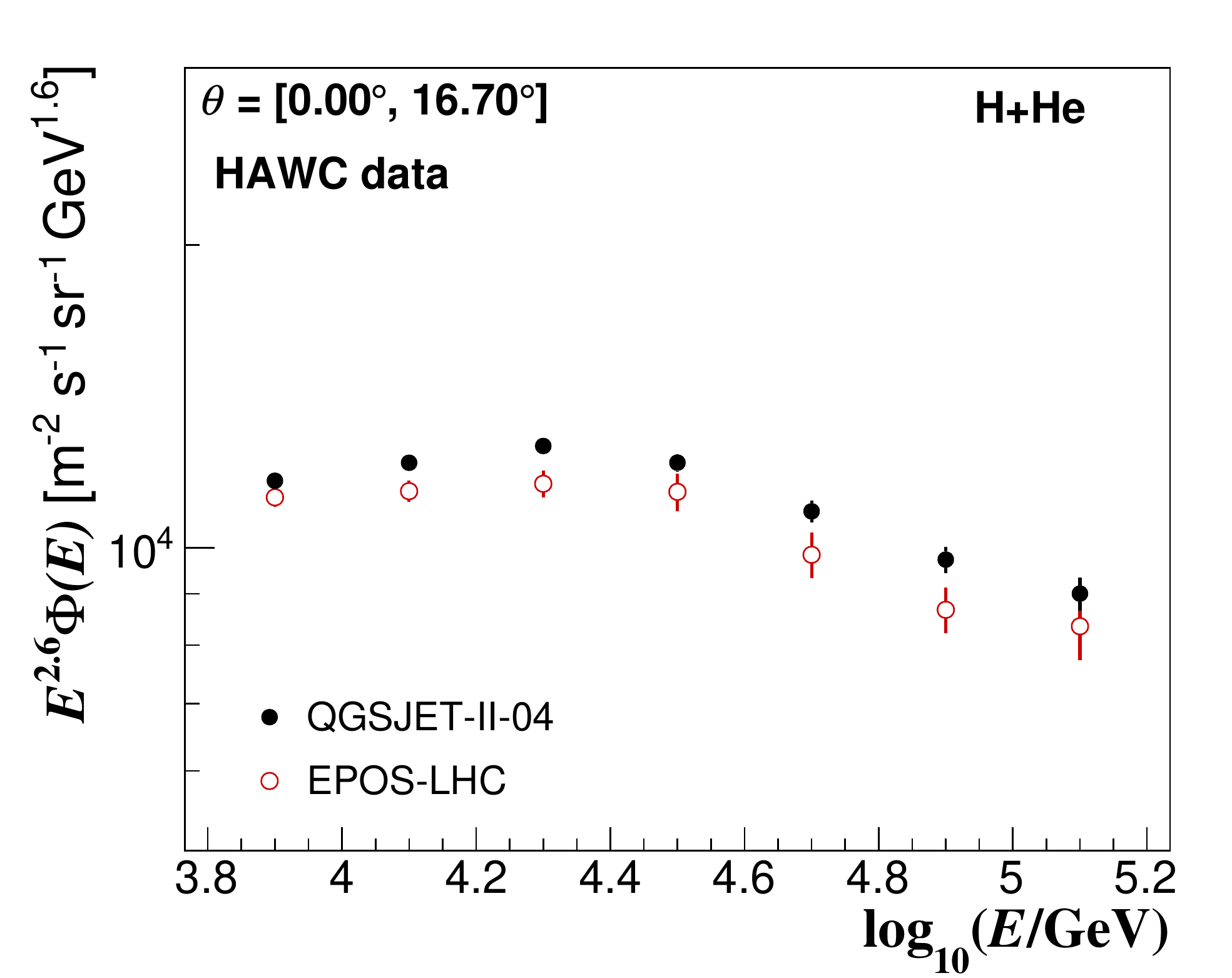} 
    \caption{\textit{Left:} expected values of the mean age parameter of vertical EAS for HAWC according to EPOS-LHC versus the reconstructed energy in comparison with the measured data (black open squares). For clarity, not all the elemental nuclei simulated in this work were included in the plot. Model predictions are only shown for H (circles) and Fe (triangles) primaries. Vertical lines in each data point represent the errors on the mean. The $s_{He-C}$ cut derived from EPOS-LHC simulations is shown with the segmented red line. \textit{Right:} the energy spectrum of protons plus helium cosmic ray nuclei as measured with HAWC and reconstructed within the framework of the QGSJET-II-04 (black circles) and EPOS-LHC (hollowed red circles) hadronic interaction models. Error bars represent the statistical uncertainties. They are larger for the reconstruction with the EPOS-LHC model due to the smaller size of the MC sample used in this case within the unfolding procedure.}
  	\label{eposlhcreco}
    \end{figure*}

      \textit{Hadronic interaction model.}
    To evaluate the influence of the uncertainties in the physics of the high energy hadronic collisions, we produced a small set of MC simulations using EPOS-LHC \cite{eposlhc} and repeated the analysis procedure. Following Sec. \ref{MCsection}, we generated $4 \times 10^{5}$ MC events for each of the following nuclei: H, He and C, and $2 \times 10^{5}$ MC showers for each of the other primaries: O, Ne, Mg, Si and Fe. The mean lateral shower age as computed with EPOS-LHC for some selected cosmic ray primaries is presented in Fig.~\ref{eposlhcreco}, left, as a function of $E_{rec}$ in comparison with the measured data. In this figure, we also display the line that represents the  $s_{He-C}$ cut as calculated with the EPOS model. Fig.~\ref{eposlhcreco} (right) shows the energy spectrum of H$+$He as estimated using HAWC data calibrated with EPOS-LHC compared with
    QGSJET-II-04.
    The magnitude of the spectrum decreases when using the EPOS model. 
    At $\log_{10}(E/\gev) = 3.9$ the error is $\sim -3.7  \%$. The minimum value of the error ($-10.9  \%$) was found at  $\log_{10}(E/\gev) = 4.9$.

\begin{figure*}[!t]
     \includegraphics[width=3.0in]{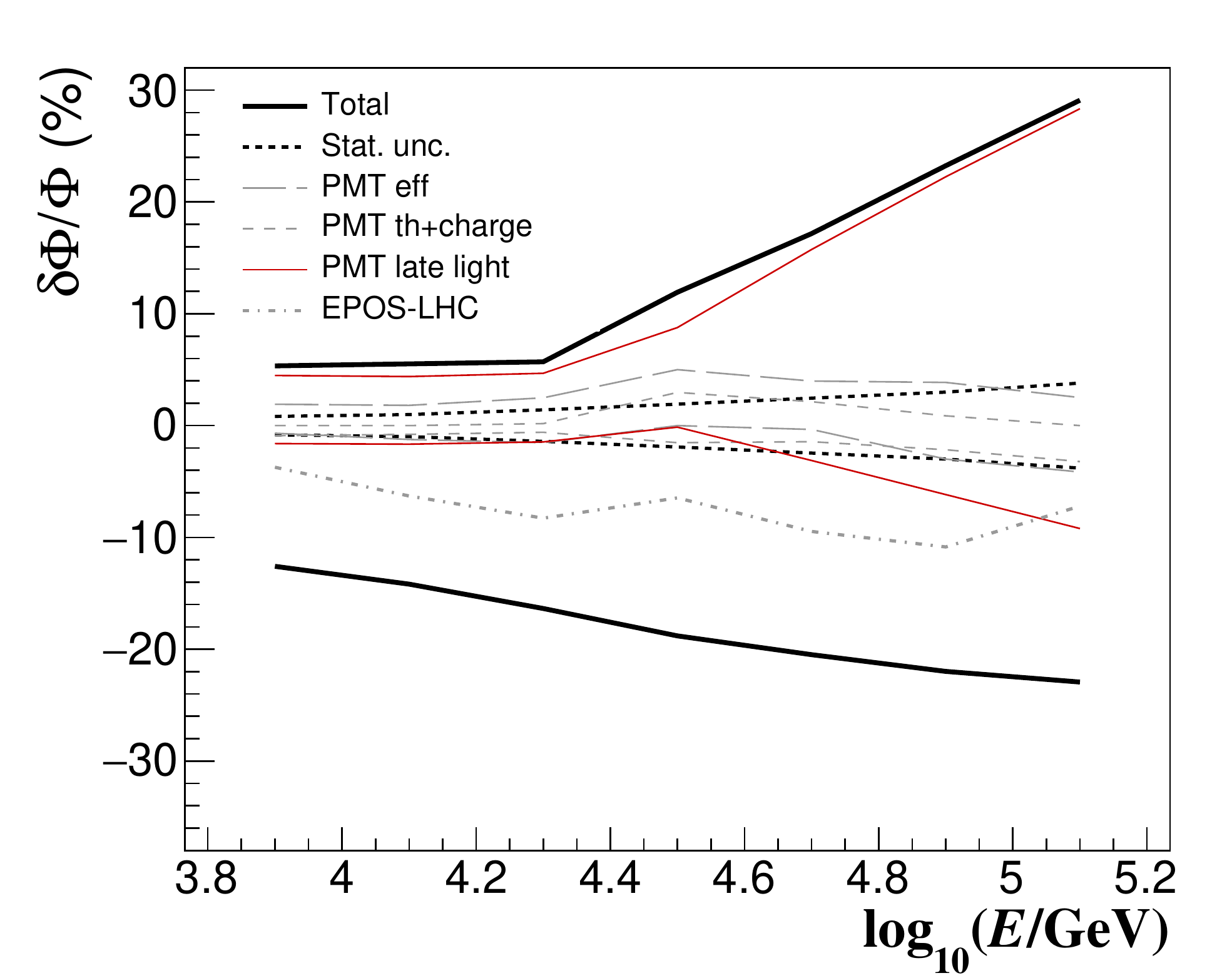} 
     \includegraphics[width=3.0in]{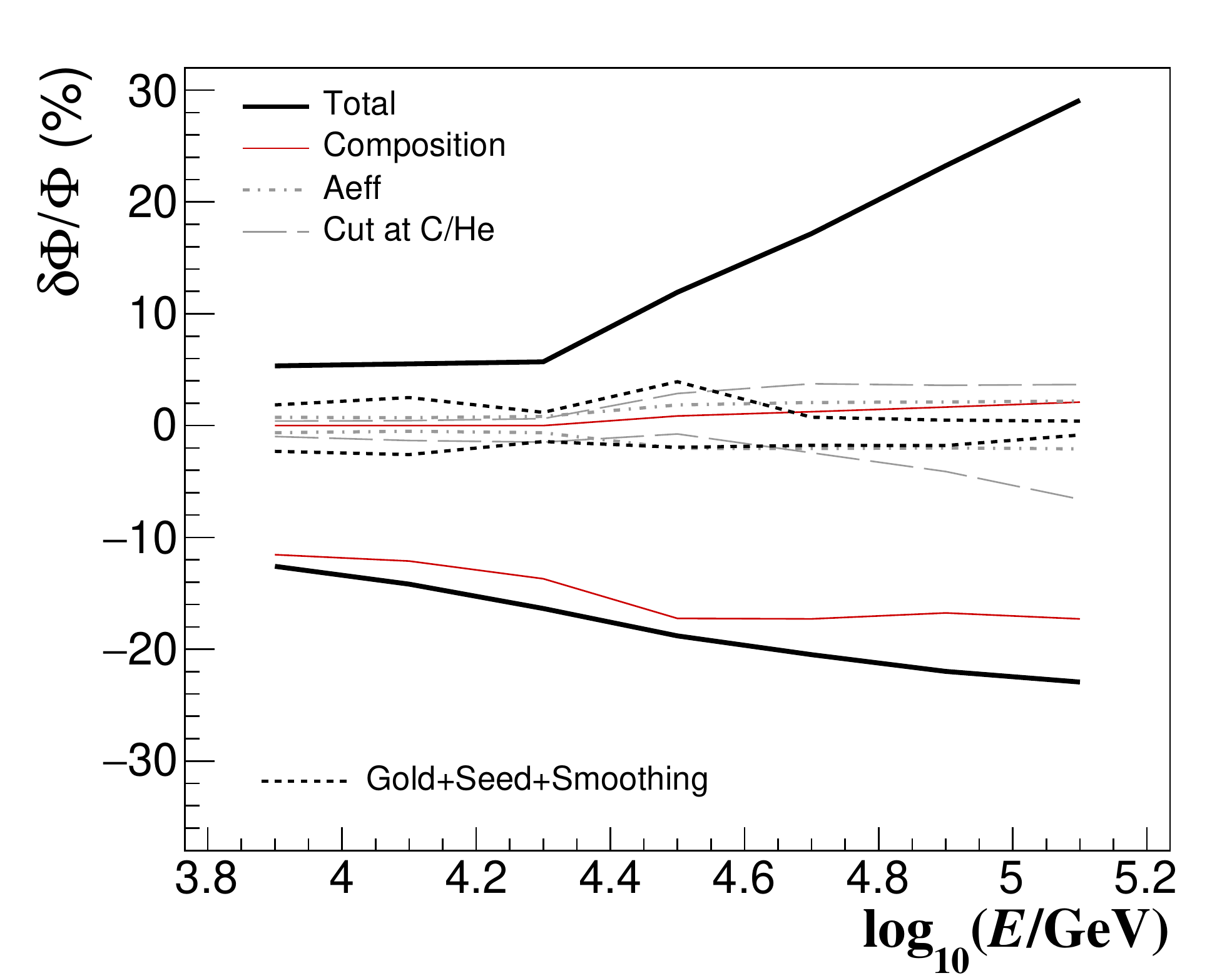} 
    \caption{Energy dependence of the relative 
    statistical and systematic uncertainties of the energy spectrum presented in this work. The total error is calculated as the sum in quadrature of the statistical and systematic uncertainties. For a description of each  uncertainty source, see Appendix \ref{apperrors}. On the plot of the right, systematic uncertainties  due to  smoothing in unfolding procedure, seed for the unfolding method and unfolding algorithm were added in quadrature and are shown as a single contribution with a dotted line.}
  	\label{spectrum_error2}
    \end{figure*}

\begin{figure*}[!t]
     \includegraphics[width=3.0in]{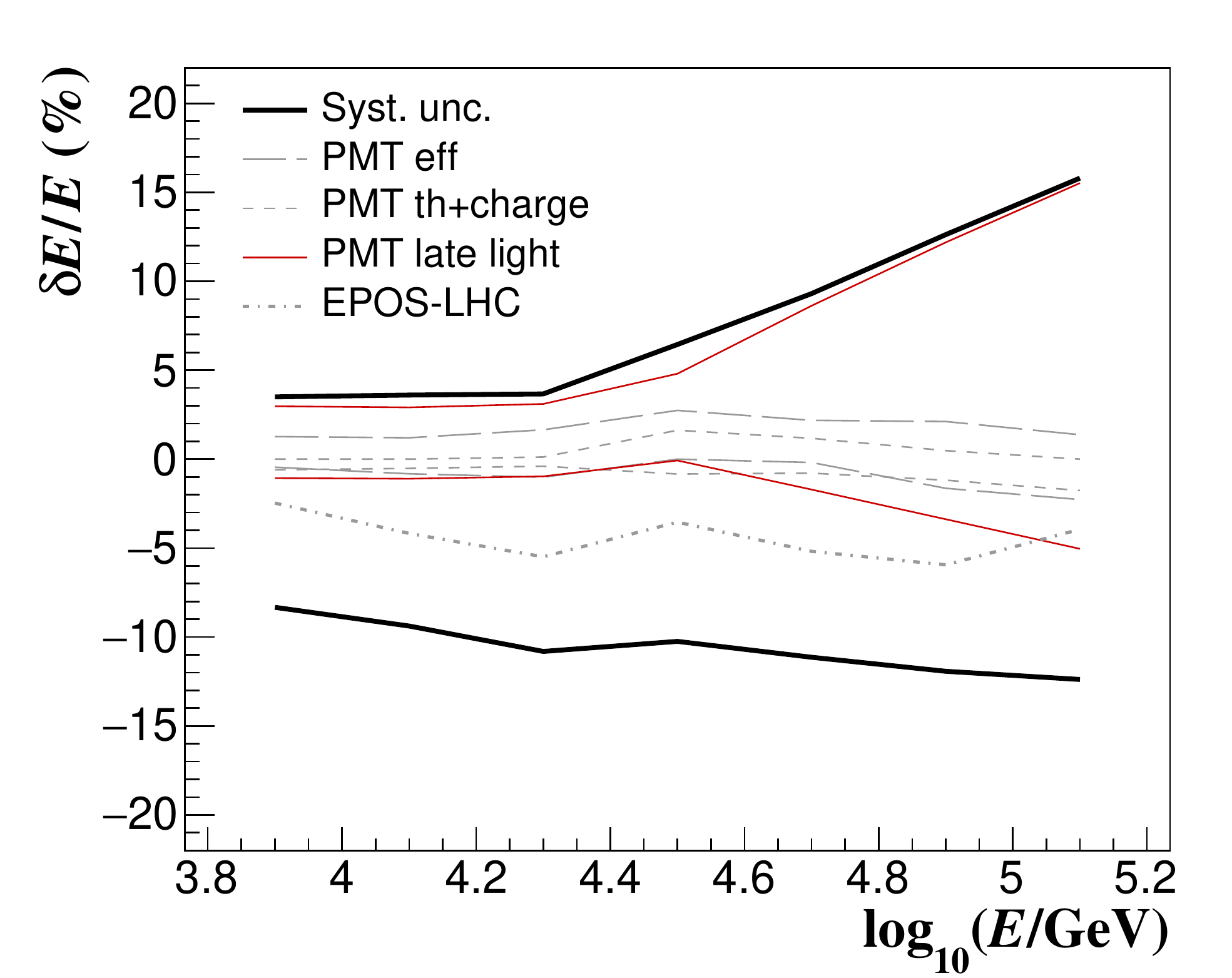} 
     \includegraphics[width=3.0in]{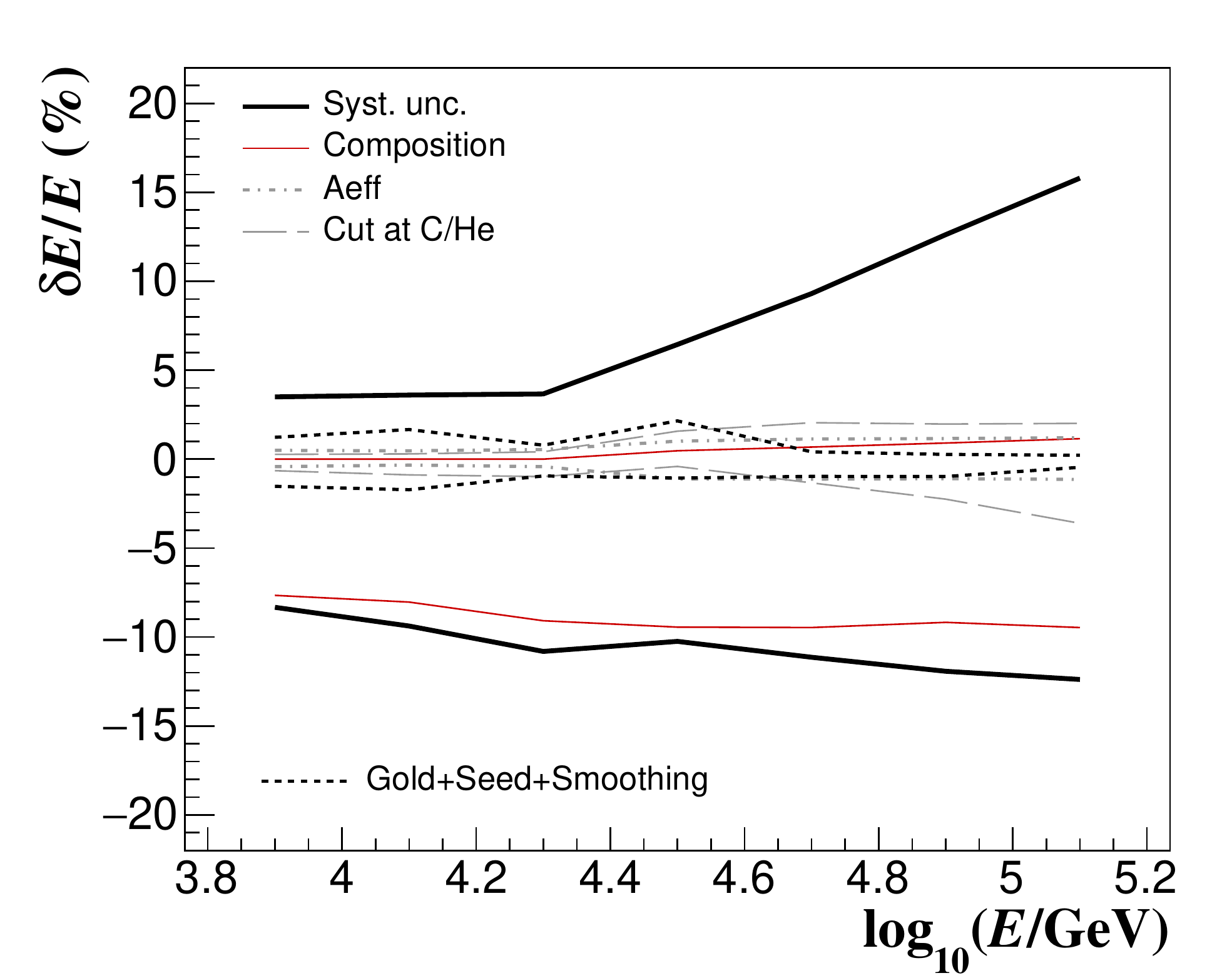} 
    \caption{The total systematic uncertainties on the energy scale against the primary energy. The results are shown along with the contributions from the different systematic sources listed in  Appendix \ref{apperrors}. }
  	\label{energy_systematic_error_detailed}
    \end{figure*}

  Finally, Fig.~\ref{spectrum_error2} 
  shows the relative uncertainty 
  of different sources of statistical and systematic errors as a function of the primary energy, along with the 
  total fractional uncertainty. 
  The dominant contributions to the error come from the uncertainty in the PMT-late-light simulation, the hadronic interaction model and uncertainties in the cosmic ray composition. To end this section, we have estimated the systematic uncertainties in the energy scale associated with the different systematic sources listed in this section. We have employed formula $\delta \Phi/\Phi = -(\gamma + 1) \delta E/E$ \cite{PAO2021}, where $\gamma$ is the local value of the spectral index in the energy spectrum as obtained from the fit to the data with Eq.~(\ref{eq10}). The results are shown in the plots of Fig.~ \ref{energy_systematic_error_detailed}.
  
  \section{Systematic checks}
  \label{appchecks} 
  We performed several checks to see whether systematic effects could produce the observed change in slope in our H $+$ He spectrum.

\begin{figure*}[!b]
     \includegraphics[width=3.0in]{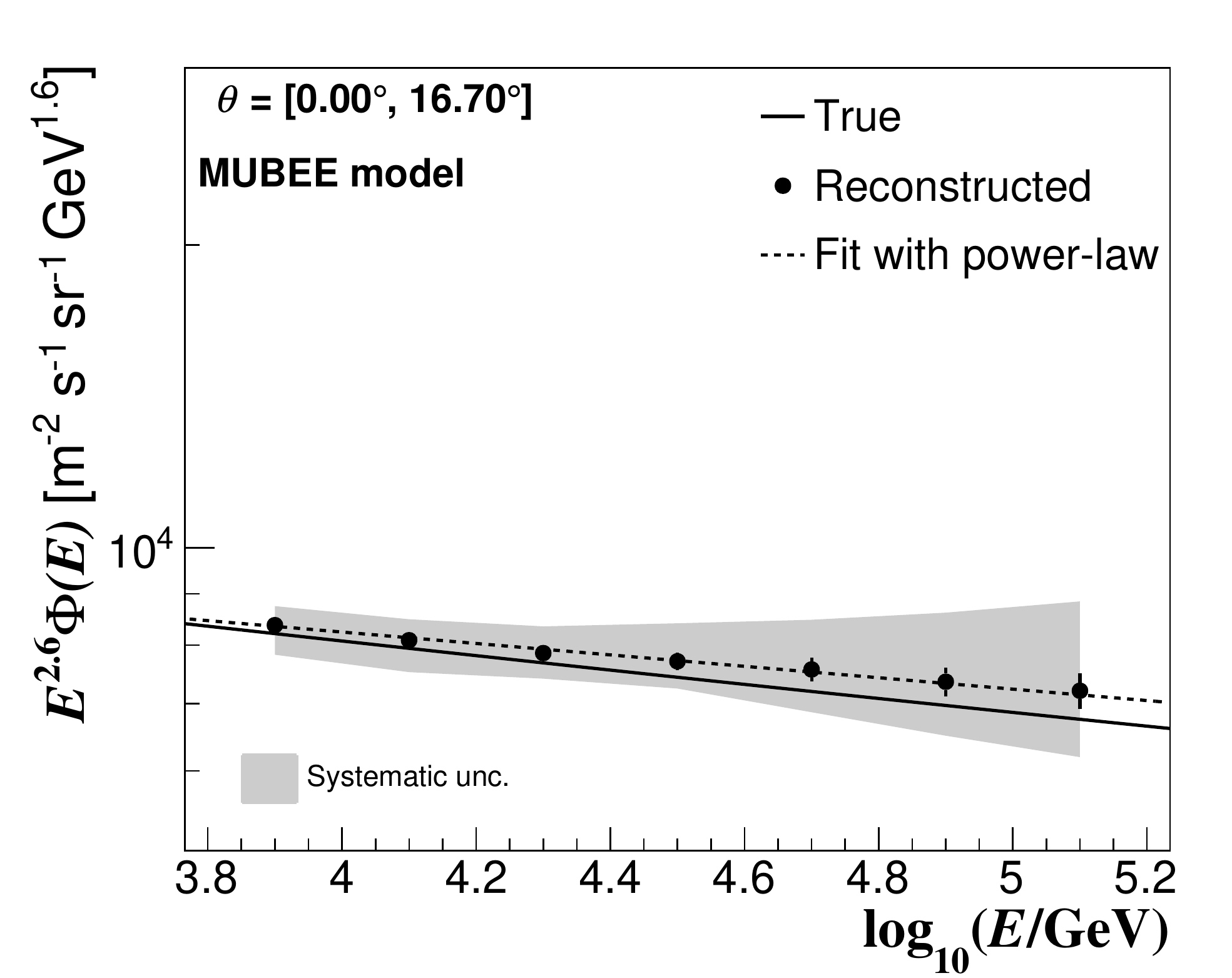} 
     \includegraphics[width=3.0in]{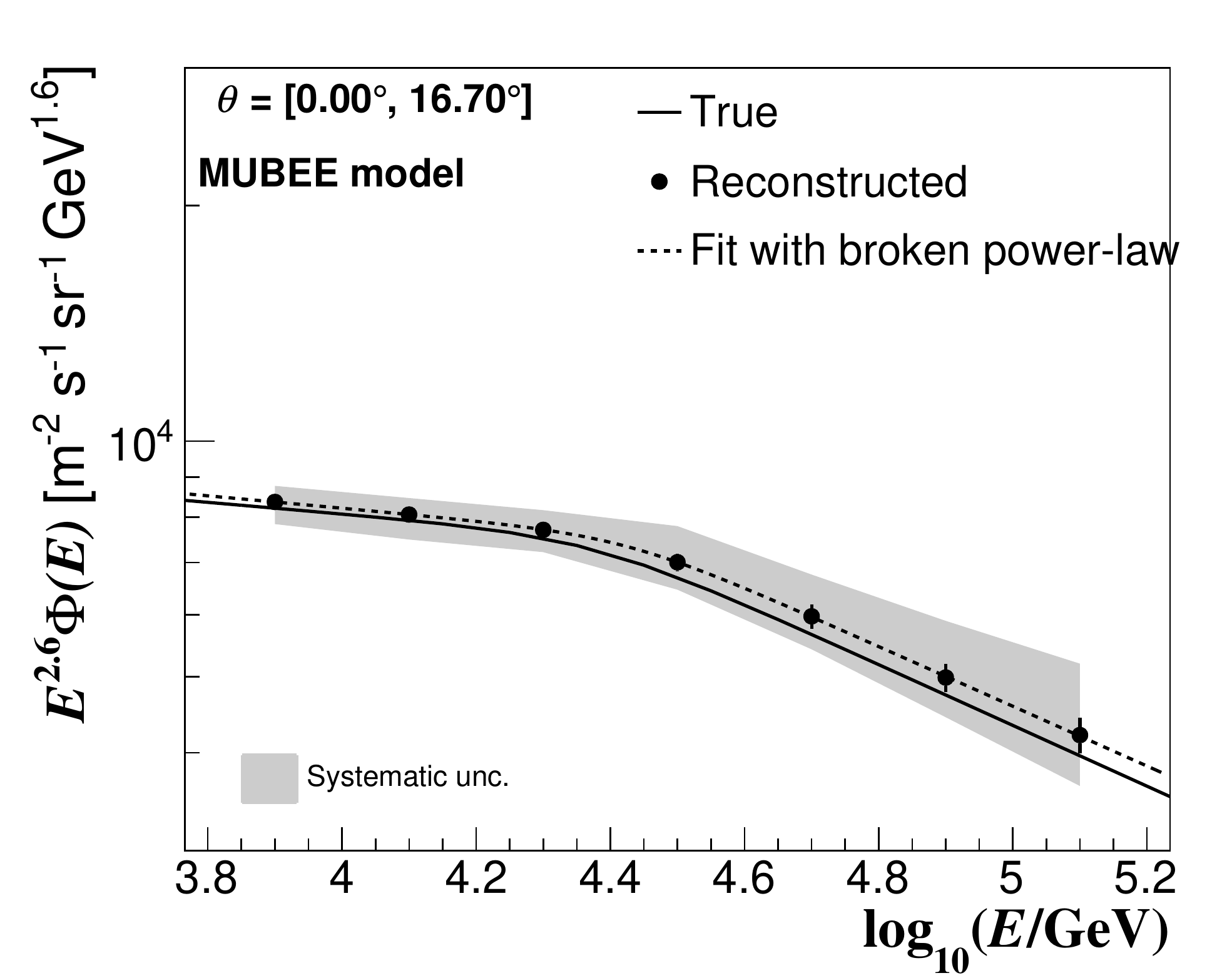} 
    \caption{\textit{Left:} the cosmic ray energy spectrum for protons plus helium primaries according to the MUBEE composition model as reconstructed with the unfolding technique described in this work (black data points). The gray error band and the vertical error bars represent the systematic and statistical uncertainties, respectively. The systematic error due to the hadronic interaction model was not included. The true energy spectrum is shown with a continuous line, while the result of the fit with a power-law formula like Eq.~(\ref{eq9}) to the reconstructed spectrum is shown with a short-dashed line. The true spectral index of the spectrum is $\gamma = -2.67$, while the fitted one is $\gamma = -2.66 \pm 0.01$. \textit{Right:} the reconstructed energy spectrum of light primaries (black circles) for the same model as before, but with the difference that the true spectrum of the light component (continuous line) has a break at $E_{c} = 24  \, \tev$ with $\Delta \gamma_c = -0.32$. The fit with the broken power-law function of Eq.~(\ref{eq10}) (represented with a short-dashed line) gave a smaller value for the change of spectral index ($\Delta \gamma = -0.30 \pm 0.06$) due to the contamination of heavy primaries, which is larger in this model than in the nominal one at high energies. }
    \label{spectrumMCtest}
  \end{figure*}

  First, we tested the reliability of the reconstruction method by applying it to MC simulations produced with QGSJET-II-04, which we treated as fake experimental data. The idea behind this test was to check that the analysis procedure does not introduce artificial features in the spectrum under study and that the reconstructed spectrum of light primaries is in agreement with the true one within systematic uncertainties. As input data, we have used the different composition models of Appendix \ref{appmodels}, which predict distinct $\Phi_{H+He}/\Phi_{Z\geq 3}$ ratios. For each test, we used two alternative spectra of H$+$He: a single power-law formula or a broken power-law behavior, the latter with the same change in spectral index and position of the spectral feature as observed in the  measured spectrum.  In all cases, we found that the shape of the studied spectrum did not have any dramatic modifications due to the analysis method (see, for instance, Fig.~\ref{spectrumMCtest}) and that the reconstructed spectrum agrees with the true one within the systematic errors. In each test, we quantified  the systematic uncertainties listed in Appendix \ref{apperrors}, but the one corresponding to the hadronic interaction model.

  On the other hand, it is interesting to point out that the correction factor $f_{corr}$ has a strong feature at $\log_{10}(E/\gev)=4.5$, close to the location of the break in the measured spectrum of H$+$He. The factor  $f_{corr}$ does not contribute to the formation of the cutoff in the spectrum, on the contrary it tends to flatten the feature, which appears already in the unfolded distribution. If we assume that the feature in the spectrum is due to an underestimation of the relative abundance of the heavy mass group and, thereby, of $f_{corr}$ by the cosmic ray composition models, we would need an enormous amount of heavy nuclei in the models  around $\log_{10}(E/\gev)=4.5$  because of the small fraction of heavy primaries in the selected data subset. This particular overabundance of heavy primaries would be in contradiction with direct measurements and with HAWC data on the shower age. Hence, this possibility is discarded. In consequence, the origin of the break in the spectrum of light primaries is not connected to the feature of the correction factor observed at $\log_{10}(E/\gev)=4.5$.  
    
  We also performed other systematic checks, which are not included as systematic errors, but that are important to discard that the observed feature in the measured spectrum is induced by the reconstruction method. 
 
  We started by studying the possibility that an unknown spectral break in the intensity for heavy primaries produces the above mentioned feature.  To rule out that scenario, we introduced a cut at $E_{c} = 24  \, \tev$ with a change of spectral index $\Delta \gamma_c = -0.32$ in the spectrum of the heavy component of our MC composition models and kept the single power-law behavior for the intensity of light elements unchanged. The unfolded results produced a small bump in the light spectrum, but with a $\Delta \gamma$ too small ($\lesssim -0.03$) to explain the observations. A sharp cut at $E_{c}$  has been discarded as it would be in contradiction with the measured shower age distributions but when investigated anyway,  
  we observed a bump in the light mass group spectrum with $\Delta \gamma \lesssim -0.01$. A recovery in the intensity of the heavy component of cosmic rays at $E_{c} = 24  \, \tev$ was also investigated. We found that it can neither 
  explain the observations, as it would produce a positive change in  $\Delta \gamma$ in the spectrum of light primaries in contradiction with the HAWC measurement.

  We also  ruled out systematic effects from either the calibration of large induced signals or the cut $Q_\mathrm{eff} \gtrsim 10^{4} \, PE$  as the reason behind the slope change.
  That conclusion was achieved by reconstructing the spectrum of light primaries for inclined air showers (in particular, with $\theta$ close to $45^{\circ}$) and by observing that the break in the spectrum is still present for events with large zenith angles. These EAS suffer a stronger attenuation in the atmosphere and, in consequence, have lower $Q_\mathrm{eff}$ values than vertical EAS. Thereby the calibration errors are expected to be smaller and so decrease the effects from the $Q_\mathrm{eff}$ cut on the composition analysis. This point offers the possibility of extending HAWC studies on the composition of cosmic rays up to $1 \, \pev$ using inclined EAS.  A complete study with inclined air shower events is in progress and will be presented in an upcoming paper. 
 
  \begin{figure}[!b]
  \centering
  \includegraphics[width=3.3in]{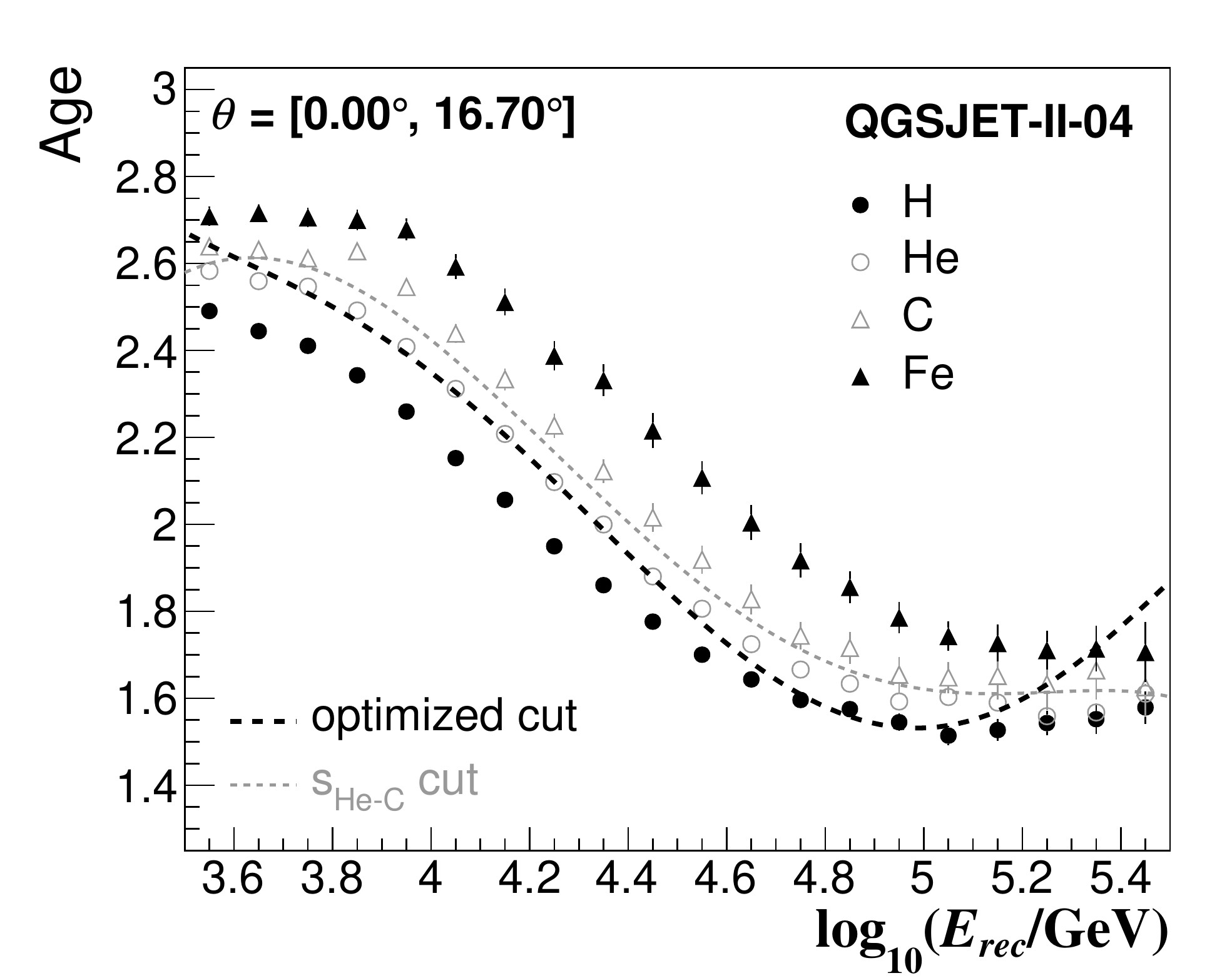}
  \caption{Predictions of the QGSJET-II-04 model for the energy dependence of the mean lateral age  in vertical air showers initiated by four cosmic ray species at HAWC. From top to bottom, the MC points correspond to Fe (solid triangles), C (hollowed triangles), He (hollowed circles) and H (solid circles) primaries, respectively. For clarity, not all the elemental nuclei simulated in this work were included in the plot.  The standard cut $s_{He-C}$ employed to extract the enriched subsample of light nuclei used for the main analysis of the paper is plotted using a gray dotted line, while the optimized age cut based on the maximization of the $Q$ factor and used in one of the systematic checks of Appendix \ref{appchecks} is shown with a dashed black line.}
  \label{FigoptAge}
  \end{figure}
  
      \begin{figure}[!t]
     \centering
     \includegraphics[width=3.3in]{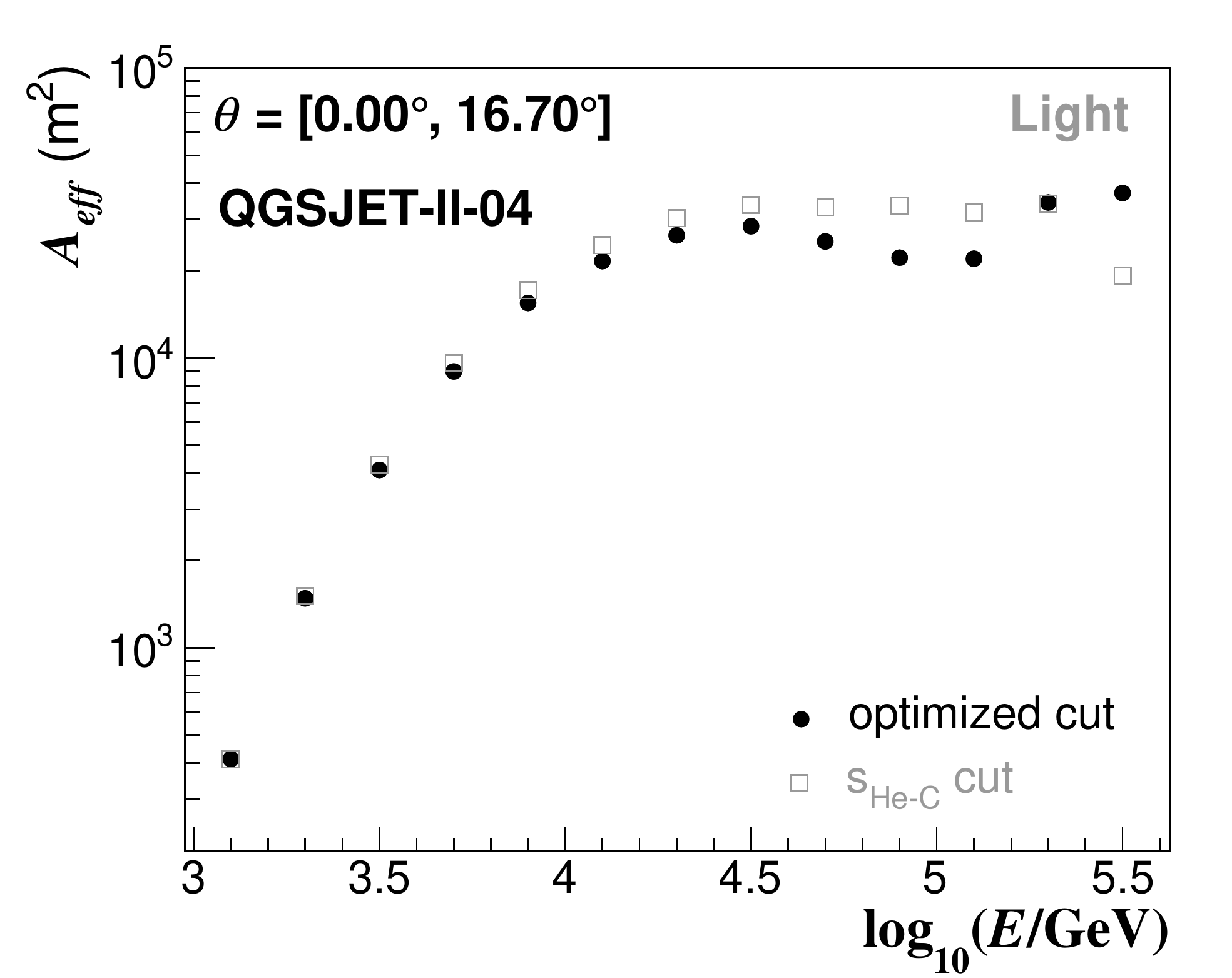}
    \caption{ The corrected effective area for light primaries estimated with Eq.~(\ref{eq7}) for HAWC using  QGSJET-II-04  simulations and the optimized age cut based on the maximization of the $Q$ factor (black data points). For comparison, also the corrected effective area for light primaries obtained after applying the standard age cut $s_{He-C}$ is displayed (gray open squares). Statistical  uncertainties  are smaller than the size of the data points.}
  	\label{Effareaopt}
    \end{figure}

   \begin{figure}[!b]
    \centering
    \includegraphics[width=3.3in]{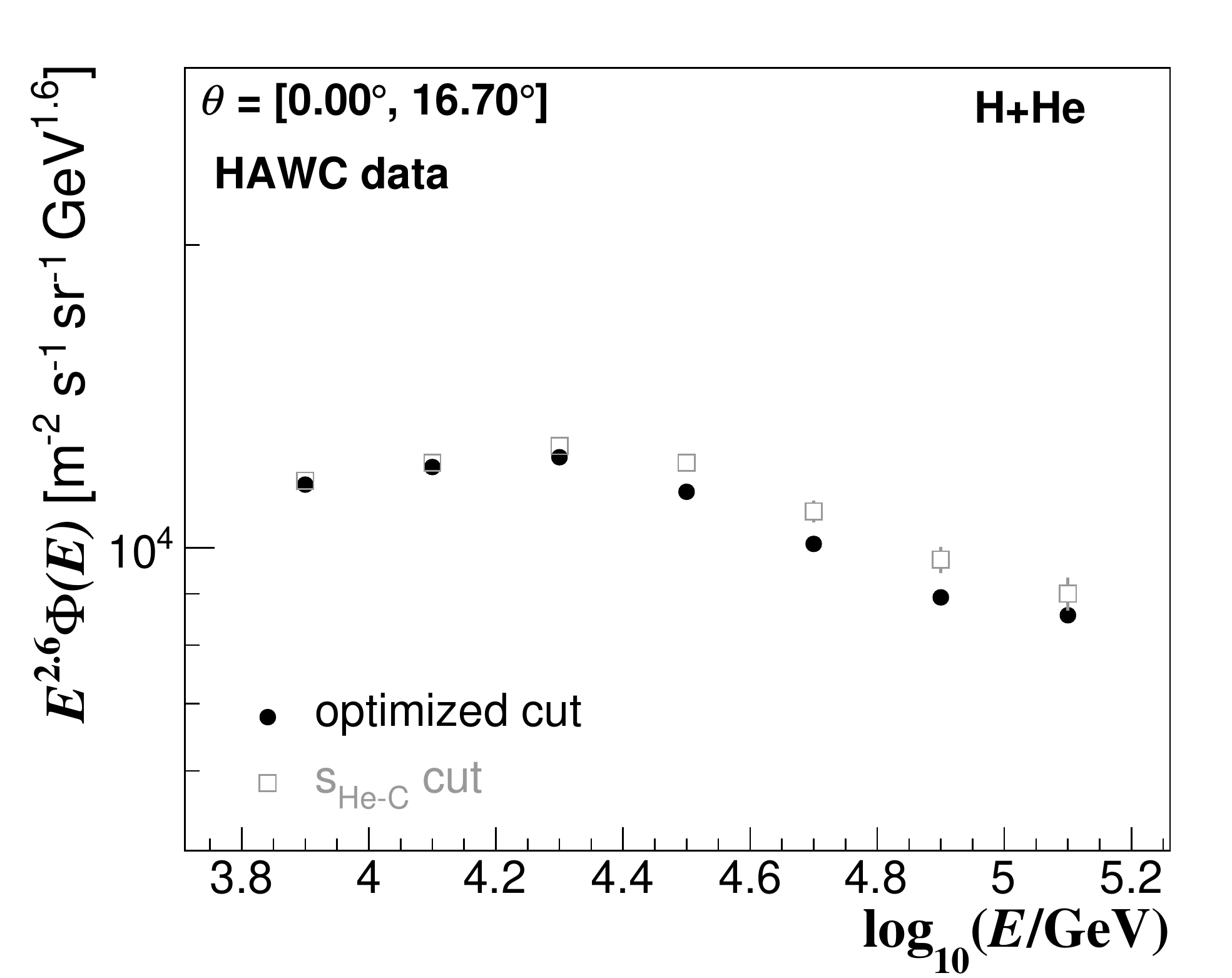}
    \caption{The energy spectrum of protons plus helium cosmic ray nuclei as measured with HAWC and reconstructed within the framework of QGSJET-II-04 using a subset of data selected with the standard age cut of Fig.~\ref{FigAge} (gray open squares) and another experimental subsample obtained by applying the optimized cut of Fig.~\ref{FigoptAge} (black circles). Vertical error bars represent the statistical uncertainties. }
    \label{spectrumopt}
  \end{figure}

    As a systematic check, we have also reconstructed the energy spectrum for H+He primaries from a data subset obtained by applying an age cut derived from the maximization of the purity of the subsample and by using the unfolding procedure describe in Sec. \ref{Analysis}. For maximizing the purity of the data subset, we define the $Q$ factor. This parameter is just the ratio between the number of remaining events of light nuclei (signal) obtained after applying the cut to the squared root of the number of heavy primaries (background) that passed the age selection. Then we found the maximum value of the $Q$ factor for each bin of reconstructed energy with the constraint that we keep a retention factor of at least  $50 \%$ for H and  $50 \%$ for He in each bin. The optimized shower age cut obtained with this procedure is presented in Fig.~ \ref{FigoptAge} in comparison with the selection cut $s_{He-C}$ used in our standard analysis and with the QGSJET-II-04 predictions for the mean shower age of different  primary nuclei. From Fig.~ \ref{FigoptAge}, we observed that the optimized cut and the standard one have a similar tendency up to $10^5 \, \gev$, but at higher energies they have distinct behaviors. With the optimized cut the purity of the subsample for $\log_{10}(E_{rec}/\gev) \leq 5.2$ is larger than the one achieved with the standard age cut. For the optimized selection, it is $> 88.5 \%$, while for the standard cut is $> 86 \%$. At higher energies, the optimized age cut has a poorer performance. In this case,  the purity of the selected data subset can decrease up to a value of $78 \%$, while for the standard selection is reduced up to  $82 \%$.

    Using the selected dataset with the optimized cut and the unfolding procedure presented in Sec. \ref{Analysis}, we have obtained the corresponding energy spectrum of light cosmic ray primaries, which is displayed in  Fig.~\ref{spectrumopt} together with the standard result. The corrected effective area employed for the computation of the spectrum with the optimized cut is shown in  Fig.~\ref{Effareaopt}, in comparison with $A_{eff}(E)$ for the standard data subsample. We can see that the effective area corresponding to the optimized cut is not flat. Despite  of that the energy spectrum of H+He has the same shape whether we use the standard age cut or  the optimized one, as we can see from  Fig.~\ref{spectrumopt}. Below   $20 \, \tev$, both spectra are in agreement.  However, we observe that  at higher energies the spectrum with the optimized cut is  softer than the original one and  that its intensity is smaller by at most $8  \%$.

   In addition, we have also investigated the impact of seasonal variations in the data. For this analysis, we have divided our data in four subsets corresponding to different periods of the year, which cover the following months: March-May, June-August, September-November and  December-February. For each period, we reconstructed the energy spectra of H$+$He and compared them with our nominal result. We found negligible variations with regard to the spectrum of Fig.~\ref{spectrum}. At low energies [$\log_{10}(E/\gev)=4.9$], the changes in the cosmic ray intensity due to seasonal variations are of the order of $+0.7\%/$$-0.5\%$. They increase with the primary energy and reach values up to $+1.26\%/$$-0.85\%$ for the bin $\log_{10}(E/\gev)=5.1$.
    
   We also studied the influence of the uncertainties in the EAS core position on the energy spectrum by using only cores reconstructed inside the array; again the  slope change did not disappear. Finally, we  investigated the combined effect of fluctuations in the signals at the lateral distribution of EAS along with the effects of shower core resolution, but that is not able neither to explain the presence of the spectral break.

\providecommand{\noopsort}[1]{}\providecommand{\singleletter}[1]{#1}%


\begin{thebibliography}{10}

\bibitem{Hawc17b}
A.~U. Abeysekara et~al.
\newblock {\em Astrophys. J.}, 843:39, 2017.

\bibitem{Hawc18}
A.~U. Abeysekara et~al.
\newblock {\em Astrophys. J.}, 865:57, 2018.

\bibitem{Hawcdaq18}
A.~U. Abeysekara et~al.
\newblock {\em Nucl. Instrum. Methods A}, 888:138, 2018.

\bibitem{Hawccrab19}
A.~U. Abeysekara et~al.
\newblock {\em Astrophys. J.}, 881:134, 2019.

\bibitem{HAWC_2013}
A.U. Abeysekara et~al.
\newblock {\em Astrop. Phys.}, 50-52:26--32, 2013.

\bibitem{PAO2021}
P.~Abreu et~al.
\newblock {\em Eur. Phys. J. C}, 81:966, 2021.

\bibitem{pamela}
O.~Adriani et~al.
\newblock {\em Science}, 332:69, 2011.

\bibitem{calet19}
O.~Adriani et~al.
\newblock {\em Phys. Rev. Lett.}, 122:181102, 2019.

\bibitem{Adye11a}
T.~Adye.
\newblock Corrected error calculation for iterative bayesian unfolding.
\newblock \url{http://hepunx.rl.ac.uk/~adye/software/unfold/bayes_errors.pdf},
  2011.

\bibitem{Eastop04}
M.~Aglietta et~al.
\newblock {\em Astrop. Phys.}, 21:223, 2004.

\bibitem{Geant4}
S.~Agostinelli et~al.
\newblock {\em Nucl. Instrum. Methods A}, 506:250, 2003.

\bibitem{ams14}
M.~Aguilar et~al.
\newblock {\em Phys. Rev. Lett.}, 114:171103, 2015.

\bibitem{ams15}
M.~Aguilar et~al.
\newblock {\em Phys. Rev. Lett.}, 115:211101, 2015.

\bibitem{cream09}
H.~S. Ahn et~al.
\newblock {\em Astrophys. J.}, 707:593, 2009.

\bibitem{cream10}
H.~S. Ahn et~al.
\newblock {\em Astrophys. J. Lett.}, 714:L89, 2010.

\bibitem{dampe21}
F.~Alemanno et~al.
\newblock {\em Phys. Rev. Lett.}, 126:201102, 2021.

\bibitem{Hawc17}
R.~Alfaro et~al.
\newblock {\em Phys. Rev. D}, 96:122001, 2017.

\bibitem{Tibetasgamma19}
M.~Amenomori et~al.
\newblock {\em EPJ Web of Conferences}, 208:03001, 2019.

\bibitem{dampe19}
Q.~An et~al.
\newblock {\em Science Advances}, 5, No. 9:eaax3793, 2019.

\bibitem{Antoni05}
T.~Antoni et~al.
\newblock {\em Astropart. Phys.}, 24:1, 2005.

\bibitem{Kascade06}
W.~D. Apel et~al.
\newblock {\em Astropart. Phys.}, 24:467, 2006.

\bibitem{Hawc19}
J.~C. Arteaga-Vel\'azquez et~al.
\newblock In {\em Proceedings of the 36th ICRC (Madison, U.S.A), PoS (ICRC2019)
  176}, 2019.

\bibitem{HAWC_CR_2021}
J.C. Arteaga-Velázquez et~al.
\newblock In {\em Proceedings of the 37th ICRC (Berlin, Germany), PoS
  (ICRC2021) 374}, 2021.

\bibitem{nucleon}
E.~V. Atkin et~al.
\newblock {\em JCAP}, 1707:020, 2017.

\bibitem{nucleon2}
E.~V. Atkin et~al.
\newblock {\em JETP Lett.}, 108:5, 2018.

\bibitem{nucleon19}
E.~V. Atkin et~al.
\newblock {\em Astron. Rep.}, 63:66, 2019.

\bibitem{Baade34}
W.~Baade and F.~Zwicky.
\newblock {\em Phys. Rev.}, 46:76, 1934.

\bibitem{lhaaso19}
X.~Bai et~al.
\newblock e-print arXiv:astro-ph.HE/1905.02773, 2019.

\bibitem{Bartoli15}
B.~Bartoli et~al.
\newblock {\em Phys. Rev. D}, 92:092005, 2015.

\bibitem{Argo15}
B.~Bartoli et~al.
\newblock {\em Phys. Rev. D}, 91:112017, 2015.

\bibitem{PDG19}
J.~J. Beatty, J.~Matthews, and S.~P. Wakely.
\newblock {\em Cosmic Rays, in M. Tanabashi et al. (Particle Data Group), Phys.
  Rev. D}, 98:030001, 2018.

\bibitem{Bell1978}
A.~R. Bell.
\newblock {\em Mon. Not. R. Astron. Soc.}, 182:147, 1978.

\bibitem{Bell13}
A.~R. Bell.
\newblock {\em Astropart. Phys.}, 43:56, 2013.

\bibitem{Blandford1978}
R.~D. Blandford and J.~P. Ostriker.
\newblock {\em Astrophys. J.}, 221:L29, 1978.

\bibitem{Blumer09}
J.~Bluemer, R.~Engel, and J.~R. Hoerandel.
\newblock {\em Prog. Part. Nucl. Phys.}, 63:293, 2009.

\bibitem{root}
R.~Brun and F.~Rademakers.
\newblock {\em Nucl. Instrum. Methods A}, 389:81, 1997.

\bibitem{Cesar1980}
C.J. Cesarsky.
\newblock {\em Ann. Rev. Astron. Astrophys.}, 18:289, 1980.

\bibitem{Cowen98}
G.~Cowan.
\newblock {\em Statistical Data Analysis}.
\newblock Clarendon Press, 1998.

\bibitem{PDG19stat}
G.~Cowan.
\newblock {\em Statistics, in M. Tanabashi et al. (Particle Data Group), Phys.
  Rev. D}, 98:030001, 2018.

\bibitem{PDG19MC}
G.~Cowan.
\newblock {\em Monte Carlo techniques, in M. Tanabashi et al. (Particle Data
  Group), Phys. Rev. D}, 98:030001, 2018.

\bibitem{agostini}
G.~D'Agostini.
\newblock {\em Nucl. Instrum. Methods A}, 362(2-3):487, 1995.

\bibitem{Fluka}
A.~Ferrari, P.~R. Sala, A.~Fass\`o, and J.~Ranft.
\newblock Fluka: a multi-particle transport code.
\newblock Report INFN/TC\_05/11, SLAC-R-773, CERN-2005-10, 2005.

\bibitem{smoothing}
J.~Friedman.
\newblock In {\em Proceedings of the 3rd CERN School of Computing (Norway)
  (CERN, Geneva, 1974)}, page 271, 1974.

\bibitem{Fuhrmann12}
D.~Fuehrmann.
\newblock {\em KASCADE-Grande Measurements of Energy Spectra for Elemental
  Groups of Cosmic Rays}.
\newblock {Ph.D.} thesis, University of Wuppertal, 2012.

\bibitem{DanielKG2013}
D.~Fuehrmann et~al.
\newblock e-print arXiv:astro-ph.HE/1309.4295, 2013.

\bibitem{Gaisser17}
T.~Gaisser.
\newblock {\em EPJ Web of Conferences}, 145:18003, 2017.

\bibitem{Giacinti15}
G.~Giacinti, M.~Kachelriess, and D.~V. Semikoz.
\newblock {\em Phys. Rev. D}, 91:083009, 2015.

\bibitem{Gold64}
R.~Gold.
\newblock An iterative unfoding method for response matrices.
\newblock Report ANL-6984, Argonne National Laboratory, USA, 1964.

\bibitem{NKG3}
K.~Greisen.
\newblock The extensive air showers.
\newblock In J.G. Wilson, editor, {\em Progress inCosmic Ray Physics}, volume
  III, page~1. North Holland, Amsterdam, 1956.

\bibitem{Zig17}
Z.~Hampel-Arias.
\newblock {\em Cosmic Ray Observations at the TeV Scale with the HAWC
  Observatory}.
\newblock {Ph.D.} thesis, University of Wisconsin-Madison, 2017.

\bibitem{Haungs03}
A.~Haungs, R.~Heinigerd, and R.~Markus.
\newblock {\em Rep. Prog. Phys.}, 66:1145, 2003.

\bibitem{Heck:1998vt}
D.~Heck, J.~Knapp, J.~N. Capdevielle, G.~Schatz, and T.~Thouw.
\newblock Corsika: A monte carlo code to simulate extensive air showers.
\newblock Report FZKA 6019, Forschungszentrum Karlsruhe-Wissenhaltliche
  Berichte, Karlsruhe, Germany, 1998.

\bibitem{Polygonato}
J.~R. Hoerandel.
\newblock {\em Astropart. Phys.}, 19:193, 2003.

\bibitem{Hoerandel04}
J.~R. Hoerandel.
\newblock {\em Astropart. Phys.}, 21:241, 2004.

\bibitem{Kachelriess15}
M.~Kachelriess, A.~Neronov, and D.~V. Semikoz.
\newblock {\em Phys. Rev. Lett.}, 115:181103, 2015.

\bibitem{Kachelriess18}
M.~Kachelriess, A.~Neronov, and D.~V. Semikoz.
\newblock {\em Phys. Rev. D}, 97:063011, 2018.

\bibitem{kachelriess19}
M.~Kachelriess and D.~V. Semikoz.
\newblock {\em Prog. Part. Nucl. Phys.}, 109:103710, 2019.

\bibitem{NKG2}
K.~Kamata and J.~Nishimura.
\newblock {\em Prog. Theor. Phys. Suppl.}, 6:93, 1958.

\bibitem{Krymsky1977}
G.~F. Krymsky.
\newblock {\em Dokl. Akad. Nauk. SSSR}, 234:1306, 1977.

\bibitem{Liu19}
W.~Liu, Y.~Q. Guo, and Q.~Yuan.
\newblock {\em JCAP}, 10:010, 2019.

\bibitem{lucy}
L.~Lucy.
\newblock {\em Astron. J.}, 79:745, 1974.

\bibitem{Maestro15}
P.~Maestro.
\newblock In {\em Proceedings of the 34th ICRC (the Hague, Netherlands), PoS
  (ICRC2015) 016}, 2015.

\bibitem{Dampe19a}
Ivan~De Mitri et~al.
\newblock In {\em Proceedings of the 36th ICRC (Madison, U.S.A), PoS (ICRC2019)
  148}, 2019.

\bibitem{Roulet17}
S.~Mollerach and E.~Roulet.
\newblock {\em Prog. Part. Nucl. Phys.}, 98:85, 2018.

\bibitem{Hawcldf19}
J.~A. Morales-Soto et~al.
\newblock In {\em Proceedings of the 36th ICRC (Madison, U.S.A), PoS (ICRC2019)
  359}, 2019.

\bibitem{NKG1}
J.~Nishimura.
\newblock Theory of cascade showers.
\newblock In K.~Sitte, editor, {\em Handbuch der Physik}, volume 9/46/2, pages
  1--114. Springer, Berlin, Heidelberg, 1967.

\bibitem{qgsjetii4}
S.~Ostapchenko.
\newblock {\em Phys. Rev. D}, 83:014018, 2011.

\bibitem{atic07}
A.~D. Panov et~al.
\newblock {\em Bull. Russ. Acad. Sci. Phys.}, 71:494, 2007.

\bibitem{atic09}
A.~D. Panov et~al.
\newblock {\em Bull. Russ. Acad. Sci. Phys.}, 73, No. 5:564, 2009.

\bibitem{atic092}
A.~D. Panov et~al.
\newblock {\em Izv. Ross. Akad. Nauk Ser. Fiz.}, 73:602, 2009.

\bibitem{Peters61}
B.~Peters.
\newblock {\em Il Nuovo Cimento}, XXII:800, 1961.

\bibitem{Picozza18}
P.~Picozza, P.~Spillantini, and L.~Marcellia.
\newblock {\em Nucl. Part. Phys. Proc.}, 297-299:207, 2018.

\bibitem{eposlhc}
T.~Pierog, I.~Karpenko, J.~M. Katzy, E.~Yatsenko, and K.~Werner.
\newblock {\em Phys. Rev. C}, 92:034906, 2015.

\bibitem{taiga19}
V.~V. Prosin et~al.
\newblock {\em Bull. Russ. Acad. Sci. Phys.}, 83, No. 8:1016, 2019.

\bibitem{Ptuskin13}
V.~Ptuskin, V.~Zirakashvili, and E.~S. Seo.
\newblock {\em Astrophys. J.}, 763:47, 2013.

\bibitem{Ptuskin1993}
V.~S. Ptuskin et~al.
\newblock {\em Astron. Astroph.}, 268:726, 1993.

\bibitem{Ptuskin10}
V.~S. Ptuskin, V.~N. Zirakashvili, and E.~S. Seo.
\newblock {\em Astrophys. J.}, 718:31, 2010.

\bibitem{Rice10}
John~A. Rice.
\newblock {\em Mathematical Statistics and Data Analysis}.
\newblock Thomson Brooks/Cole, Belmont, 3rd edition, 2010.

\bibitem{richardson}
W.~H. Richardson.
\newblock {\em J. Opt. Soc. Am.}, 62:55, 1972.

\bibitem{sibyll23c}
F.~Riehn, R.~Engel, A.~Fedynitch, and T.~K. Gaisser.
\newblock {\em Phys. Rev. D}, 102:063002, 2020.

\bibitem{Savchenko15}
V.~Savchenko, M.~Kachelriess, and D.~V. Semikoz.
\newblock {\em Astrophys. J. Lett.}, 809:L23, 2015.

\bibitem{Stanev14}
T.~Stanev, T.~K. Gaisser, and S.~Tilav.
\newblock {\em Nucl. Instrum. Methods A}, 742:42, 2014.

\bibitem{Strong97}
A.~W. Strong, I.~V. Moskalenko, and V.~S. Ptuskin.
\newblock {\em Annu. Rev. Nucl. Part. Sci.}, 57:285, 2007.

\bibitem{jacee}
Y.~Takahashi et~al.
\newblock {\em Nucl. Phys. B (Proc. Suppl.)}, 60:83, 1998.

\bibitem{BPL}
S.~V. Ter-Antonyan and L.~S. Haroyan.
\newblock e-print arXiv: arXiv:hep-ex/0003006, 2000.

\bibitem{Thoudam16}
S.~Thoudam, J.~P. Rachen, A.~van Vliet, A.~Achterberg, S.~Buitink, H.~Falcke,
  and J.~R. Hoerandel.
\newblock {\em Astron. And Astrophys.}, 595:A33, 2016.

\bibitem{Ulrich01}
H.~Ulrich et~al.
\newblock In {\em Proceedings of the 27th ICRC (Hamburg, Germany) (Copernicus
  Gesellschaft e.v., Munich, 2001)}, page~97, 2001.

\bibitem{Axford1977}
E.~Leer W.I.~Axford and G.~Skadron.
\newblock The acceleration of cosmic rays by shock waves.
\newblock In {\em Proceedings of the 15th ICRC (Plovdiv, Bulgaria) (Publishing
  House of the Bulgarian Academy of Sciences, Bulgaria)}, volume~11, page 132,
  1977.

\bibitem{cream11}
Y.~S. Yoon et~al.
\newblock {\em Astrophys. J.}, 728:122, 2011.

\bibitem{cream17}
Y.~S. Yoon et~al.
\newblock {\em Astrophys. J.}, 839:5, 2017.

\bibitem{Yue20}
C.~Yue et~al.
\newblock {\em Front. Phys.}, 15:24601, 2020.

\bibitem{mubee}
V.~I. Zatsepin et~al.
\newblock In D.~A. Leahy, R.~B. Hickws, and D.~Venkatesan, editors, {\em
  Proceedings of the 23rd ICRC (Calgary, Canada), Vol. 2}, page~13. World
  Scientific, 1993.

\bibitem{Zatsepin}
V.~I. Zatsepin and N.~V. Sokolskaya.
\newblock {\em Astron. Astrophys.}, 458:1, 2006.

\bibitem{Zatsepin2}
V.~I. Zatsepin and N.~V. Sokolskaya.
\newblock {\em Astron. Lett.}, 33:25, 2007.

\end{thebibliography}
\end{document}